\RequirePackage{color}

\documentclass[11pt,a4paper]{JHEP3}

\usepackage{amssymb,amsmath}
\usepackage{euscript}
\usepackage{slashed}
\usepackage{amsfonts}
\usepackage{verbatim}
\usepackage[latin9]{inputenc}
\usepackage{amsthm}
\def\be{\begin{eqnarray}}
 \def\ee{\end{eqnarray}}
 \def\0{\nonumber}
\def\tr{\rm tr}
\def\Tr{\rm Tr}
\def\det{\rm det}

\def\EW{\EuScript{W}}
\def\EF{\EuScript{F}}\def\EP{\EuScript{P}}
\def\EA{\EuScript{A}}
\def\ER{\EuScript{R}}

\def\mm{{\mathfrak m}}

\def\bfd{{\bf d}}

\def\bfh{{\bf h}}
\def\bfG{{\bf G}}
\def\bfD{{\bf D}}

\def\det{\rm det}
\usepackage{slashed}
\usepackage{verbatim}
\usepackage{latexsym}
\usepackage{euscript}
\newcommand\EE{\EuScript{E}}
\newcommand\ED{\EuScript{D}}
\newcommand\EK{\EuScript{K}}

\newcommand\EJ{\EuScript{J}}
\normalfont\large
\preprint{SISSA/55/2018/FISI\\ {ZTF-EP-18-06}\\{\tt hep-th/1812.05030 }\\ {\it  
new version,  Jan. 2020} }

\title{HS in flat spacetime. YM-like models}

\author{ L.~Bonora$^{a}$, M.~Cvitan$^{b}$, P.~Dominis
Prester$^{c}$, S.~Giaccari$^{d}$, T.~$\bf
\check{S}$temberga$^{b}$
\\\textit{${}^{a}$ International School for Advanced Studies (SISSA),\\Via
Bonomea 265, 34136 Trieste, Italy, and INFN, Sezione di
Trieste\\}%
\textit{${}^{b}$ Department of Physics, Faculty of Science, University
of Zagreb, \\ 
Bijeni\v{c}ka cesta 32, 10000 Zagreb, Croatia\\}%
\textit{${}^{c}$ Department of Physics, University of Rijeka,\\
Radmile Matej\v{c}i\'{c} 2, 51000 Rijeka, Croatia\\
\textit{${}^{d}$ Department of Sciences,
Holon Institute of Technology (HIT),\\
52 Golomb St., Holon 5810201, Israel}
}%

E-mail: \email{bonora@sissa.it}, \email{mcvitan@phy.hr}, 
\email{pprester@phy.uniri.hr}, \email{stefanog@hit.ac.il},
\email{mateo.paulisic@phy.uniri.hr}, 
\email{tstember@phy.hr}}

\abstract{
 We introduce and analyse a few examples of massless
higher spin theories in Minkowski spacetime. They are defined in terms of master
fields, i.e. fields defined in the whole phase space.  More specifically we
introduce the HS YM-like theories in any dimension and HS CS-like ones in any
odd dimension, in both Abelian and non-Abelian cases. These theories are 
invariant under gauge transformations that include ordinary gauge
transformations, diffeomorphisms and HS gauge transformations.
They are not 
at first sight invariant under local Lorentz transformations, but we show how
this 
invariance can be recovered.We explicitly
write down the actions, the eom's as well as the (infinite many) conservation
laws in both HS YM and HS CS cases. Then we focus in particular on the HS YM
models, we illustrate their $L_\infty$ structure and perform their BRST
quantization.  We also introduce HS scalar and fermion master fields and show
that the Higgs
mechanism can be realized also in the case of HS YM theories. Next we start the
discussion of the perturbative approach to quantization by means of Feynman
diagrams. We show that the dependence on the conjugate momentum can be absorbed
in a redefinition of the component fields, the coupling and the coordinates. 
In such a new {\it frozen momentum} framework, we argue that 
only physical states propagate in physical amplitudes and carry out a sample
calculation. We show that the theory, beside the HS gauge symmetry, has a hidden symmetry which can be unfolded in a modified theory with explicit nonlocalities.}

\keywords{}

\begin{document}

\section{Introduction}

It is accepted nowadays that higher spin (HS) theories in dimension larger than
2, except for 3d examples, must involve an infinite number of (local)
fields (for HS theories, see \cite{HSgeneral}). This characteristic, which was
seen in the past as unattractive (to say
the least), may actually be an inevitable feature of any theory with the
ambition of unifying all the forces of nature. It is not yet clear why this is
so, but there exist several hints in this direction. First and foremost
(super)string theory, which is still the most authoritative example, has this
feature. But also the AdS/CFT correspondence has shown that we may well limit
ourselves to a (conformal or quasi-conformal) field theory on the boundary of
AdS, but if we wish to resolve its singularities we had better consider the dual
theory, which is a (super)string theory (and, so, has infinite many fields).
Other arguments suggest that, when gravity is involved, infinite many local
fields of increasing spins are necessary in order to avoid a conflict with
causality, \cite{Maldacena,Kundu}, see also \cite{Steinacker}. On the other hand  the infinite number of
fields
with increasing spins is related to the good UV behavior of string theory.
Therefore HS theories are at the crossroad of many important themes: locality,
causality, calculability.

The previous considerations are the general underlying motivation for our
research on HS theories. 
However in this as well as in the previous paper, { \cite{I}
referred to as I}, we
concentrate on a specific problem, for which there is no answer yet in the
literature: can one formulate a sensible local massless HS theory in a Minkowski
space-time (the issue of masslessness is fundamental here, being related to
gauge invariance)? Actually the general attitude in the literature, 
{for dimensions larger than 3\footnote{For 3d models see \cite{3d}.}, is
skeptical.}
This is due to two reasons. The first is the so-called no-go
theorems, which prevent the existence of such theories under rather general
conditions. The second is experimental theory: the construction of  {fully
interacting} HS theories
has been so far successful in AdS spaces {(but see \cite{Taronna})}, 
but unsuccessful in flat spacetimes
\cite{BBvD,FV,Vasiliev}\footnote{Complete models, which however seem to be
characterized by a trivial S matrix, have been 
formulated by means of chiral fields in a light-cone framework, 
see \cite{lightcone1}.  In a more general context, the consequences of unbroken 
HS gauge symmetry on the S matrix have been also discussed in 
\cite{Sleight:2016xqq}. Partial attempts to construct consistent interacting 
vertices are numerous, see \cite{3vertex}   and also \cite{Taronna:2017wbx,Roiban} for 
a recent discussion. Conformal
HS models have been introduced in ref.\cite{segal,Tseytlin}.}.
In this paper we present examples of theories defined in flat spacetime in any
dimension, which are massless, {HS gauge invariant, Poincar\'e invariant,}
classically consistent and fully interacting,  and seem {to be ghost free} and
not unmanageable from the quantum point of view.

Let us be more specific. In I we have improved the analysis started in
\cite{BCDGPS} of the effective action produced by integrating out the fermions
in a theory of free fermions coupled to external potentials and quantized
according to the worldline quantization
\cite{strassler,segal,bekaert,schmidt,dai,Bekaert1,Bekaert2,bonezzi}. 
In particular we have developed methods to compute
current correlators; we have clarified the relation with
the analogous effective action obtained by integrating out the scalar field
coupled to external sources; finally we have analysed the possible obstructions
(anomalies) in the construction of the effective action. At this juncture we are
faced with two possibilities. The first is to explicitly compute the 
above mentioned current correlators, much in the same way as in
\cite{BCLPS,BCDGLS,BCDGS}, and explicitly determine the effective action. We are
guaranteed by the Ward identities (barring anomalies) that the resulting
effective action will be HS gauge invariant and will lead to a realization of
an $L_\infty$ symmetry. However the experience made in
\cite{BCLPS,BCDGLS,BCDGS} tells us that most likely the resulting effective
action will be non-local.
This in itself is not a negative feature, because we know that the completion of
the Fronsdal program \cite{Fronsdal}, at least at the linear level, requires
non-local terms in the action \cite{FS} {(although such non-locality 
involves only non-physical terms)}. The problem is the lack of an evident
symmetry pattern in the two and three-point correlators (the most accessible
ones), which makes it very difficult to reconstruct the effective action to all
perturbative orders. 

This first program may still be viable and worth pursuing, but there is perhaps
a second possibility, a sort of shortcut. It consists in exploiting the (already
remarked in I) analogy of the HS gauge transformations with the gauge
transformations in ordinary non-Abelian gauge theories, to construct 
analogous HS invariants and covariant objects and in particular actions. As we
show below this is rather elementary and allows us to directly `integrate the
$L_\infty$ algebra', that is to find explicit equations of motion that satisfy
the
$L_\infty$ axioms\footnote{For a first introduction to $L_\infty$, see
\cite{HZ}, 
for the mathematical aspects see \cite{Linfmath}, for physical applications
\cite{Linfappl}.}. 
They are derived from HS gauge invariant {\it primitive
actions} represented by integrals of local polynomials of {\it master fields} in
the phase space.  In this way we can define perturbatively local HS Chern-Simons
in flat spacetimes in any odd dimensions, as well as HS Yang-Mills theories in
any dimension, of which we can also easily carry out the BRST quantization. We
can also define matter master field models and couple them to the previous
theories and, for instance, reproduce the Higgs mechanism in the HS context. It
is very likely possible to define other types of theory as well.

In this paper we focus in particular on the HS YM models. They 
are defined by means of  a master 
field whose first two component can be identified with an ordinary gauge field
and a 
vielbein fluctuation, respectively. They are characterized 
by a unique coupling constant, like the ordinary YM theories, and by invariance
under HS gauge 
transformations which include in particular ordinary gauge  transformations and
diffeomorphisms.
However covariance is not attained by replacing
ordinary derivatives by covariant ones like in many earlier attempts. In fact
the way gravity appears in these models is different from the familiar
Einstein-Hilbert theory, it is rather similar to teleparallel gravity,
\cite{teleparallel}. The formalism lends itself to interpret 
the first two component fields of the relevant master field as gauge and gravity
fluctuations, an interpretation reinforced by the disclosure of a hidden 
local Lorentz covariance. {This is what we refer to as the gravitational (G) interpretation.
This interpretation will be further {analyzed} in { a subsequent paper}. However in HS YM-like models there
is room also for a non-gravitational interpretation, referred to as a-G.}

Then we tackle the problem of perturbative quantization by means of Feynman 
diagrams. 
The way we do it, by expanding the master fields around the zero momentum and 
integrating out the worldline momenta, brings about the appearance of a mass 
scale. This comes down naturally from the momentum space integration and 
follows from the worldline quantization. In this paper we only broach
this problem, we outline some sample computations, which are 
however enough to show that the perturbative approach  is  
viable. In particolar we argue that  although, like in any HS gauge 
theory, there is abundance of non-physical modes,  a well-defined prescription can be given so that  in physical amplitudes only physical modes contribute.  {The prescription is implemented by inserting projectors in the internal lines of the Feynman diagrams. On the other hand such projectors can be incorporated in the initial theory by modifying it in a precise way. Such modification  renders the theory explicitly non-local but more symmetric, it is in fact endowed with  an additional  gauge symmetry (which we refer to as {\it hidden gauge symmetry}). It is almost useless to stress the similarity of all this with the results of \cite{FS}.}
The final issue we consider here is the problem of no-go
theorems for massless particles in flat background, 
\cite{Weinberg,WW,Porrati,Bekaert3, Bekaert} (see also \cite{ColMand,benin,Ponomarev}).  In the last part of the
paper we show that a few hypotheses on which the no-go theorems rely, notably locality, are
not respected in quantized HS YM theories.

The paper is organized as follows. In section 2 we review the effective action
method, focusing in particular on the HS gauge transformations and their
interpretation. In section 3 we define the HS YM and HS CS theories in both
Abelian and non-Abelian case. Section 4 is devoted to the eom's and  the
conservation laws in both HS YM and HS CS case. We discuss also the $L_\infty$
structure and the BRST quantization of the HS YM models. In section 5 we
introduce HS scalar {and fermion} master fields and show that the Higgs
mechanism can be
reproduced also in the HS theories. Section 6 is devoted to a general discussion
of the action principle in this kind of HS models. In section 7
we discuss
the issue of local Lorentz covariance. In section 8 we start the
discussion of the perturbative approach based on Feynman diagrams. We work out
the example in which only the first two field are present (the gauge and
vielbein fields). Finally we show that the dependence on the conjugate momentum 
$u_\mu$ can be absorbed in a redefinition of the component fields, the coupling
and the coordinate $x^\mu$. The consequence is that a mass scale becomes 
explicit.
In section 9 we outline a sample calculation (the 2-pt function at one loop)
in such a new {\it frozen momentum} framework and in section 10 
we invert the kinetic operator and determine the propagator. In section 11 we 
present our argument for getting rid of ghosts, as announced above, and work out several explicit examples of physical and unphysical mode counting. {In section 12 we unfold the hidden gauge symmetry in the action, at the expense of locality.} Section 13
is devoted to a
discussion of the above-mentioned no-go theorems
and section 14 to our conclusions. Several cumbersome details and formulas are
deferred to a few final appendices.

\section{The method of effective action}

The original matter model, {analysed} in \cite{BCDGPS}, is 
\be 
S_{matter}&=& \int d^dx \, \overline \psi(i \gamma\!\cdot\! \partial-m)\psi
+\sum_{s=1}^\infty \int  {d^d x}\, J^{(s)}_{a\mu_1\ldots\mu_{s-1}}(x)\,
h_{(s)}^{a\mu_1\ldots\mu_{s-1}}(x)\label{S}\\
&=& S_0 + S_{int}\0
\ee
The interaction part $S_{int}$ is\footnote{ {There are in the literature also 
supersymmetric generalizations of $S_{matter}$, which we will not consider here.
See for instance
 \cite{gates} and {references} therein}.}
\be
S_{int}= \langle\!\langle J_a, h^a\rangle\!\rangle \equiv \int   d^dx\,
\frac {d^du}{(2\pi)^d}\, J_a (x,u) h^a(x,u)\label{Sint}
\ee
The (external) gauge fields are collectively represented by \footnote{
The position in the phase space are denoted by couples of letters $(x,u),
(y,v),(z,t),(w,r)$, the first 
letter being for the space-time coordinate and the second for the momentum of
the
worldline particle. 
The letters $k,p,q$ will be reserved for the momenta of the
(Fourier-transformed) physical amplitudes.} 
\be
h^a(x,u)=\sum_{n=0}^\infty \frac 1{n!}\, 
h_{(s)}^{a\mu_1\ldots\mu_n} (x)\, u_{\mu_1}\ldots u_{\mu_n},\label{hmmm}
\ee 
and
\be
J_a(x,u)&=& \frac {\delta S_{int}}{\delta h^a(x,u)}= \int d^dz \, e^{iz\cdot
u} \overline \psi\left(x+\frac z2\right)  \gamma_a\psi\left(x-\frac
z2\right)\label{Jmu}\\
&=& \sum_{n,m=0}^\infty \frac {(-i)^n i^m}{2^{n+m} n! m!} \,\, \partial^n
\overline \psi(x) \gamma_a \partial^m \psi(x) \,\,\frac
{\partial^{n+m}}{\partial u^{n+m}} \delta(u)\0\\
&=& \sum_{s=1}^\infty (-1)^{s-1} J^{(s)}_{a\mu_1 \ldots \mu_{s-1}}(x)\,\,
\frac
{\partial^{s-1}}{\partial u_{\mu_1} \ldots \partial u_{\mu_{s-1}}}\delta(u)\0
\ee
which is obtained by expanding $ e^{iu\cdot z}$. In order to extract
$J^{(s)}_{a\nu_1 \ldots u_{s-1}}(x)$ from $J_a(x,u)$ one must multiply it by
$u_{\nu_1}\ldots u_{\nu_{s-1}}$, integrate over $u$ and divide by $(s-1)!$.
Also
\be
J^{(s)}_{a\mu_1\ldots\mu_{s-1}}(x) = \frac {i^{s-1}} {(s-1)!} \frac
{\partial}{\partial z^{(\mu_1}
}\ldots  \frac {\partial}{\partial z^{\mu_{s-1})}} \overline \psi \left(x+\frac
z2\right)  \gamma_a \psi\left(x-\frac z2\right)
\Big{\vert}_{z=0}.\label{jmmm}
\ee

A generic field, like $h_a(x,u)$, depending both on coordinates and momenta,
will  be called {\it master field}.
 
The gauge transformation of $h^a$ is
\be
\delta_\varepsilon h_a(x,u) = \partial_a^x 
\varepsilon(x,u)-i [h_a(x,u) \stackrel{\ast}{,} \varepsilon(x,u)] 
\equiv {\cal D}^{\ast x}_a  \varepsilon(x,u),
\label{deltahxp}
\ee
where we introduced the covariant derivative
\be
{\cal D}^{\ast x}_a = \partial_a^x- i  [h_a(x,u) \stackrel{\ast}{,}\quad]
.\0
\ee 

The effective action is denoted $\EW[h]$ and takes the form
\be
\EW[h] = \sum_{n=0}^\infty\, \frac 1{n!}\int \prod_{i=1}^n d^dx_i\,
\frac {d^du_i}{(2\pi)^d}\,  \EW_{a_1\ldots a_n}^{(n)}(x_1,u_1,\ldots, x_n,
 u_n)\, h^{a_1}(x_1,u_1) \ldots  h^{a_n}(x_n,u_n) \label{EW}
\ee 
where
\be
&&\EW_{a_1\ldots a_n}^{(n)}(x_1,u_1,\ldots, x_n, u_n)= 
\langle J_{a_1}(x_1,u_1) \ldots   J_{a_n}(x_n,u_n)\rangle\label{corrJ1-Jn}\\
&=& {\left(\sum^\infty_{s_1=1} \frac {\partial}{\partial
u_{1\mu^{(1)}_1}}\ldots 
\frac {\partial}{\partial u_{1\mu^{(1)}_{s_1-1}} }\delta(u_1)\right) \ldots
\left(\sum^\infty_{s_n=1} \frac {\partial}{\partial u_{n\mu^{(n)}_1}}\ldots 
\frac {\partial}{\partial u_{n\mu^{(n)}_{s_n-1}}} \delta(u_n)\right)}\0\\
&&\quad\quad \times \, \langle J^{(s_1)}_{a_1 \mu^{(1)}_{1} \ldots
\mu^{(1)}_{s_1-1} }\ldots
J^{(s_n)}_{a_n \mu^{(n)}_{1}(x_1) \ldots \mu^{(n)}_{s_n-1}}(x_n)\rangle \0
\ee

The statement of invariance under \eqref{deltahxp} is the global Ward identity
(WI)
\be
\delta_\varepsilon \EW[h]=0\label{WI}
\ee
Taking the variation with respect to $\varepsilon(x,u)$ this becomes 
\be
 &&\sum_{n=1}^\infty\, \frac 1{n!}\int \prod_{i=1}^n d^dx_i\, \frac
{d^du_i}{(2\pi)^d}\, {\cal D}_x^{\ast\mu} \EW^{(n+1)}_{a a_1\ldots,
a_n}\!(x,u,x_1,u_1\ldots, x_n, 
u_n)\, h^{a_1}(x_1,u_1) \ldots h^{a_n}(x_n,u_n)=0\0\\
&&\label{EWI}
\ee
This must be true order by order in $h$, i.e.
 \be
 0&=&\int \prod_{i=1}^n d^dx_i\, \frac
{d^du_i}{(2\pi)^d}\, {\partial}_x^{a} \EW^{(n+1)}_{a a_1\ldots
a_n}\!(x,u,x_1,u_1\ldots, x_n, 
u_n)\, h^{a_1}(x_1,u_1) \ldots h^{a_n}(x_n,u_n)\0\\
&&-i\, n \int \prod_{i=1}^{n} d^dx_i\, \frac
{d^du_i}{(2\pi)^d}\, \Big{[}h^{a}(x,u)\stackrel{\ast}{,}
\EW^{(n)}_{a a_1\ldots a_{n-1}}\!(x,u,x_1,u_1\ldots, x_{n-1}, 
u_{n-1})\Big{]}\0\\
&&\quad\quad\quad\times \, h^{a_1}(x_1,u_1) \ldots
h^{a_{n-1}}(x_{n-1},u_{n-1})\label{EWWI}
\ee

 { 
\subsection{The gauge transformation in the fermion model}
\label{ss:gaugetransf}

In the fermion model the gauge transformation of the master field $h_a(x,u)$ is,
\cite{BCDGPS},
\be
\delta_\varepsilon h_a(x,u) = \partial_a^x 
\varepsilon(x,u)-i [h_a(x,u) \stackrel{\ast}{,} \varepsilon(x,u)] 
\equiv {\cal D}^{x\ast}_a \varepsilon(x,u)\label{deltahxpcov}
\ee
Now, the expansion of $h_a(x,u)$ is 
\be
h_a(x,u) &=& A_a(x) +\chi_a^\mu(x) u_\mu 
+ \frac 12 b_a^{\mu\nu}(x) u_\mu u_\nu+\frac 16 c_a^{\mu\nu\lambda}(x) u_\mu
u_\nu
u_\lambda
+ \frac 1{4!}d_a^{\mu\nu\lambda\rho}(x)u_\mu u_\nu u_\lambda
u_\rho\0\\
&&+\frac 1{5!} f_a^{\mu\nu\lambda\rho\sigma}(x)u_\mu u_\nu u_\lambda
u_\rho u_\sigma+\ldots\label{haunn}
\ee
Notice that in the expansion \eqref{hmmm} the indices $\mu_1,\ldots,\mu_n$ are
upper (contravariant), as it should be, because in the Weyl quantization
procedure the momentum has lower index, since it must satisfy $[x^\mu, p_\nu] =i
\,\delta_\nu^\mu$. The index $a$ is different in nature. As we will justify
below, $h_a$ will be referred to as a {\it frame-like master field}.
Of course when the background metric {is flat} all indices
are on the same footing, but
writing in this way leads to the correct interpretation.

We also recall
\be
\varepsilon(x,u)&=& \epsilon(x) +\xi^\mu(x) u_\mu+\frac 12
\Lambda^{\mu\nu}(x)u_\mu
u_\nu+\frac 1{3!} \Sigma^{\mu\nu\lambda}(x)u_\mu
u_\nu u_\lambda\0\\
&&+\frac 1{4!} P^{\mu\nu\lambda\rho}(x)u_\mu
u_\nu u_\lambda u_\rho+\frac 1{5!}\Omega^{\mu\nu\lambda\rho\tau} (x)u_\mu
u_\nu u_\lambda u_\rho u_\tau+\ldots\label{epsxu}
\ee
To the lowest order the transformation \eqref{deltahxpcov} reads
\be
&& \delta^{(0)} A_a= \partial_a \epsilon\0\\
&&\delta^{(0)} \chi_a^{\nu} = \partial_a \xi^\nu  \0\\
&&\delta^{(0)}b_a{}^{\nu\lambda} = \partial_a\Lambda^{\nu\lambda}
\label{deltaAhb}
\ee

To first order we have
\be
\delta^{(1)} A_a &=& \xi\!\cdot\!\partial A_a - \partial_\rho \epsilon
\,\chi_a^{\rho} \label{delta1Ahb}\\
\delta^{(1)} \chi_a^{\nu} &=& \xi\!\cdot\!\partial \chi_a^\nu-\partial_\rho
\xi^\nu \chi_a^\rho 
+ \partial^\rho A_a \Lambda_{\rho}{}^\nu   - \partial_\lambda \epsilon
\,b_a{}^{\lambda\nu}\0\\
\delta^{(1)} b_a^{\nu\lambda} &=& \xi\!\cdot\!\partial b_a{}^{\nu\lambda}
-\partial_\rho \xi^\nu b_a{}^{\rho\lambda}- \partial_\rho \xi^\lambda 
b_a{}^{\rho\nu }+{\partial_\rho \chi_a^{\nu} \Lambda^{\rho\lambda}
+\partial_\rho
\chi_a^{\lambda} \Lambda^{\rho\nu}}
- \chi_a^{\rho} \partial_\rho \Lambda_{\nu\lambda} \0
\ee
The next orders contain three and higher derivatives.
Let us denote by $\tilde A_a, \tilde e_a^\mu = \delta_a^\mu -\tilde \chi_a^\mu$
the standard gauge and vielbein fields. The standard gauge and diff
transformations, are
\be
\delta \tilde A_a&\equiv& \delta \left(\tilde e_a^\mu \tilde A_\mu\right)\equiv
\delta  \left((\delta_a^\mu -\tilde \chi_a^\mu) \tilde
A_\mu\right)\label{standardtransf}\\
&=&\left(-\xi\!\cdot\!\partial \tilde \chi_a^\mu +\partial_\lambda \xi^\mu
\tilde \chi_a^\lambda\right) \tilde A_\mu+(\delta_a^\mu -\tilde
\chi_a^\mu)\left(\partial_\mu\epsilon + \xi\!\cdot\! \tilde A_\mu\right)\approx
\partial_a\epsilon + \xi\!\cdot\! \tilde A_a- \tilde \chi_a^\mu
\partial_\mu\epsilon \0
\ee
and
\be
\delta \tilde e_a^\mu \equiv  \delta  (\delta_a^\mu -\tilde \chi_a^\mu) =
\xi\!\cdot\! \tilde e_a^\mu -\partial_\lambda \xi^\mu \tilde e_a^\lambda = -
\xi\!\cdot\! \tilde \chi_a^\mu -\partial_a \xi^\mu +\partial_\lambda \xi^\mu
\tilde \chi_a^\lambda\label{deltaeamu}
\ee
so that
\be
\delta \tilde \chi_a^\mu= \xi\!\cdot\! \tilde \chi_a^\mu +\partial_a \xi^\mu
-\partial_\lambda \xi^\mu \tilde \chi_a^\lambda\label{deltaeamu1}
\ee
where we have retained only the terms at most linear in the fields.
From the above we see that we can make the identifications
\be
A_a= \tilde A_a, \quad\quad \chi_a^\mu = \tilde \chi_a^\mu\label{identAAee}
\ee

In terms of components $h_a(x,u)$ contains more 
than symmetric tensors: beside the completely symmetric   $h^{(a \mu_1\ldots
\mu_n)}$ it includes also Lorentz representations in 
which the index $a$ and one of the other indices are antisymmetric. A particular 
mention deserves
the field $\chi_a^\mu$. The transformations \eqref{deltaAhb}, \eqref{delta1Ahb}
allow us to interpret it as the fluctuation of the inverse 
vielbein, therefore the effective action may 
accommodate gravity. However, as it is, this action is not invariant under 
local 
Lorentz transformations. We shall prove in section 7 that actually this 
invariance can be recovered 
by suitably modifying the action. When this is so, we refer to it as the G
interpretation, or gravitational interpretation, of the HS YM-like theory. However there is 
another possibility, namely we may leave the action unchanged. In 
this case local Lorentz covariance is not present and the gravitational interpretation 
is impossible. The  theory {can nevertheless be made consistent with global Lorentz covariance. We shall refer to it as 
the a-G or a-gravitational interpretation. In this case the field $\chi_a^\mu$ 
has a different physical meaning and the parameter $\xi^\mu$ does not represent 
diffeomorphisms. We will argue that in both interpretations ghosts do not contribute to physical amplitudes.

\subsection{Analogy with gauge transformations in gauge theories}

It should be remarked that in eq.\eqref{deltahxpcov} and \eqref{deltaAhb} the
derivative $\partial_a$means $\partial_a = \delta_a^\mu \partial_\mu,$ not
$ \partial_a = e_a^\mu \partial_\mu= \left(\delta_a^\mu
-\chi_a^\mu+\ldots\right)\partial_\mu.$
In fact the linear correction $ -\chi_a^\mu\partial_\mu$ is
contained in the term $ -i [h_a(x,u) \stackrel{\ast}{,} \varepsilon(x,u)]$, see
for instance the second term in the RHS of the first equation
\eqref{delta1Ahb}. 
From this point of view the transformation \eqref{deltahxpcov} looks similar to
the ordinary gauge transformation of a non-Abelian gauge field
\be
\delta_\lambda A_a = \partial_a \lambda + [A_a, \lambda]\label{gaugetransf}
\ee
where $A_a= A_a^\alpha T^\alpha, \lambda = \lambda^\alpha T^\alpha$, $T^\alpha$
being the Lie algebra generators.

In gauge theories it is useful to represent the gauge potential as a connection
one form ${\bf A}= A_a dx^a$, so that \eqref{gaugetransf} becomes
\be
\delta_\lambda {\bf A} = {\bf d} \lambda + [{\bf A}, \lambda]\label{gaugetrform}
\ee
As was done in I, we can do the same for \eqref{deltahxpcov}
\be
\delta_\varepsilon {\bf h}(x,u) ={\bf d} 
\varepsilon(x,u)-i [{\bf h} (x,u) \stackrel{\ast}{,} \varepsilon(x,u)] \equiv 
\bfD \varepsilon (x,u) \label{deltahxpbf}
\ee
where ${\bf d}= \partial_a\, dx^a, {\bf h}= h_a dx^a$, $x^a$ are coordinates
in the tangent spacetime and it is understood that 
\be
  [{\bf h} (x,u) \stackrel{\ast}{,} \varepsilon(x,u)]=  [h_a (x,u)
\stackrel{\ast}{,} \varepsilon(x,u)]dx^a\0
\ee

We will apply this formalism to the construction of CS- and YM-like
actions\footnote{In I we have used this analogy to construct HS anomalies.}.

\section{Integrating $L_\infty$}

As we know from previous works \cite{BCDGLS,BCDGS}, the effective field theory
method yields HS gauge invariant results, which in general are non-local. From
them one may be able to eventually extract sensible, even though non-local
actions. However the previously noticed analogy with the non-Abelian gauge
theories formalism, suggests a shortcut: one can directly `integrate' the
$L_\infty$
relations and find (perturbatively) local actions. Based on the analogy with
ordinary local field theory it is not hard to construct HS gauge invariant
action terms. Such terms are defined by means of an integral over the phase
space. For this reason, we will call them at times, not simply actions, but {\it
primitive action functionals}. 

This section is devoted to the
construction of HS Chern-Simons (HS-CS) actions and HS Yang-Mills (HS-YM)
theories, both Abelian and non-Abelian.

\subsection{Preliminaries}

In the sequel we will use for the curvature the notation, introduced in (I),
\be
{\bf G} = {\bf d} {\bf h}
 -\frac i 2 [ {\bf h}\stackrel {\ast}{,}{\bf h}],\label{curv1} 
\ee
with the transformation property
\be
\delta_\varepsilon {\bf G} = -i [{\bf G}  \stackrel {\ast}{,}\varepsilon]
\label{deltaG}
\ee
The functionals we will consider are  integrated polynomials of $\bf G$ or of
its components $G_{ab}$. In order to exploit the transformation property
\eqref{deltaG} in the construction we need the `trace property', analogous to
the trace of polynomials of Lie algebra generators in ordinary non-Abelian gauge
theories.  
The only object with trace properties we can define in the HS context is 
\be
\langle\!\langle f\ast g\rangle\!\rangle\equiv \int d^dx \int \frac
{d^du}{(2\pi)^d}
f(x,u)\ast g(x,u) = \int d^dx \int \frac {d^du}{(2\pi)^d} f(x,u) g(x,u)=
\langle\!\langle g\ast f\rangle\!\rangle  \label{trace}
\ee
From this, plus associativity, it follows that
\be
&&\langle\!\langle f_1 \ast f_2\ast \ldots \ast f_n\rangle\!\rangle=
\langle\!\langle f_1 \ast (f_2\ast \ldots \ast f_n)\rangle\!\rangle\0\\
&&=(-1)^{\epsilon_1(\epsilon_2+\ldots+\epsilon_n)} \langle\!\langle  (f_2\ast
\ldots \ast f_n)\ast
f_1\rangle\!\rangle=(-1)^{\epsilon_1(\epsilon_2+\ldots+\epsilon_n)} 
\langle\!\langle  f_2\ast \ldots \ast f_n\ast f_1\rangle\!\rangle\label{cycl}
\ee
where $\epsilon_i$ is the Grassmann degree of $f_i$. In particular
\be
\langle\!\langle [f_1 \stackrel{\ast}{,} f_2\ast \ldots \ast
f_n\}\rangle\!\rangle=0\label{comm0}
\ee
where $[\quad  \stackrel{\ast}{,}\quad\}$ is the $\ast$-commutator or
anti-commutator, as appropriate.

This property holds also when the $f_i$ are valued in a Lie algebra, provided
the symbol
$\langle\!\langle   \quad  \rangle\!\rangle$ includes also the trace over the
Lie algebra generators.

\subsection{CS primitive action}
 
In (I) we have shown that the action
\be
{\cal CS}(\bfh)=n \int_0^1 dt \langle\!\langle \bfh  \ast \bfG_t \ast \ldots
\ast\bfG_t
 \rangle\!\rangle\label{CSh}
\ee
where
\be
{\bf G}_t = {\bf d}
{\bf h}_t
 -\frac i 2 [ {\bf h}_t \stackrel {\ast}{,}{\bf h}_t],\quad\quad {\bf h}_t= t
{\bf h},\label{Gt}
\ee
 is HS gauge invariant in a space of odd dimension $d=2n-1$. We assume it as the
HS gauge invariant CS action in such dimensions.

\subsection{HS Yang-Mills action}

The curvature form-components, see \eqref{curv1}, are
\be
G_{ab}= \partial_a h_b - \partial_b h_a -i [h_a \stackrel{\ast}{,} h_b ]
\label{Gab}
\ee
Their transformation rule is
\be
\delta_\varepsilon G_{ab}=-i [G_{ab}\stackrel{\ast}{,}
\varepsilon]\label{deltaGab}
\ee
Remembering that $a$ and $b$ are flat indices, it follows that
\be 
\delta_\varepsilon \langle\!\langle G^{ab} \ast G_{ab} \rangle\!\rangle = -i
\langle\!\langle 
   G^{ab} \ast G_{ab} \ast\varepsilon -\varepsilon \ast  G^{ab} \ast G_{ab}
\rangle\!\rangle=0 \label{YMinvariance}
\ee
Therefore 
\be
{\cal YM}({\bfh})=- \frac 1{4 g^2}\langle\!\langle G^{ab} \ast G_{ab}
\rangle\!\rangle\label{YMh}
\ee
is invariant
under HS gauge transformations and it is a well defined  primitive functional
in any dimension.

{\bf Remark 1}. Observe that the dimensions of \eqref{CSh} and \eqref{YMh} are
not the ones of
an action. One should divide it by a factor ${\cal V}_u$  proportional to the
integration volume over the momentum space. For the time being, for the sake of
simplicity, we disregard this factor. We will resume it later on. 

\subsection{The non-Abelian case}

All that has been done for the Abelian $U(1)$ case up to now can be repeated for
the non-Abelian case without significant changes. One has to consider a fermion
field $\psi$ belonging to some representation of a non-Abelian Lie algebra with
generators $T^\alpha$. Then one can introduce the non-Abelian sources
\be
{\mathsf h}= {\mathsf h}^\alpha T^\alpha, \quad\quad {\mathsf h}^\alpha =
h^\alpha_a
dx^a\label{halpha}
\ee
where summation over $\alpha$ is understood. The interacting action is
\be
S_{int}= \langle\!\langle J^\alpha_a, h^{\alpha,a}\rangle\!\rangle = \int  
d^dx\,
\frac {d^du}{(2\pi)^d}\,{\rm Tr} \left( J^a (x,u) h_a(x,u)\right)\label{SintnA}
\ee
where $\Tr$ is the trace over the Lie algebra generators. More explicitly
\be
h_a^\alpha(x,u)=\sum_{n=0}^\infty \frac 1{n!}\, 
h_{a(s)}^{\alpha, \mu_1\ldots\mu_n} (x)\, u_{\mu_1}\ldots
u_{\mu_n},\label{hmmmnA}
\ee 
and
\be
J_a^\alpha(x,u)&=& \frac {\delta S_{int}}{\delta h_a^\alpha(x,u)}= \int d^dz \,
e^{iz\cdot
u} \overline \psi\left(x+\frac z2\right)  \gamma_a T^\alpha \psi\left(x-\frac
z2\right)\label{JmunA}
\ee
Then the HS gauge parameter is
\be
{\mathsf e} (x,u) = \varepsilon^\alpha(x,u) T^\alpha\label{epsilonnA}
\ee
and the transformation {of $ {\mathsf h} (x,u) $ is }
\be
\delta_{\mathsf e} {\mathsf h} (x,u) = \bfd^x 
	{\mathsf e}(x,u)-i [{\mathsf h}(x,u) \stackrel{\ast}{,}{\mathsf e}
(x,u)] ,
\label{deltahxpnA}
\ee
if the generators $T^\alpha$ are anti-hermitean. In this case the curvature is
\be
{\mathsf G} = {\bf d} {\mathsf h}
 -\frac i 2 [ {\mathsf h}\stackrel {\ast}{,}{\mathsf h}]\label{curv2}
\ee
The $\ast$-commutator includes now also the Lie algebra commutator. Of course we
have, in particular,
\be
\delta_{\mathsf e} {\mathsf G} (x,u)= -i [{\mathsf G}(x,u)
\stackrel{\ast}{,}{\mathsf e} (x,u)]
\ee
Everything works as before provided the symbol $\langle\! \langle \quad
\rangle\!\rangle$ comprises also the trace over the Lie algebra generators.
In particular
\be
{\cal YM}({\mathsf h})=- \frac 1{4 g^2}\langle\!\langle {\mathsf G}^{ab} \ast
{\mathsf G}_{ab}
\rangle\!\rangle\label{YMhnonAb}
\ee
is invariant
under the HS non-Abelian gauge transformations and it is a well defined 
primitive functional
in any dimension\footnote{ We have been made aware by a referee of this paper that an embryonic version of \eqref{YMhnonAb}, containing the lowest order, without star product,  has appeared in \cite{Savvidy}}.

\section{Covariant eom's  and conservation laws}

The expressions \eqref{CSh} and \eqref{YMh} do not have the form of the usual
field theory actions, 
because they are integrals over the phase space of the point
 particle with coordinate $x^a$. Nevertheless we can extract from them covariant
eom's by 
taking the variation with respect to $\bfh$. In other words we assume 
that the action principle holds for fields defined in the phase space. We will
justify this later on. For the time being we use this principle to extract from
the primitive
functional the relevant equations of motion.

\subsection{Covariant YM-type eom's}
\label{s:eoms}

From\eqref{YMh}  we get the following eom:
\be
\partial_b G^{ab} -i [h_b\stackrel{\ast}{,} G^{ab}]\equiv{\cal D}_b^\ast
G^{ab}=0\label{YMeom}
\ee
which is, by construction, covariant under the HS gauge transformation
\be
\delta_\varepsilon\left( {\cal D}_b^\ast G^{ab}\right)= -i[ {\cal D}_b^\ast
G^{ab},\varepsilon]\label{covYMeom}
\ee
This is analogous to eq.(2.58) of \cite{BCDGPS}.
In components this equation splits into an infinite set according to the powers
of $u$. Let us expand  $G_{ab}$ in the notation of sec.\ref{ss:gaugetransf}. We
have
\be
G_{ab}&=&  F_{ab}+ X_{ab}^\mu u_\mu +\frac 12 B_{ab}^{\mu\nu} u_\mu u_\nu
+\frac 16 C_{ab}^{\mu\nu\lambda}u_\mu u_\nu u_\lambda
+ \frac 1{4!}  D_{ab}^{\mu\nu\lambda\rho}u_\mu u_\nu u_\lambda u_\rho
+\ldots\label{Gabcomp}
\ee
An explict expansion of $F_{ab}, X^\mu_{ab},...$ in terms of component fields is
given 
in appendix \ref{ss:curvatureexp}.

The eom's from \eqref{YMeom} are
\be
0&=&\partial_a F^{ab} + {\partial_\sigma A_a\, X^{ab\sigma}
-\chi_a^\sigma \partial_\sigma
F^{ab}}+\frac 18\Big(
\partial_{\sigma_1}\partial_{\sigma_2}\partial_{\sigma_3}A^a
C_{ab}^{\sigma_1\sigma_2\sigma_3}\label{YMeom1}\\
&&+ c^{a \sigma_1\sigma_2\sigma_3}
\partial_{\sigma_1}\partial_{\sigma_2}\partial_{\sigma_3}F_{ab}+
3\partial_{\sigma_1}\partial_{\sigma_2}\chi^{b\sigma_3} \partial_{\sigma_3}
B_{ab}^{     \sigma_1\sigma_2} - 3 \partial_{\sigma_3} b^{a \sigma_1\sigma_2}
\partial_{\sigma_1}\partial_{\sigma_2}X_{ab}^{\sigma_3}\Big)+\ldots\0\\
0&=& \partial_a X^{ab\mu} + {\partial_\sigma A_a\, B^{ab\sigma\mu}
-b_a^{\sigma\mu} \partial_\sigma F^{ab} +\partial_\sigma \chi_a^\mu
X^{ab\sigma}-\chi_a^\sigma \partial_\sigma X^{ab\mu}}+\ldots
\label{YMeom2}\\
0&=&\partial_a B^{ab\mu\nu}+ { \partial_\sigma b_a^{\mu\nu}
X^{ab \sigma} +
2\partial_\sigma \chi_a^{(\mu} B^{ab\nu)\sigma}+\partial_\sigma A_a
C^{ab\sigma\mu\nu}}\0\\
&& {-\partial_\sigma F^{ab} c_a^{\sigma\mu\nu} -2\partial_\sigma X^{ab(\mu}
b_a^{\nu)\sigma}
-\chi_a^\sigma \partial_\sigma B^{ab\mu\nu}}+  \ldots\label{YMeom3}\\
&&\dots\dots\0
\ee
where the ellipses in the RHS refer to terms containing {three or more}
derivatives.

More explicitly, for instance the first eom is
\be
0&=& \square A_b -\partial_b \partial \!\cdot\!A +  \left(
\partial_\sigma \partial \!\cdot\!A \chi_b^\sigma
+\partial_\sigma A^a \partial_a \chi_b^\sigma - \partial_\sigma \partial^a A_b
\chi_a^\sigma 
- \partial_\sigma  A_b \partial
\!\cdot\!\chi^\sigma\right)\label{squareFabexpl}\\
&&+ \partial_\sigma A^a \left( \partial_a \chi_b^\sigma - \partial_b
\chi_a^\sigma
+\frac 12 \left( \partial_\lambda A_a b_b^{\lambda\sigma} -\partial_\lambda A_b
b_a^{\lambda\sigma} +
\partial_\lambda \chi_a^\sigma \chi_b^\lambda 
- \partial_\lambda \chi_b^\sigma \chi_a^\lambda\right)\right) \0\\
&& - \chi_a^\sigma \left( \partial_\sigma \partial^a A_b -
\partial_\sigma \partial_b A^a 
+ \frac 12 \left(\partial_\sigma \partial_\lambda A^a \chi^\lambda_b 
+\partial_\lambda A^a \partial_\sigma \chi^\lambda_b-  
\partial_\sigma \partial_\lambda A_b \chi^{a\lambda} 
-\partial_\lambda A_b \partial_\sigma \chi^{a\lambda}\right)\right) 
\0\\
&&+\ldots\ldots\0
\ee

\be
\square \chi_a^\mu- \partial_a \partial^b \chi_b^\mu&=&  
\partial^b( \partial_\sigma A_a \,b_b^{\sigma \mu}- \partial_\sigma A_b
\,b_a^{\sigma \mu}+\partial_\sigma \chi_a^\mu \chi_b^\sigma- \partial_\sigma
\chi_b^\mu \chi_a^\sigma)\label{eomchi}\\
&&+ \partial_\tau A^b \partial_a b_b^{\mu\tau} -  \partial_\tau A^b \partial_b
b_a^{\mu\tau}+ 
\partial_\tau \chi^{b\mu} \partial_a\chi_b^\tau - \partial_\tau \chi^{b\mu}
\partial_b\chi_a^\tau\0\\
&&- \partial_\tau\partial_a A_b\, b^{b\tau\mu}+\partial_\tau\partial_b A_a\,
b^{b\tau\mu}-  \partial_\tau\partial_a \chi_b^\mu
\chi^{b\tau}+\partial_\tau\partial_b \chi_a^\mu \chi^{b\tau} +\ldots
\ee

Let us see a few elementary examples. Consider the case of a pure U(1) gauge
field $A$ alone. The equation of motion is 
\be
\partial_a F^{ab}=\square A^b - \partial_b \partial\!\cdot \!A=0\label{daFab}
\ee
In the Feynman gauge  $ \partial\!\cdot \!A=0$ this reduces to $\square A^b
=0$.

Let us suppose next that $\chi_a^\mu$ is present. Eq.\eqref{YMeom2} becomes
\be
\partial_a X^{ab\mu}= \square \chi_b^\mu - \partial_b \partial\!\cdot\! \chi^\mu
=0\label{daXab}
\ee
In the `Feynman gauge'  $\partial\!\cdot\! \chi^\mu =0$, \eqref{daXab} reduces
to 
$\square \chi_b^\mu=0$\footnote{In ordinary gravity ($R_{\mu\nu}=0$) we have to
impose 
the DeDonder gauge in order to obtain the same result.}.

Finally, keeping only the spin 3 field \eqref{YMeom3} becomes 
\be
\partial_a B^{ab\mu\nu}= \square b_{b}{}^{\mu\nu} - \partial_b \partial^a
b_a^{\mu\nu} =0\label{daBab}
\ee
Again in the `Feynman gauge' $\partial^a b_a^{\mu\nu} =0$ we get $\square
b_{b}{}^{\mu\nu}=0$.

In general we can impose for all the fields the Feynman gauge
\be
\partial^a h_a(x,u)=0\label{Feynman}
\ee
{and obtain the same massless Klein-Gordon equation.}

{\bf Remark 2}. As is clear from \eqref{squareFabexpl}, for instance, the above
eom's 
are characterized by the fact that at each order, defined by the number of
derivatives, there is a finite number of terms. {We adopt the
terminology of  \cite{Bekaert} and call a theory with this characteristic {\it
perturbatively local}
}.

{\bf Remark 3}. {If we stick to the gravitational (G) 
interpretation} the above is an entirely new approach to covariance. The gauge
transformation 
\eqref{deltahxp} reproduces both ordinary U(1) gauge transformations and
diffeomorphisms, but 
the primitive action functional is defined in the phase space. It gives rise to
local
equations of motion that reproduce the ordinary YM eoms, but not completely the
metric 
equations of motion: the linear eom coincide with the ordinary one after gauge
fixing, 
but there is a huge difference with ordinary gravity because in the latter the
interaction terms are 
infinite and include all powers in the fluctuating field, while in the
\eqref{YMh} there are at most 
quartic interactions (at most cubic in the relevant eom). It should be noted,
however, that  the ordinary 
gravity is formulated in terms of the fluctuating field $h_{\mu\nu}$ where 
$g_{\mu\nu} =\eta_{\mu\nu}+ h_{\mu\nu}$, while the `HS gravity' is formulated in
terms of
$\chi_a^\mu$, where $e_a^\mu=\delta_a^\mu-\chi_a^\mu$. What is the relation
between $\chi_a^\mu$ and
 $h_{\mu\nu}$? As it is well known one has to make a `gauge' choice in order to
find this relation. Choosing a symmetric `gauge' for $\chi_a^\mu$ it is given by
\be
e_a^\mu&=&\delta_a^\mu-\chi_a^\mu\0\\
e^a_\mu&=& \delta^a_\mu+\chi_a^\mu+\chi^a_b\chi^b_\mu+\ldots\0\\
g^{\mu\nu}&=& \eta^{\mu\nu}-2\chi^{\mu\nu} +\chi_a^\nu\chi^{a\nu}\0\\
g_{\mu\nu}&=&\eta_{\mu\nu}+2 \chi_{\mu\nu}+3\chi_{\mu}^a
\chi_{a\nu}+\ldots\label{expansions}
\ee
So that
\be
h_{\mu\nu}&=& 2 \chi_{\mu\nu}+3\chi_{\mu}^a \chi_{a\nu}+\ldots\label{hchi}\\
\chi_a^\mu &=&\frac 12h_a^\mu -\frac 34 h_a^\nu h_\nu^\mu+\ldots\label{chih}
\ee
It follows that, expressed in terms of the fluctuation $h_{\mu\nu}$, the cubic 
and quartic powers in $\chi_a^\mu$ turn out to contain powers of any order.

This is not yet enough to clarify what kind of gravity is described by the
equations \eqref{YMeom3}. In a companion paper \cite{III} we show that
it resembles the  so-called {\it teleparallel gravity}, \cite{teleparallel}.

\subsection{CS covariant eom's}

For CS we start from the primitive functional
\be
{\cal CS}(\bfh)=n \int_0^1 dt\int d^dx\, \langle\!\langle  \bfh  \ast \bfG_t
\ast \ldots \ast\bfG_t\rangle\!\rangle,
\quad\quad d=2n-1,  \label{csh}
\ee
Taking a generic variation $\delta \bfh$, with the usual manipulations, we get
\be
\delta {\cal CS}(\bfh)&=&n \int_0^1 dt\int d^dx\,\langle\!\langle\frac d{dt}  
\Big( t \delta\bfh  \ast \bfG_t \ast \ldots
\ast\bfG_t\Big)\rangle\!\rangle\label{deltahCS}\\
&=& n\, \langle\!\langle  \delta\bfh  \ast \bfG  \ast \ldots \ast\bfG
\Big)\rangle\!\rangle\0
\ee
It follows that the overall CS eom is
\be
 \bfG  \ast \ldots \ast\bfG =0\label{CSeom}
\ee
where an exterior product of $n-1=\frac {d-1}2$ factors of $\bfG$ is understood.
Since $\delta_\varepsilon \bfG = -i[\bfG\stackrel{\ast}{,}\varepsilon]$, it is
evident that this equation is HS gauge covariant. For instance, in 3d in
components this means
\be
F_{ab}=0, \quad\quad X_{ab}{}^\mu=0,\quad\quad B_{ab}{}^{\mu\nu}=0,\quad\quad
\ldots
\label{CSeomcomp}
\ee
These equations are covariant because a HS gauge transformation maps them 
to (infinite) linear combinations of themselves.

\subsection{Conservation laws}

The conservation laws of the HS models can be found following the analogy of a
current in {an ordinary} gauge theory or the energy momentum
tensor in gravity theories. For instance the latter is identified with the eom
itself:
\be
T_{\mu\nu} = \frac 2{\sqrt{g}} \frac {\delta S}{\delta g^{\mu\nu}}, \0
\ee
which, in absence of matter, vanishes on shell. It is singled out from the
invariance relation of the
action under diffeomorphisms
\be
0=\delta_\xi S = \int d^dx \,\frac { \delta S}{\delta g_{\mu\nu}}\delta_\xi
g_{\mu\nu}=
-2 \int d^dx \, \xi_\mu \nabla_\nu  \frac { \delta S}{\delta g_{\mu\nu}}\0
\ee
where $ \delta_\xi g_{\mu\nu}= \nabla_\mu \xi_\nu+ \nabla_\nu \xi_\mu$.

Let us proceed in an analogous way, for instance, for HS YM. If we express the
invariance of the action under the HS gauge transformation we can write
\be
0=-\frac 14 \delta_\varepsilon \langle\!\langle G_{ab} \ast G^{ab}
\rangle\!\rangle
= \langle\!\langle \delta_\varepsilon h_a  \ast {\cal D}^\ast_b  G^{ab}
\rangle\!\rangle=
\langle\!\langle{\cal D}^\ast_a\varepsilon   \ast {\cal D}^\ast_b  G^{ab}
\rangle\!\rangle=
- \langle\!\langle \varepsilon   \ast {\cal D}^\ast_a{\cal D}^\ast_b  G^{ab}
\rangle\!\rangle,\label{conslaw}
\ee
which implies the off-shell relation or conservation law
\be
 {\cal D}^\ast_a{\cal D}^\ast_b  G^{ab} =0\label{offshell}
\ee
from which we identify the conserved master current
\be
{\EJ}_a= {\cal D}^\ast_b  G^{ab} \label{mastercurrent}
\ee
In other words the conserved currents are the first members of the eoms derived
above. They vanish on shell and are conserved off-shell. Expanding in $u$
\be
{\EJ}_a= \sum_{n=0}^\infty \frac 1{n!} {\EJ}_a^{\mu_1\ldots\mu_n}
u_{\mu_1}\ldots u_{\mu_n}\label{Jmu1mun}
\ee
we find the component generators. 

\subsection{The $L_\infty$ structure}

As was shown in \cite{BCDGPS} (see also I), the effective action obtained by
integrating out a fermion field coupled to external sources hides an algebraic
structure, which is revealed once we consider the relevant equations of motion.
The basic relations in this game are the eoms
 \be
\EF_\mu(x,u)=0\label{GenEoM}
\ee
where
\be \label{gEoMt}
\EF_\mu(x,u) &\equiv&  \sum_{n=0}^\infty\, \frac 1{n!}\int \prod_{i=1}^{n}
d^dx_i\,
\frac
{d^du_i}{(2\pi)^d}\,  \EW_{\mu,\mu_1\ldots, \mu_{n}}^{(n+1)}(x,u,x_1,u_1,\ldots,
x_{n},  u_{n}) \0\\
&&\quad \times \, h^{\mu_1}(x_1,u_1) \ldots  h^{\mu_{n}}(x_{n},u_{n})
\0
\ee 
$ \EW_{\mu_1,\mu_2\ldots, \mu_{n}}$ being the n-point correlators of the master
fermion currents \eqref{jmmm}, and the covariance relation
\be
\delta_\varepsilon \EF_\mu (x,u)=i [ \varepsilon(x,u) \stackrel{\ast}{,} 
\EF_\mu(x,u)]\label{dvareEF}
\ee

It was shown in section 3 of  \cite{BCDGPS}, that this allows us to define
$j$-linear maps (products) 
$L_j$, $j=1,\ldots,\infty$ of degree $d_j=j-2$ among vector spaces $X_i$ of
degree $i$, defined by the assignments  $\varepsilon\in X_0$, $h^\mu\in X_{-1}$
and  ${\EF}_\mu \in X_{-2}$, which satisfy  the relations
\be
\sum_{i+j=n+1} (-1)^{i(j-1)} \sum_\sigma (-1)^\sigma \epsilon(\sigma;x)\, L_j
(L_i(x_{\sigma(1)},\ldots, x_{\sigma(i)}),x_{\sigma(i+1)},\ldots
,x_{\sigma(n)})=0
\label{Ln}
\ee
In this formula $\sigma$ denotes a permutation of the entries so that
$\sigma(1)<\ldots\sigma(i)$ and $\sigma(i+1)<\ldots\sigma(n)$, and
$\epsilon(\sigma;x)$ is the Koszul sign.

The obvious question is whether the HS-CS or HS-YM models eoms are
representations of this algebra, as one would expect. This is indeed so, and we
do not need to prove it in detail because it is enough
to remark that the analog of the basic relation \eqref{dvareEF} have already
been proven above, and this is all one needs in order to prove \eqref{Ln}.
For instance, in the HS-YM case we have the eom \eqref{YMeom} and the covariance
relation \eqref{covYMeom}, which we rewrite here
\be
\delta_\varepsilon\left( {\cal D}_b^\ast G^{ab}\right)= -i[ {\cal D}_b^\ast
G^{ab},\varepsilon]\label{covYMeom1}
\ee
In this case we define the $X_i$ just as above,  with $\varepsilon(x,u)\in X_0$,
$h_a(x,u)\in X_{-1}$ and  $  {\cal D}_b^\ast G^{ab}(x,u) \in X_{-2}$. The basic
definitions are
\be
L_1 (\varepsilon)_a&=& \partial^x_a \varepsilon(x,u)\label{ell1e}\\
L_2 (\varepsilon,h)_a&=& -i [h_a(x,u)\stackrel{\ast}{,}
\varepsilon(x,u)]=- L_2 (h,\varepsilon)_a\label{ell2eh}\\
L_2(\varepsilon_1,\varepsilon_2)&=&  i\, [{\varepsilon_1}\stackrel{\ast}{,}
{\varepsilon_2}] \label{l2e1e2}
\ee
all the other products, involving at least one $\varepsilon$ factor and any
number of $h$ factors, being identically 0. In order to define the products
involving only $h$ factors we have to decompose
$ {\cal D}_b^\ast G^{ab}(x,u) $ into the sum of three addends according to the
number of $h_a(x,u)$ factors (which are at most three):
\be
 {\cal D}_b^\ast G^{ab}(x,u) = F_1^a(h)+F^a_2(h,h)+ F^a_3(h,h,h)\label{DGdeomp}
\ee
where, of course, $F_i$ contain, beside $h_a(x,u)$, also space differential
operators.
We then define
\be
L_1(h) ^a& =&F_1^a(h), \quad L_2(h_1,h_2)^a= \frac 12\bigl(  F_2^a
(h_1,h_2)+F_2^a(h_2,h_1)\bigr)\label{Li}\\
L_3(h_1,h_2,h_3) &=& \frac 1{3!} \left(F_3^a(h_1,h_2,h_3) + {\rm
perm}(h_1,h_2,h_3)\right), \quad \quad L_i^a=0 \quad i\geq 3\0
\ee
The proof of the relations \eqref{Ln} is the same as in \cite{BCDGPS} and is
based on \eqref{covYMeom1}. The only difference is that in \cite{BCDGPS} there
is an infinite number of nonvanishing $L_n(h_1,\ldots,h_n)$.

A similar construction holds for HS-CS. 
 
\subsection{BRST quantization of HS Yang-Mills}
\label{ssec:BRST}

To BRST quantize the action \eqref{YMh} we have to fix the gauge and apply the
Faddeev-Popov approach.
We impose the Lorenz gauge with parameter $\alpha$ and apply the standard
approach, so
the quantum action becomes
\be
{\cal Y}{\cal M}(h_a,c,B)=\frac 1{g^2} \langle\!\langle  -\frac 1{4 } G_{ab}
\ast G^{ab} -
h^a\ast \partial_a B-i \partial^a \overline c\ast {\cal D}_a^\ast c +\frac
{\alpha}2 B\ast B  \rangle\!\rangle\label{YMhquantum}
\ee
where $c, \overline c$ and $B$ are the ghost, antighost and Nakanishi-Lautrup
master fields, respectively. $c, \overline c$ are anticommuting fields, while
$B$ is commuting.

The action \eqref{YMhquantum} is symmetric under the BRST transformations
\be
s h_a &=& {\cal D}_a^\ast c\label{BRSTtr}\\
s c&=& i c\ast c = \frac i2 [c \stackrel{\ast}{,} c] \0\\
s \overline c &=& i B\0\\
s B&=&0\0
\ee
which are nilpotent. In particular
\be
s( {\cal D}_a^\ast c)=0, \quad\quad s (c\ast c)=0\0
\ee

From the point of view of the $u$ dependence $c, \overline c$ and $B$ are to be
expanded like a scalar master field, see eq.\eqref{Phixu} below.

 Integrating out $B$ in \eqref{YMhquantum} we obtain the standard gauge-fixed
action. 
\be
{\cal Y}{\cal M}(h_a,c)=\frac 1{g^2} \langle\!\langle  -\frac 1{4 } G_{ab} \ast
G^{ab} -
\frac 1{2\alpha} \partial_a h^a\ast\partial_b h^b -i \partial^a \overline c\ast
{\cal D}_a^\ast c  \rangle\!\rangle\label{YMhquantumint}
\ee

\section{Adding bosonic matter}

So far we have treated only gauge fields of any spin. We can couple to the
previous theories matter-type fields of any spin. Let us add, for instance, 
a complex multi-boson field 
\be
\Phi(x,u)= \sum_{n=0}^\infty \frac 1{n!}\Phi^{\mu_1\mu_2\ldots \mu_n}(x)
 u_{\mu_1}u_{\mu_2}\ldots u_{\mu_n}\label{Phixu}
\ee
which transforms like
\be
\delta_\varepsilon \Phi = i \varepsilon\ast\Phi, \quad\quad 
\delta_\varepsilon \Phi^\dagger =- i \Phi^\dagger\ast \varepsilon,
\label{deltaPhi}
\ee
Let us define the covariant derivative
\be
\ED^\ast_a \Phi= \partial_a \Phi -i h_a \ast \Phi\label{covder}
\ee
and its hermitean conjugate
\be
\left(\ED^\ast_a \Phi\right)^\dagger= \partial_a \Phi^\dagger +i \Phi^\dagger
\ast h_a\label{covderadj}
\ee
They have the properties
\be
\delta_\varepsilon \ED^\ast_a \Phi&=& i\,\varepsilon \ast \ED^\ast_a \Phi,\0\\
\delta_\varepsilon(\ED^\ast_a \Phi)^\dagger&=& -i (\ED^\ast_a \Phi^\dagger)\ast
\varepsilon\label{deDPhi}
\ee
As a consequence we have in particular
\be
\delta_\varepsilon (\ED^\ast_a \ED^\ast_b\Phi)= i\varepsilon \ast (\ED^\ast_a
\ED^\ast_b\Phi)\label{deDDPhi}
\ee

It follows that 
\be
{\cal S}(\Phi,h)=\frac 12 \langle\!\langle(\ED_\ast^a
\Phi)^\dagger\ast\ED^\ast_a \Phi\rangle\!\rangle
+\sum_{n=1}^\infty \frac{\lambda_{2n}}{n!} \langle\!\langle(\Phi^\dagger \ast
\Phi)^n_\ast\rangle\!\rangle\label{SPhih}
\ee
is gauge invariant. We remark that this action is real because 
$\Phi^\dagger\ast
\Phi$ is. The generalized eom is
\be
\ED^\ast_a \ED^{\ast a}\Phi+
\sum_{n=1}^\infty\frac{\lambda_{2n}}{(n-1)!}\Phi\ast(\Phi^\dagger \ast
\Phi)^{n-1}_\ast =0\label{HSKG}
\ee
which could be called interacting HS Klein-Gordon equation.

\subsection{A Higgs mechanism}
\label{ss:higgs}

Let us expand $\Phi$
\be
\Phi(x,u)&=& \varphi_0(x)+\varphi_1^\mu (x) u_\mu +\frac 12
\varphi_2^{\mu\nu}(x)u_\mu u_\nu +\frac 16 \varphi_3^{\mu\nu\lambda}u_\mu u_\nu
u_\lambda+\frac 1{4!} \varphi_4^{\mu\nu\lambda\rho}u_\mu u_\nu u_\lambda u_\rho+
 \ldots\0\\
&& \label{Phivarphi}
\ee
Explicit formulas for the component transformations under \eqref{deltaPhi} are
given in 
Appendix \ref{ss:deltaPhi}.

Let us consider the case in which in \eqref{SPhih}, $\lambda_2=\mu^2,
\lambda_4=-\lambda$, while the other couplings vanish, so that the potential is
\be
V(\Phi) = \langle\!\langle \frac {\mu^2}2 \Phi^\dagger \ast
\Phi - \frac {\lambda}{4!}  (\Phi^\dagger \ast
\Phi)^2_\ast \rangle\!\rangle\label{potential}
\ee
Let us suppose that  only $\varphi_0$ takes on a nonvanishing vacuum expectation
value, say $v$, so that
\be
\varphi_0(x)= v + \phi_0(x),\quad\quad v = \mu \frac
{6}{\sqrt{\lambda}}\label{varphiphi}
\ee
Looking at \eqref{deltavarphi}, it is easy to see that this vev breaks the
symmetry completely, for the HS gauge transformations on the vacuum take the
form
\be
\delta_\varepsilon\langle\varphi_0\rangle &=& i\, v \,\epsilon 
\label{vacuumtranf}\\
\delta_\varepsilon \langle\varphi_1^\lambda\rangle &=&  i
\, v\,\xi^\lambda   \0\\
\delta_\varepsilon\langle \varphi_2^{\lambda\rho}\rangle &=&  i\, v\,
\Lambda^{\lambda\rho}\0\\
\delta_\varepsilon \langle\varphi_3^{\mu\nu\lambda}\rangle &=& i\,
v\,\Sigma^{\mu\nu\lambda} \0\\
\ldots &=& \ldots\0
\ee

Next it is convenient to use finite transformations:
\be
\Phi \longrightarrow e^{i \varepsilon}_\ast \ast \Phi,\quad\quad 
\Phi^\dagger  \longrightarrow \Phi^\dagger \ast e^{-i \varepsilon}_\ast
\label{finitePhitrans}
\ee
 and
\be
h_a \longrightarrow i\, e^{i \varepsilon}_\ast  \ast \ED_a^\ast e^{-i
\varepsilon}_\ast\label{finitehatrans}
\ee
Since $e^{i \varepsilon}_\ast  \ast e^{-i \varepsilon}_\ast =e^{-i
\varepsilon}_\ast \ast e^{i \varepsilon}_\ast =1$, it follows, in particular,
that  
\be
 \ED_a^\ast \Phi \longrightarrow  e^{i \varepsilon}_\ast \,\ast\, \ED_a^\ast
\Phi\label{finitetransDPhi}
\ee
So \eqref{SPhih} is invariant under finite HS gauge transformations as well.

{For simplicity let us restrict ourselves to the case where we can parametrize} a
generic configuration of $\Phi$ as
\be
\Phi = e^{i \omega}_\ast \ast \varphi_0 \label{parametrization}
\ee
where 
\be
\omega = \omega_1^\mu (x) u_\mu +\frac 12 \omega_2^{\mu\nu}(x)u_\mu u_\nu
+\ldots \label{omegaomegai}
\ee

Since the RHS of \eqref{parametrization} is formally a HS gauge transformation,
the terms of the action are form invariant. In particular, if $\varphi_0$ is 
real, the potential becomes
\be
V(\phi_0) &=& \langle\!\langle \frac {\mu^2}2 \varphi_0^2 - \frac {\lambda}4 
\varphi_0^4\rangle\!\rangle\label{potential1}\\
&=&  \langle\!\langle \frac {\mu^4}{4\lambda} - \mu^2 \phi_0^2
-\frac 34 \mu\sqrt{\lambda} \phi_0^3 -\frac \lambda 4 \phi_0^4\rangle\!\rangle\0
\ee
where $\varphi_0$ is given by \eqref{varphiphi}. The term linear in $\phi_0$
vanishes, while there is a constant term and  a quadratic, cubic and quartic
term in $\phi_0$. 

The kinetic term in  \eqref{SPhih} reduces to
\be
K(\varphi_0,h')=\frac 12 \langle\!\langle(\ED_\ast^{'a}
\varphi_0)^\dagger\ast\ED^{'\ast}_a \varphi_0\rangle\!\rangle\label{kinetic}
\ee
where $\ED^{'\ast}$ is the covariant $\ast$-derivative with respect to $h_a'$
\be
h'_a =  e^{i \omega}_\ast  \ast \ED_a^\ast e^{-i \omega}_\ast\label{h'a}
\ee
In this way all the matter field components, except $\varphi_0$, are `eaten' by
the gauge fields. Moreover, since
\be
\ED_a^{'\ast} (v+\phi_0)= -i h_a'\, v+ \ED_a^{'\ast} \phi_0\0
\ee
the kinetic term becomes
 \be
K(\varphi_0,h')=\frac 12 \langle\!\langle(\ED_\ast^{'a}
\phi_0)^\dagger\ast\ED^{'\ast}_a \phi_0+v^2\, h^{'a}
h'_a\rangle\!\rangle\label{kinetic'}
\ee
The second term is a mass term for the gauge field components of $h_a'$, whose
kinetic term, obtained from \eqref{YMh} is 
\be
{\cal YM}(h')=-\frac 1{4g^2}\langle\!\langle G^{'ab} \ast G'_{ab}
\rangle\!\rangle\label{YMh'}
\ee
Therefore the second term in  \eqref{kinetic'} provides a mass term for the
gauge fields, which become all massive by `eating' the matter fields. The field
$\phi_0$ survives and due to \eqref{potential1} it is massive.

{\it Warning.} The constant term in \eqref{potential1} is divergent due to the
$x$ integration. This is a well-known fact in the ordinary Higgs mechanism. In
addition the terms in  \eqref{potential1} are infinite due to the momentum
integration. This has to be seen in relation with the primitive functional and
will be discussed later.

\section{The action principle}

As pointed out above the primitive functional \eqref{YMh} and \eqref{CSh} are 
integrals in the phase space. Our definition of effective action, \eqref{EW}, is
also an integral in the
phase space. The examples of equations of motion we obtain, \eqref{YMeom},
\eqref{CSeom} are 
nevertheless space-time local (the leading term in the \eqref{YMeom} equations
is quadratic, that is a (pseudo)elliptic operator). The natural question is:
does
the action principle make sense in this case?  

To answer {this} question let us recall that an ordinary
field theory action
is an abstract expression, a spacetime integral of a polynomial of the fields
and its derivatives. For a generic field configuration one cannot say whether
the integral is convergent or not: there are plenty of field configurations for
which the integral converges and plenty of field configurations for which the
integral is divergent. The action principle determines an extremum, which
requires a calculus of variations, i.e. it requires a topology in the manifold
of fields. Thus it is clear that the action principle is based on the assumption
that the space of field configurations that give rise to a finite action
integral {is dense} enough to  define a topology in the space of
fields.

Now, let us consider a primitive functional like \eqref{YMh} or \eqref{CSh}. We
can interpret the $u$ integrand as a series in $u^2$ (because of (global)
Lorentz
invariance, see below), the coefficient of each powers of $u^2$ being a
spacetime integral. We suppose of course that the latter are convergent and
small enough so that the series in $u^2$ is convergent and integrable. We
suppose that the fields configurations that give rise to an overall convergent
result are dense enough to define a topology in the space of master fields, so
as to
allow for a variational calculus.  

Another point to be remarked is that in the primitive action $x$ and $u$ do not
play the same role. 
While $x$ spans the dynamics of the fields (the dynamical derivatives are the
spacetime ones), $u$ 
plays the role of auxiliary variable or bookkeeping device (much like a
discrete summation {over the fields would do in a theory with
infinite many of them}). 

These considerations are at the basis of the discussion in the following
subsection.

\subsection{Primitive functionals}

Let us denote a primitive functional by 
\be
{\cal S}=\langle\!\langle{\cal L}(\Phi) \rangle\!\rangle=\int d^dx \frac
{d^du}{(2\pi)^d}
{\cal L}(\Phi(x,u))\label{calS}
\ee 
where $\Phi(x,u)$ represents any master field (i.e. function of $x$ and $u$). 
We assume that ${\cal L}$ is a $\ast$-polynomial in  $\Phi$ and its
space-time derivatives. As just said we
interpret the $u$ integration as a bookkeeping device: we could replace the
functional 
with a sum of spacetime integrals over the component fields, provided the
expansion 
in $u$ is integrable; in this case the action can be written as an infinite
series 
of spacetime integrals. Let us apply the action principle to the latter series.
There is still a question to be answered: { is the eom to be
identified with  the master
field variation or with} the variation of each component field separately? Let
us derive 
the eom's by taking the variation with respect to any component field 
and equating it to 0. Since the primitive functional is, so to speak,
$\ast$-analytic
in $\Phi$ and its spacetime derivatives, this is equivalent to taking the
variation with respect
not to a 
single component field, but with respect to $\delta \Phi$:
\be
\delta \Phi =\delta  \phi_0(x) +\delta \phi_1^\mu(x) u_\mu 
+ \frac 12 \delta \phi_2^{\mu\nu}(x) u_\mu u_\nu+\frac 1{3!} 
\delta \phi_3^{\mu\nu\lambda}(x) u_\mu u_\nu u_\lambda+\ldots\ldots\0
\ee
The action principle in this case takes the form
\be
0=\delta{\cal S}= \langle\!\langle\delta\Phi(x,u)\, \EF(x,u) 
\rangle\!\rangle\label{deltacalS1}
\ee
where
\be
\EF(x,u)=\EF_0(x) +\EF_1^\mu(x)u_\mu +\frac 12 \EF_2^{\mu\nu}(x)u_\mu u_\nu+
\frac 1{3!} 
\EF_3^{\mu\nu\lambda}(x) u_\mu u_\nu u_\lambda
+\ldots\0
\ee
So \eqref{deltacalS1} has the explicit form
\be
0&=& \langle\!\langle\delta\phi_0(x) \EF_0(x) +
\big(\delta\phi_0(x)\EF_1^\mu(x)
 + \phi_1^\mu(x) \EF_0(x)\big)  u_\mu\label{deltacalS2}\\
&&+ \frac 12\big(\delta \phi_2^{\mu\nu}(x)\EF_0+2 \delta \phi_1^\mu(x)
\EF_1^\nu(x)+\delta  \phi_0(x) \EF_2^{\mu\nu}(x)\big)u_\mu u_\nu \0\\
&&+ \frac 16 \big(\delta \phi_3^{\mu\nu\lambda}(x)\EF_0+3\delta
\phi_2^{\mu\nu}(x)
\EF_1^\lambda(x)+ 3 \delta \phi_1^{\mu}(x)\EF_2^{\nu\lambda}(x)+
\delta  \phi_0(x) \EF_3^{\mu\nu\lambda}(x)\big)u_\mu u_\nu
u_\lambda+\ldots\dots\0\\
&&+ \frac 1{n!} \big(\delta \phi_n^{\mu_1\mu_2 \ldots \mu_n}(x)\EF_0
+\left(\begin{matrix}n\\ 1\end{matrix}\right)\delta
\phi_{n-1}^{\mu_1\mu_2\ldots \mu_{n-1}}(x) \EF_1^{\mu_n}(x)
+ \left(\begin{matrix}n\\ 2\end{matrix}\right) 
\delta \phi_{n-2}^{\mu_1\ldots
\mu_{n-2}}(x)\EF_2^{\mu_{n-1}\mu_{n}}(x)+\ldots\ldots\0\\
&&+\dots\ldots +\delta  \phi_0(x) \EF_n^{\mu_1\mu_2\ldots
\mu_n}(x)\big)u_{\mu_1} u_{\mu_2}\ldots 
 u_{\mu_n}\big)+\ldots\dots
\rangle\!\rangle\0
\ee
The integration over $u$ simplifies due to (global) Lorentz covariance. So
\eqref{deltacalS2} becomes

\be
0&=& \langle\!\langle\delta\phi_0(x) \EF_0(x) 
+ \frac 12\big(\delta \phi_2^{\mu\nu}(x)\EF_0+2 \delta \phi_1^\mu(x)
\EF_1^\nu(x)+\delta  \phi_0(x) \EF_2^{\mu\nu}(x)\big)\eta_{\mu\nu} \frac
{u^2}d\0\\
&&+\frac 1{4!}\big(\delta \phi_4^{\mu\nu\lambda\rho}(x)\EF_0+4 \delta
\phi_3^{\mu\nu\lambda}(x)
\EF_1^\rho(x)+6\delta  \phi_2^{\mu\nu}(x) \EF_2^{\lambda\rho}(x)
+4 \delta  \phi_1^{\mu}(x) \EF_3^{\nu\lambda\rho}(x)\0\\
&&\quad+ \delta  \phi_0(x) \EF_4^{\mu\nu\lambda\rho}(x)
\big)\big(\eta_{\mu\nu}\eta_{\lambda\rho} + 
\eta_{\mu\lambda}\eta_{\nu\rho}+ \eta_{\mu\rho} \eta_{\nu \lambda}\big) \frac
{u^4}{d(d+2)}+\ldots \ldots\0\\
&&+\frac 1{(2n)!}\Big(\delta \phi_{2n}^{\mu_1\mu_2\ldots\mu_{2n}}(x)\EF_0 
+ \left(\begin{matrix}2n\\ 1\end{matrix}\right)  \delta
\phi_{2n-1}^{\mu_1\mu_2\ldots\mu_{2n-1} }(x) \EF_1^{\mu_{2n}}(x)\0\\
&&+  \left(\begin{matrix}2n\\ 2\end{matrix}\right) \delta 
\phi_{2n-2}^{\mu_1\mu_2\ldots\mu_{2n-2}}(x) 
\EF_2^{\mu_{2n-1}\mu_{2n}}(x)
+ \delta  \phi_0(x) \EF_{2n-2}^{\mu_1\mu_2\ldots\mu_{2n} }(x)
\Big)\0\\
&&\times\big(\eta_{\mu_1\mu_2}\eta_{\mu_3\mu_3}\dots \eta_{\mu_{2n-1}\mu_{2n}}
+ 
{\rm perm}\big) \frac
{u^{2n}}{d(d+2)(d+4)\ldots(d+2n-2)}+\ldots 
\ldots\rangle\!\rangle\label{deltacalS3}
\ee
where `perm' means all distinct permutations of $\mu_1,\ldots,\mu_{2n}$.  
Since the variations $\delta\phi_i$ are arbitrary it follows that this can only
be true if 
all the terms vanish separately. So
\be
\EF_0(x)=0,\quad\quad \EF_1^\mu(x)=0\quad\quad  \EF_2^{\mu\nu}(x)=0, 
\dots\label{eomcomponents}
\ee
i.e.
\be
 \EF(x,u) =0 \label{eomF}
\ee
which is the eom for the master field.

{\bf Remark 4.} It would seem that the eoms are not \eqref{eomcomponents} but
the quantities proportional to a component field variations equated to 0. For
instance, looking in \eqref{deltacalS3}
at the term proportional to $\delta\phi_0$, the corresponding eom looks
\be
0=\EF_0(x) + \frac 12 \frac {u^2}d \EF_{2\mu}{}^\mu + \frac {3u^4}{d(d+2)
}\EF_{4\mu\nu}{}^{\mu\nu}+ \ldots\label{falseeom}
\ee
But this is not the case, because the vanishing must be true for any value of
$u$, which is impossible unless each separate $x$-dependent coefficient
vanishes. Thus we are back to \eqref{eomcomponents}.

\section{Local Lorentz symmetry}
\label{s:LLI}

As pointed out before the HS YM action is fully invariant in particular under
diffeomorphism. This has prompted us to interpret 
{(G interpretation)} the second component of
$h_a(s,u)$ in the $u$ expansion, $\chi_a^\mu$,  as  a vielbein fluctuation, and
$\delta_a^\mu- \chi_a^\mu$ as a vielbein or local frame. However this implies
that $_a$ is a flat index and must transform appropriately under local Lorentz
transformations. But, at least at first sight, the local Lorentz invariance
does not seem to be there.
Consider simply the case in
which only the field $A_a$ is non-vanishing, the form of the Lagrangian is
\be
L_A \sim F_{ab} F^{ab}, \quad\quad F_{ab} =\partial_a A_b-\partial_b
A_a\label{LA}
\ee
This is not invariant under a local Lorentz transformation, because, under $A_a \to 
A_a+\Lambda_a{}^b A_b$, terms $\left(\left(\partial_a
\Lambda_b{}^c\right) A_c -\left( \partial_b \Lambda_a{}^c
\right)A_c\right)F^{ab}$ are generated, that do not vanish. This is a simple
example of a
general problem in HS YM. It is crucial to clarify it.

\subsection{Inertial frames and connections}

Let us start from the definition of trivial frame. A trivial (inverse) frame
$e_a^\mu (x)$ is a frame that can be reduced to a Kronecker delta by means of a
local Lorentz transformation (LLT), i.e.  
such that there exists a (pseudo)orthogonal transformation $O_a{}^b(x)$ for
which
\be
O_a{}^b(x) e_b{}^\mu (x) = \delta_a^\mu\label{trivialframe}
\ee
As a consequence $e_b{}^\mu (x)$ contains only inertial (non-dynamical)
information.
A full gravitational (dynamical) frame is the sum of a trivial frame and
nontrivial piece
\be
\tilde E_a^\mu(x) = e_a{}^\mu (x)-\tilde\chi_a^\mu(x) \label{fullframe}
\ee
By means of a suitable LLT it can be cast in the form
\be
E_a^\mu(x) = \delta_a^\mu -\chi_a^\mu(x) \label{fullframetr}
\ee
This is the form we have encountered above in HS theories. But it should not 
be forgotten that the Kronecker delta represents a trivial frame. If we want to
recover local Lorentz covariance,
instead of $\partial_a=\delta_a^\mu \partial_\mu$ we must understand 
\be
\partial_a = e_a{}^\mu(x) \partial_\mu, \label{truepartiala}
\ee
where $e_a{}^\mu(x)$ is a trivial (or purely inertial) vielbein. In particular,
under an infinitesimal LLT, it transforms according to
\be
\delta_\Lambda  e_a{}^\mu(x) = \Lambda_a{}^b(x) e_b{}^\mu(x)\label{deltaLea}
\ee 

 A trivial connection (or inertial spin connection) is defined by
\be
\EA^a{}_{b\mu} = \left(O^{-1} (x) \partial_\mu O(x)\right)^a{}_b\label{teleA}
\ee
where $O(x)$ is a generic local (pseudo)orthogonal transformation (finite local 
Lorentz transformation).  As a consequence its curvature
vanishes
\be
\ER^a{}_{b\mu\nu} = \partial_\mu \EA^a{}_{b\nu}- \partial_\nu \EA^a{}_{b\mu} 
+ \EA^a{}_{c\mu}\EA^c{}_{b\nu}- \EA^a{}_{c\nu}\EA^c{}_{b\mu}=0\label{teleER}
 \ee
Let us recall that the space of connections is affine. We can obtain any
connection from a fixed one by adding to it adjoint-covariant tensors, i.e.
tensors that transform according to the adjoint representation. When the
spacetime is topologically trivial we can choose as origin of the affine space
the 0 connection. The latter is a particular member in the class of trivial
connections. This is done as follows. Suppose we start with the spin connection 
\eqref{teleA}. A Lorentz transformation  of a spin connection $ \EA_\mu =
\EA_\mu{}^{ab}\Sigma_{ab}$ is
\be
\EA_\mu(x) \rightarrow L(x) D_\mu L^{-1} (x)= L(x) (\partial_\mu + \EA_\mu)
L^{-1} (x)\label{LLL}
\ee
where $L(x)$ is a (finite) LLT. If we choose $L=O$ we get
\be
\EA_\mu(x) \rightarrow 0\label{Lfixing}
\ee
But at this point the LL symmetry gets completely concealed: choosing the zero
spin connection amounts to fixing the local Lorenz gauge.

The connection $\EA_\mu$ contains only inertial and no gravitational
information. It will be referred to as the {\it inertial connection}. It is a
{\it non-dynamical} object (its content is pure gauge). It  plays a role
analogous
to a trivial frame $e_a^\mu(x)$. The dynamical degrees of freedom will be
contained in the
adjoint tensor to be added to $\EA_\mu$ in order to form a fully dynamical spin 
connection\footnote{The splitting of vierbein and spin connection into an
inertial
 and a dynamical part is characteristic of teleparallelism, see
\cite{teleparallel}}.
$\EA_\mu$ is nevertheless a connection and it makes sense to define the inertial
covariant derivative
\be
D_\mu =  \partial_\mu -\frac i2\EA_\mu\label{inertialcovder}
\ee
which is Lorentz covariant.

In ordinary Riemannian geometry the vielbein is annihilated by the covariant
derivative provided we use it to build the metric  and consequently the
Christoffel symbols.
A trivial frame and a trivial connection have an analogous relation provided the
(pseudo)orthogonal transformation $O$ in \eqref{trivialframe} and \eqref{teleA}
is the same in both cases. For we have
\be
D_\mu e_a^\nu&= &\left(\partial_\mu \delta_a^b+ \EA_{\mu a}{}^b \right) e_b^\nu
=
\partial_\mu e_a^\nu +  \left( O^{-1} \partial_\mu O\right) _a{}^b O_b^{-1}{}^c
\delta_c^\nu\0\\
&=& \partial_\mu O_a^{-1}{}^c \delta_c^\nu - \partial_\mu O_a^{-1}{}^c
\delta_c^\nu =0
\label{metriclike}
\ee
From now on we assume that this is the case.

\vskip 1cm

It is clear that the results ensuing from the effective action method as well as
the HS YM 
and HS CS theories are all formulated in a trivial frame setting,
eq.\eqref{fullframetr}, with
a trivial spin connection. In other words the local Lorenz gauge is completely
fixed.
However from this formalism it is not difficult to recover local Lorentz
covariance.

\subsection{How to recover local Lorentz symmetry}

Let us restart from the definition of $J_a(x,u)$
\be
J_a(x,u)\!\! &=& \!\!\sum_{n,m=0}^\infty \frac {(-i)^n i^m }{2^{n+m} n!m!}
\partial_{\mu_1} \ldots \partial_{\mu_m} \bar \psi(x) 
\gamma_a \partial_{\nu_1}\ldots \partial_{\nu_n}\psi(x)\0\\
&& \quad\quad\quad \times 
\frac  {\partial^{n+m}}{\partial{u_{\mu_1}}\ldots\partial{u_{\mu_m}} 
\partial{u_{\nu_1}}\ldots\partial{u_{\nu_n}}} \delta(u)\0\\
&=& \sum_{s=1}^\infty (-1)^{s-1} J^{(s)}_{a\mu_1\ldots \mu_{s-1} }(x) 
\frac  {\partial^{s-1}}{\partial{u_{\mu_1}}\ldots\partial{u_{\mu_{s-1}}}}
\delta(u)\label{Jaxuexpl}
\ee 
from which we derive
\be
J^{(s)}_{a\mu_1\ldots \mu_{s-1} }(x) = \sum_{n=0}^{s-1} \frac {(-1)^n}{{2^{s-1}}
(s-1)!}
\partial_{(\mu_1}\ldots \partial_{\mu_n} \bar \psi(x) 
\gamma_a \partial_{\mu_{n+1} }\ldots \partial_{\mu_{s-1})}\psi(x)\label{Jas}
\ee

Assume now the following LLT
\be
\delta_\Lambda \psi&=& -\frac i2 \Lambda \psi,\quad\quad \Lambda= \Lambda^{ab}
\Sigma_{ab}, \quad\quad\Sigma_{ab}=\frac i4[\gamma_a,\gamma_b]
\label{deltapsi}\\
\delta_\Lambda\bar \psi&=&\frac i2 \bar \psi \Lambda\0
\ee
and replace in \eqref{Jas} the ordinary derivative on $\psi$ with the inertial
covariant derivative
\be
\partial_\mu\psi  \rightarrow D_\mu\psi= \left(\partial_\mu -\frac i2
\EA_\mu\right)\psi\label{dmupsi}
\ee
and on $\bar \psi $ with
\be
\partial_\mu \bar \psi \rightarrow D_\mu^\dagger \bar\psi= \partial_\mu \bar
\psi +\frac i2 
\bar\psi \EA_\mu\label{dmubarpsi}
\ee

Eq.\eqref{Jas} becomes
\be
J^{(s)}_{a\mu_1\ldots \mu_{s-1} }(x)= \sum_{n=0}^{s-1} \frac {(-1)^n}{{2^{s-1}}
(s-1)!}
D^\dagger_{(\mu_1}\ldots D^\dagger_{\mu_n} \bar \psi(x) 
\gamma_a D_{\mu_{n+1} }\ldots D_{\mu_{s-1})}\psi(x)\label{JasD}
\ee
Now, given 
\be
\delta_\Lambda \EA_\mu =- \partial_\mu \Lambda +\frac i2 [\EA_\mu, \Lambda]
\label{deltaEA}
\ee
and \eqref{deltapsi}, it is easy to prove that
\be
\delta_\Lambda (D_\mu\psi)=- \frac i2 \Lambda  (D_\mu\psi), \quad\quad \delta
{(D^\dagger_\mu\bar \psi)= \frac i2   (D^\dagger_\mu\bar
\psi)\Lambda}\label{deltaDmupsi}
\ee
The same holds for multiple covariant derivatives
\be
\delta_\Lambda (D_{\mu_1}\ldots D_{\mu_n}\psi)= {-} \frac i2 \Lambda
(D_{\mu_1}\ldots D_{\mu_n}\psi),\quad {\rm etc.}\0
\ee
It follows that
\be
 \delta_\Lambda J^{(s)}_{a\mu_1\ldots \mu_{s-1} }(x)
& =& -{\frac{i}{2}}\sum_{n=0}^{s-1} \frac
{(-1)^n}{{2^{s-1}} (s-1)!}
D^\dagger_{(\mu_1}\ldots D^\dagger_{\mu_n} \bar \psi(x) 
[\gamma_a,\Lambda] D_{\mu_{n+1} }\ldots D_{\mu_{s-1})}\psi(x)\0\\
&=& \Lambda_a{}^b(x)\, J^{(s)}_{b\mu_1\ldots \mu_{s-1} }(x)\label{deltaJasD}
\ee
Therefore the interaction term
\be
S_{int}=\sum_{s=1}^\infty \int d^dx \, J^{(s)}_{a\mu_1\ldots \mu_{s-1} }(x)
h^{a\mu_1\ldots \mu_{s-1}}\label{Sintexp}
\ee
is invariant under \eqref{deltapsi} and \eqref{deltaEA} provided
\be
\delta_\Lambda  h^{a\mu_1\ldots \mu_{n}}(x)= \Lambda^a{}_b(x)\, 
h^{b\mu_1\ldots \mu_{n}}(x)\label{delatLha}
\ee
On the other hand, writing
\be
S_0 = \int d^dx \, \bar \psi \left(i \gamma^a \left(\partial_a -\frac i2
\EA_a\right)-m\right)\psi\label{S0}
\ee
also $S_0$ turns out to be invariant under LLT. So, provided we define LLT via
\eqref{deltapsi} and
\eqref{deltaEA}, $S=S_0+S_{int}$ is invariant.

Replacing simple spacetime derivatives $\partial_\mu$ with the inertial ones
$D_\mu$ everywhere
is not enough. There is also another apparent inconsistency. Let us take the HS 
field strength $G_{ab}$, \eqref{Gab}.
If we follow the above recipe we have to replace everywhere, also in the $\ast$
product, the ordinary derivatives with covariant ones (covariant with respect to
the spin connection $\EA_a$)\footnote{Replacing $\partial_\mu$ with $D_\mu$ 
does not create any ordering problem because $[D_\mu,D_\nu]=0$.}. This gives
different transformation properties
for the various pieces. $  D_a h_b $ transforms differently from 
\be
\delta_\Lambda (  h_a \ast h_b) = \Lambda_a{}^c (h_c \ast h_b) + \Lambda_b{}^c 
h_a \ast h_c
\label{deltahahb}
\ee

The inertial frame fixes this inconsistency. Instead of writing
$\partial_a=\delta_a^\mu \partial_\mu$ we should write $\partial_a =
e_a{}^\mu(x) \partial_\mu$, 
where $e_a{}^\mu(x)$ is a purely inertial frame. In particular, under a LLT, it
transforms
according to
\be
\delta_\Lambda  e_a{}^\mu(x) = \Lambda_a{}^b e_b{}^\mu(x)\label{deltaLea2}
\ee 
Moreover, whenever a flat index $O_a$ is met we should rewrite it $O_a
=e_a{}^\mu  O_\mu $.

Finally in spacetime integrated expression we must introduce in the integrand
the factor $e^{-1}$, where $e= \det \left(e_a^\mu\right)$, the determinant of
the inertial frame.

With this new recipes all inconsistencies disappear. For instance
\be
\delta_\Lambda (D_a J_b)= \Lambda_a{}^c  (D_c J_b)+ \Lambda_b{}^c  (D_a J_c)\0
\ee
Therefore $\delta_\Lambda(\eta^{ab}D_a J_b)=0$.  

Likewise 
\be
\delta_\Lambda G_{ab} = \Lambda_a{}^c G_{cb} +  \Lambda_b{}^c
G_{ac}\label{deltaLGab}
\ee
which implies the local Lorentz invariance of $G_{ab} G^{ab}$.

\vskip 1cm

{\bf Summary}. {\it The HS effective action approach breaks completely the
symmetry 
under the local Lorentz transformations. This is due the fact that in its
formalism 
(and, in particular, in the HS YM and CS formalism)
the choice $e_a^\mu= \delta_a^\mu$ and $\EA_a=0$ for the inertial frame and
connection, 
is implicit. However the same formalism offers the possibility to recover the LL
invariance
by means of a simple recipe: 
\begin{enumerate}
\item replace any spacetime derivative, even in the $\ast$ product,
with the inertial covariant derivative, 
\item interpret any flat index $_a$ attached to any
object $O_a$ as $e_a^\mu(x) O_\mu$,
\item in any spacetime integrand insert $e^{-1}$.

\end{enumerate}
}
In the process of quantization
$e_a^\mu(x)$ and $\EA_a(x)$ should be treated as classical backgrounds. 
But in the rest of this paper, for simplicity, we stick to the gauge
$e_a^\mu= \delta_a^\mu$ and $\EA_a=0$.

\subsection{Coupling to fermion master fields}

A teleparallel framework allows us to introduce a coupling of the master field
$h_a(x,u)$ to fermion master fields.
Let us start by defining the latter 
\be
\Psi(x,u) = \sum_{n=0}^\infty\frac 1{n!} \Psi_{(n)}^{\mu_1\ldots \mu_n}(x)
u_{\mu_1} \ldots u_{\mu_n}, \label{Psi}
\ee
Under HS gauge transformations it transforms according to
\be
\delta_\varepsilon \Psi = i \varepsilon\ast\Psi, \quad\quad 
\delta_\varepsilon \Psi^\dagger =- i \Psi^\dagger\ast \varepsilon,
\label{deltaPsi}
\ee
and let us define the covariant derivative
\be
\ED^\ast_a \Psi= \partial_a \Psi -i h_a \ast \Psi\label{covderPsi}
\ee
together with its hermitean conjugate
\be
\left(\ED^\ast_a \Psi\right)^\dagger= \partial_a \Psi^\dagger +i \Psi^\dagger
\ast h_a\label{covderbarPsi}
\ee
We get, in particular,
\be
\delta_\varepsilon (\ED^\ast_a \Psi)=i\varepsilon \ast (\ED^\ast_a
\Psi)\label{deltaeDpsi}
\ee
It is evident that the action
\be
S(\Psi,h) = \langle\!\langle \overline \Psi i\gamma^a \ED_a \Psi
\rangle\!\rangle
=  \langle\!\langle \overline \Psi \gamma^a\left(i\partial_a+ h_a \ast\right)
\Psi \rangle\!\rangle
\label{Spsih}
\ee
is invariant under the HS gauge transformations \eqref{deltahxp}. However, as it
is, it is not invariant under the local Lorentz transformations
\be
\delta_\Lambda \Psi = -\frac i2 \Lambda \Psi, \quad\quad \delta_\Lambda
\overline\Psi = \frac i2 \Psi\Lambda, \quad\quad \delta_\Lambda h_a =
\Lambda_a{}^b h_b \label{deltaLPsi}
\ee
But we know how to recover the LL invariance. We must replace in \eqref{Spsih}
$i\partial_a$ with
$ e_a^\mu \left(i\partial_\mu+\frac 12 \EA_\mu\right)$ and add $e^{-1}$ in the
spacetime integral. Then \eqref{Spsih} becomes
\be
S(\Psi, h, \EA) &=& S_1+S_2\label{SpsihEA}\\
S_1 &=& \langle\!\langle \overline \Psi \gamma^a\left(i\partial_a+\frac 12
\EA_a\right) \Psi \rangle\!\rangle=\langle\!\langle \overline \Psi \gamma^a
e_a^\mu \left(i\partial_\mu+\frac 12 \EA_\mu\right) \Psi \rangle\!\rangle
\label{S_1}\\
S_2&=& \langle\!\langle \overline \Psi \gamma^a h_a \ast\Psi
\rangle\!\rangle\label{S2}
\ee
Notice that in $S_1$ there is no $\ast$ product. It is not hard to prove that
$S_1$ and $S_2$ are separately invariant under \eqref{deltaLPsi}. Let us recall
that in order to prove this one must
replace the ordinary derivatives with inertial covariant ones also in the $\ast$
product. This allows 
to prove, for instance, the intermediate result
\be
\delta_\Lambda \left( h_a \ast \Psi\right) =  -\frac i2 \Lambda  \left( h_a \ast
\Psi\right)+ \Lambda_a{}^b\left( h_b\ast \Psi\right) \label{deltaLhaPsi}
\ee 
The rest of the proof is straightforward.

Having modified the form of the actions according to the rules contained in the
summary above, it is not clear a priori that they remain invariant under the HS
gauge transformations. But this is so. The proof is postponed to the Appendix
\ref{s:compatible}.

\subsubsection{The covariantization recipe}

The covariantization procedure we have just illustrated is not the familiar one
in gravity theories. Thus, before proceeding to quantization, it is worth
reviewing it here once more. Let us focus on the action \eqref{YMh}.
 We have seen that it  is invariant under the HS gauge
transformations \eqref{deltahxp}. The latter contain in
particular the diffeomorphisms. So \eqref{YMh} is invariant,
in particular, under diffeomorphisms, but it is clearly not invariant under
local Lorentz transformations. In this section  we have
shown, however, that one can easily recover local Lorentz invariance by means of
the recipe presented above. It induces some changes in the formulas, so one may
suspect that at the end the latter, and in particular \eqref{YMh}, 
{ are no longer invariant under the HS gauge transformations.  It is
not so, as we have shown in Appendix B: the full covariantization process is compatible 
with HS gauge invariance (and so, in particular, with diff invariance). A
crucial step in recovering local Lorentz invariance is the replacement of the
ordinary space-time derivative with the {\it inertial} covariant derivative. Let
us stress that, contrary to the usual covariantization procedure, we do not
replace ordinary derivatives with (ordinary) covariant derivatives. This would
lead to a total mess. Instead we replace ordinary derivatives with (zero
curvature) inertial covariant ones. This operation, so to speak, separates the
{invariance} under HS gauge  transformations (in particular, diffeomorphisms)
from local Lorentz transformation, and guarantees Local Lorentz invariance
without compromising diffeomorphism invariance.  This property is characteristic
of teleparallelism \cite{teleparallel}.

\section{Perturbative approach}

Let us consider the gauge-fixed HS YM
\be
{\cal Y}{\cal M}(h_a,c)=\frac 1{g^2} \langle\!\langle  -\frac 1{4 } G_{ab} \ast
G^{ab} -
\frac 1{2\alpha} \partial_a h^a\ast\partial_b h^b -i \partial^a \overline c\ast
{\cal D}_a^\ast c  \rangle\!\rangle\label{YMhgaugef}
\ee
as an ordinary field theory and  try to apply to it an ordinary 
perturbative quantization procedure. To this end we rely on the 
master field 
expansion around $u=0$. This is not the only possible choice, but it is the 
most 
obvious one to start a discussion of perturbative quantization.  
{From now on we will refer to the action \eqref{YMhgaugef}. The 
equations of motion will be the ones extracted from this action functional and 
the HS gauge symmetry is understood to be replaced by the BRST symmetry defined 
in subsection \ref{ssec:BRST}.}

\subsection{The propagator}

We use the results of the section 6, replacing
$\Phi$  with
$h_a$, the index $a$ being 
contracted with that of $\EF^a$, The linear part of the eom's, 
see (\ref{YMeom},\ref{YMeom1},\ref{YMeom2},\ref{YMeom3}), 
once the gauge is fixed\footnote{{We refer to the gauge fixed action
\eqref{YMhgaugef},
 but, as we shall see later, this gauge fixing is partial.}},
becomes simply $\square h_a{}^{\mu_1\ldots}=0$. If we
wish to proceed to 
quantization we have to know the propagators. One would think it is $\frac
1{p^2}$, i.e. the inverse 
of $\square$. But, in fact, the situation is more complicated. 

Let us  specialize to the HS YM case\footnote{ When considering explicit 
expressions of the action, such as the kinetic term, the cubic and quartic 
terms 
(see below) it is important to specify that we drop altogether total 
derivatives 
with respect to $u$. The latter correspond to boundary terms, much like the 
total derivatives with respect to $x$ in ordinary field theories. Like in the 
latter case one may or may not drop total derivatives, such a choice is part of 
the definition of the theory (for instance, boundary terms are crucial in 
distinguishing between open and closed strings). Here we make the simplest 
choice and drop total $u$-derivatives. }
and consider only the quadratic part in \eqref{YMhgaugef}.  To start with we
absorb the coupling $g$
in $h_a$ and $c,\bar c$.
In the general Lorenz gauge (the Feynman gauge
corresponds to $\alpha=1$) of \eqref{YMhquantum} the kinetic operator takes the
form\footnote{{From now on, for simplicity, we will drop the FP
ghosts. We will get back to them later on.}}
{\scriptsize
\be
&&K_{ab}^{\{\mu\}\{\nu\}}(x,u)=\left( \eta_{ab} \square_x - \frac
{\alpha-1}\alpha
\partial^x_a\partial^x_b\right) 
\label{kineticop}\\ 
&&\times \left(\begin{matrix}
       1 & 0 &\eta^{\nu_1\nu_2} \frac {u^2}{2d} &0& \Pi^{\nu_1\nu_2\nu_3\nu_4}
\frac
{u^4}{4!d(d+2)}&0\\
0 &\eta^{\mu_1\nu_1} \frac {u^2}{d}& 0 &\Pi^{\mu_1\nu_1\nu_2\nu_3} \frac
{u^4}{3!d(d+2)} &0&\ldots\\
\eta^{\mu_1\mu_2} \frac {u^2}{2d}& 0& \Pi^{\mu_1\mu_2\nu_1\nu_2} \frac
{u^4}{4d(d+2)}&0&\ldots&\ldots\\
0 &\Pi^{\mu_1\mu_2\mu_3\nu_1} \frac {u^4}{3!d(d+2)} &0 &\ldots&\ldots&\ldots\\
 \Pi^{\mu_1\mu_2\mu_3\mu_4} \frac {u^4}{4!d(d+2)}&0&\dots
&\ldots&\ldots&\ldots\\
0 &\dots &\ldots&\ldots&\ldots&\ldots\\
 \end{matrix}\right)\0
\ee}
where $ \Pi_{\mu\nu\lambda\rho}= \eta_{\mu\nu}\eta_{\lambda\rho} + 
\eta_{\mu\lambda}\eta_{\nu\rho}+ \eta_{\mu\rho} \eta_{\nu \lambda}$. 
We will call $N^{\{\mu\}\{\nu\}}(u)$ the matrix in the RHS.
The  matrix operator \eqref{kineticop} 
acts on the column vector $(A^b(x), \chi^b_{\nu_1}(x),
b^b_{\nu_1\nu_2}(x), c^b_{\mu_1\mu_2\mu_3}(x),\ldots)^T$, the result being
contracted with the row vector $ (A_a(x), \chi_{a\mu_1}(x), b_{a\mu_1\mu_2}(x),
c_{a\mu_1\mu_2\mu_3}(x), \ldots)$.

{If the matrix $N$ is invertible,\footnote{We will see 
below that this may not be the case.}} the propagator is given by 
the inverse  of \eqref{kineticop}. Let us denote it by
$P^{\{\mu\}\{\nu\}}_{ab}(x,y,u,v)$. It has the structure
\be
P^{\{\mu\}\{\nu\}}_{ab}(x,y,u,v) &=&  \langle h_a(x,u)\, h_b(y,v)\rangle_0\label{prophahb}\\
& =&-i
\int \frac {d^dk}{(2\pi)^d} e^{ik\cdot (x-y) } \left(\frac {\eta_{ab}}{k^2} 
+(\alpha-1) \frac {k_ak_b}{k^4} \right)\delta(u-v)
\,M^{\{\mu\}\{\nu\}}(u)\0
\ee
where $M^{\{\mu\}\{\nu\}}(u)$ is an {infinite} matrix to be determined. It is 
the
inverse of $N^{\{\mu\}\{\nu\}}(u)$, i.e.
\be
N^{\{\mu\}\{\nu\}}(u) M_{\{\nu\}\{\lambda\}}(u)=
\delta^{\{\mu\}}_{\{\lambda\}}\0
\ee
One can guess the structure of $M$ 
\be
&&M^{\{\mu\}\{\nu\}}(u)=\label{Mm}\\
&&\left(\begin{matrix}
       1 & 0 &\eta^{\nu_1\nu_2} \frac {a_{0,2}} {u^2} &0&
t^{\nu_1\nu_2\nu_3\nu_4} \frac {a_{0,4}}{u^4}&0\\
0 &\eta^{\mu_1\nu_1}\frac {a_{1,1}}{u^2}& 0 &t^{\mu_1\nu_1\nu_2\nu_3} \frac
{a_{1,3}}{u^4} &0&\ldots\\
\eta^{\mu_1\mu_2} \frac {a_{2,0}} {u^2}& 0& t^{\mu_1\mu_2\nu_1\nu_2}
\frac{a_{2,2}}
{u^4}&0&\ldots&\ldots\\
0 &t^{\mu_1\mu_2\mu_3\nu_1} \frac{a_{3,1}} {u^4} &0 &\ldots&\ldots&\ldots\\
 t^{\mu_1\mu_2\mu_3\mu_4}  \frac{a_{4,0} }{u^4}&0&\dots &\ldots&\ldots&\ldots\\
0 &\dots &\ldots&\ldots&\ldots&\ldots\\
 \end{matrix}\right)\0
\ee
where $t^{\{\mu\} \{\nu\}}$ are tensors constructed out of $\eta$, which are
symmetric in $\{\mu\}$ and $\{\nu\}$ separately. $a_{i,j}$ are constants to be
determined, with $a_{i,j}=a_{j,i}$.
\vskip 1cm

Applying this to the column vector $(j_0^b(x), j^b_{1\nu_1}(x),
j^b_{2\nu_1\nu_2}(x), j^b_{3\nu_1\nu_2\nu_3}(x),\ldots)^T$
and contracting the result with the row vector $ (j_{0a}(x), j_{1a\mu_1}(x),
j_{2a\mu_1\mu_2}(x), j_{3b\mu_1\mu_2\mu3}(x), \ldots)$ we create the expression 
\be
\langle\!\langle j^a\, P_{ab}\, j^b\rangle\!\rangle= \int d^dx d^dy \frac
{d^du}{(2\pi)^d}\, \sum_{\{\mu\}, \{\nu\}} j^a_{\{\mu\}}(x)\,
P^{\{\mu\}\{\nu\}}_{ab}(x,y,u)\,  j^b_{\{\nu\}}(y)\label{jaPabjb}
\ee

It is more convenient to go to the momentum representation:
 \be
&& \widetilde P^{\{\mu\}\{\nu\}}_{ab}(k,u,v) =-i
 \left(\frac {\eta_{ab}}{k^2} 
+(\alpha-1) \frac {k_ak_b}{k^4} \right)\delta(u-v)
\,M^{\{\mu\}\{\nu\}}(u)\label{prophahbtilde1}
\ee

Applying this to the column vector $(\tilde j_0^b(-k), \tilde
j^b_{1\nu_1}(-k),\tilde j^b_{2\nu_1\nu_2}(-k), \ldots)^T$
and contracting the result with the row vector $ (\tilde j_{0a}(k),\tilde 
j_{1a\mu_1}(k),\tilde  j_{2a\mu_1\mu_2}(k), \ldots)$ we create the expression 
\be
\langle\!\langle\tilde j^a\, \widetilde P_{ab}\,\tilde 
j^b\rangle\!\rangle=\int\frac {d^dk}{(2\pi)^d}\, \int\frac
{d^du}{(2\pi)^d}\,\sum_{\{\mu\}, \{\nu\}} \tilde j^a_{\{\mu\}}(k)\,\widetilde
P^{\{\mu\}\{\nu\}}_{ab}(k,u)\, \tilde  j^b_{\{\nu\}}(-k)\label{jaPabjbtilde}
\ee
This  and \eqref{jaPabjb} give the $\langle\!\langle j^a\, P_{ab}\,
j^b\rangle\!\rangle$ and $ \langle\!\langle\tilde j^a\, \widetilde
P_{ab}\,\tilde  j^b\rangle\!\rangle$ term, respectively, for each couple of
local fields separately. 

It is evident that the crucial object to be determined is the matrix
$M^{\{\mu\}\{\nu\}}(u)$. {In view of the $u$ integration, the inverse powers of
$u^2$ in it are hard  if not impossible to deal with. To gain some insight about
this obstacle let us consider, below, a simple
example}\footnote{For a more general discussion about the
Feynman rules, see Appendix \ref{app:funct}}.

\subsection{An example: the $A-\chi$ model, gauge field and vielbein}
\label{ss:Achi}

Let us suppose that the master field $h_a(x,u)$ contains only two fields: $A_a$
and $\chi_a^\mu$. In this case  the kinetic operator becomes
\be
K_{ab}^{\{\mu\}\{\nu\}}(x,u)=\left( \eta_{ab} \square_x - \frac {\alpha-1}\alpha
\partial^x_a\partial^x_b\right)
\left(\begin{matrix}
       1 & 0 \\
0 &\eta^{\mu\nu} \frac {u^2}{d}\end{matrix}\right)\label{kineticopred} 
\ee
Its inverse is 
\be
P^{\{\mu\}\{\nu\}}_{ab}(x,y,u) =
\int \frac {d^dk}{(2\pi)^d} e^{ik\cdot (x-y) } \left(\frac {\eta_{ab}}{k^2} 
+(\alpha-1) \frac {k_ak_b}{k^4} \right)
\left(\begin{matrix}
       1 & 0 \\
0 &\eta^{\mu\nu} \frac {d}{u^2}\end{matrix}\right)\label{propred}
\ee

To gain some insight for the general case it is useful to develop the
perturbative approach for this
$A-\chi$ model. Let us write down the interaction terms for these two fields. In
Appendix \ref{ss:S3S4}
we have collected the $u$ expansions for the interaction part of the action
\be
S= S_0 +S_{int},\quad\quad  S_{int} = S_3+S_4\label{S3+S4}
\ee 
where, in the $A-\chi$ model, $S_0$ is the free part with kernel
\eqref{kineticopred} and $S_3, S_4$ are
the cubic, quartic interaction, respectively. The latter can be obtained from
the equation 
\eqref{S3} and \eqref{S4} by suppressing all the other component fields. 

The cubic term is 
\be
S_3&=&  {- g} \langle\!\langle \partial^a A^b (\partial_\sigma A_a
\chi_b^\sigma - \partial_\sigma A_b \chi_a^\sigma)+\frac 1{d}  \partial^a
\chi^b_\nu \bigl(  \partial_\sigma \chi_a^\nu \chi_b^\sigma - \partial_\sigma
\chi_b^\nu \chi_a^\sigma\bigr) u^2 \rangle\!\rangle \label{S3red}
\ee
and the quartic is
\be
S_4 &=&  -  \frac {g^2}{{2}}\, \langle\!\langle  \bigl(\partial_\sigma A^a
\chi^{b\sigma}- \partial_\sigma A^b \chi^{a\sigma}\bigr) \partial_\tau A_a
\chi_b^\tau
+ \frac 1{d} \bigl( 
\partial_\sigma\chi^{a\nu} \chi^{b\sigma} - \partial_\sigma\chi^{b\nu}
\chi^{a\sigma} \bigr)
\partial_\tau\chi_{a\nu} \chi_{b}^\tau \,u^2 \rangle\!\rangle, \label{S4red}
\ee
exactly.

Let us define the Feynman rules in the usual way, by considering $u^2$ as a
constant 
and ignoring, for the time being the integration over $u$.
The free $AA$ propagator is of order $u^0$. Consider
then the next order, i.e. the bubble diagram with two external $A$-legs. This
can
be formed with 
two cubic vertices (the first two terms in $S_3$), one $AA$ and one $\chi\chi$
propagator. Looking at \eqref{propred}, we see that the result is of order
$\frac 1{u^2}$. Another possibility is to create a seagull term by means of a
quartic vertex (the first two terms in $S_4$) and a  $\chi\chi$ propagator. This
is also of order $\frac 1{u^2}$. 
Next let us consider the two-loop order, which is formed with three $AA\chi$ and
one $\chi\chi\chi$ vertices, three internal $\chi\chi$ plus two $AA$
propagators. This is of order $u^{-4}$.

Similarly, the vertex $AA\chi$ is of order $u^0$. The one-loop contribution to
the 3-point $AA\chi$
is of order $u^{-2}$. And so on.

Consider next the diagrams with two external $\chi$-legs. The 0-th order 
$\chi\chi$ propagator is of order $u^2$. Using two cubic $AA\chi$ vertices  and
two $AA$ propagators we obtain a contribution of order $u^0$. We can obtain a
bubble diagram with two external $\chi$-legs by means of two $3\chi$ cubic
vertices and two $\chi\chi$ propagators. The result is also of order $u^0$. We
can also create two seagull terms  using
the quartic vertices $AA\chi\chi$ and one $AA$ propagator, and the vertex
$\chi\chi\chi\chi$ with a $\chi\chi$ propagator. These results are also of order
$u^0$.

Similarly the $3\chi$ vertex is of order $u^{-2}$. The one-loop  $3\chi$
function is of order $u^0$.

We can continue by considering higher loop contributions or diagrams with more
legs. The results are somewhat disconcerting, because we obtain a different $u$ 
dependence for different loop-orders. It is rather impervious to assign a
meaning whatsoever
to the $u$ integration {in this context}. On the other hand it is
not hard to notice a regularity in them. 
Regularity that one can see by realizing that
in \eqref{S3red} and \eqref{S4red}  we could redefine fields and coupling as
follows\footnote{{The momentum square $u^2$ can be positive or 
negative according to whether it is spacelike or timelike. As we have shown 
above (see for instance \eqref{deltacalS3})   the $u$-integrand 
in 
the action can be reduced to a series in $u^2$. At that point we perform a Wick 
rotation in $u$ so that $u^2$ is always non-negative. This gives an unambiguous 
meaning to the expression $\sqrt{u^2}$ and to the $u$-integration.  }} $A\to A, 
\chi \to \sqrt {u^2}\chi$ and $g\to \frac g{\sqrt {u^2}}$ and
get
\be
S_3&=& {- g} \langle\!\langle \partial^a A^b (\partial_\sigma A_a
\chi_b^\sigma - \partial_\sigma A_b \chi_a^\sigma)+\frac 1{d}  \partial^a
\chi^b_\nu \bigl(  \partial_\sigma \chi_a^\nu \chi_b^\sigma - \partial_\sigma
\chi_b^\nu \chi_a^\sigma\bigr) \rangle\!\rangle \label{S3red1}
\ee
and
\be
S_4 &=&  -  \frac {g^2}{{2}}\, \langle\!\langle  \bigl(\partial_\sigma A^a
\chi^{b\sigma}- \partial_\sigma A^b \chi^{a\sigma}\bigr) \partial_\tau A_a
\chi_b^\tau
+ \frac 1{d} \bigl( 
\partial_\sigma\chi^{a\nu} \chi^{b\sigma} - \partial_\sigma\chi^{b\nu}
\chi^{a\sigma} \bigr)
\partial_\tau\chi_{a\nu} \chi_{b}^\tau   \rangle\!\rangle\0\\
&& \label{S4red1}
\ee
where, for simplicity, the new fields $\chi'= \sqrt {u^2}\chi$ and new coupling
$g'=\frac g {\sqrt {u^2}}$ are denoted with  the old symbols.

We see that $u$ has disappeared. But this is  a bit too much, because now, 
since $u$ has the dimension of a mass, the field $\chi$ has the same dimension
as $A$, i.e. 1. It cannot represent a {gravity}
vielbein. What we have to do, instead, is split $\sqrt{u^2}= \mathfrak{m} {\rm
u} $ where
$ {\mathfrak m}$ is a fixed mass scale and 
$\rm u$ is the dimensionless variable part, and absorb the latter in $\chi$ and
$g$, by setting  $\chi'= {\rm u}\chi$ and $g'=\frac g {\rm u}$. We obtain in
this way
 \be
S_3&=& {- g} \langle\!\langle \partial^a A^b (\partial_\sigma A_a
\chi_b^\sigma - \partial_\sigma A_b \chi_a^\sigma)+\frac {{\mathfrak m}^2}{d} 
\partial^a \chi^b_\nu \bigl(  \partial_\sigma \chi_a^\nu \chi_b^\sigma -
\partial_\sigma \chi_b^\nu \chi_a^\sigma\bigr) \rangle\!\rangle \label{S3red2}
\ee
and
\be
S_4 &=&  -  \frac {g^2}{{2}}\, \langle\!\langle  \bigl(\partial_\sigma A^a
\chi^{b\sigma}- \partial_\sigma A^b \chi^{a\sigma}\bigr) \partial_\tau A_a
\chi_b^\tau
+ \frac {{\mathfrak m}^4}{d} \bigl( 
\partial_\sigma\chi^{a\nu} \chi^{b\sigma} - \partial_\sigma\chi^{b\nu}
\chi^{a\sigma} \bigr)
\partial_\tau\chi_{a\nu} \chi_{b}^\tau   \rangle\!\rangle\0\\&& \label{S4red2}
\ee

Now the integration over $u$ yields an overall dimensionful
infinite factor
which can be eliminated by dividing the primitive functional ${\cal YM}(h)$ by
the same factor, as suggested in Remark 1. {Moreover in \eqref{propred}} { $u^2$
is replaced by ${\mathfrak m}^2$, in this way disarming the risk triggered by
the $u$ integration in the amplitudes.}

\subsection{Absorbing the $u$ dependence. A frozen momentum background}
\label{ss:absorbing}
As in the previous example it is possible to absorb the ${\rm u}$ dependence
completely in the full model. However in the
general case it is not enough to redefine
{$h_a^{\mu_1\ldots\mu_{s-1}}\to
h_a^{'\mu_1\ldots\mu_{s-1}}= {\rm u}^{s-1} h_a^{\mu_1\ldots\mu_{s-1}}$}, we must
also
redefine the coordinates as follows: $ x^\mu \to {\rm u}^{-1} x^\mu$. The coupling is
also redefined as before $g \to \frac g{\rm u}$. Under these redefinition, for
instance $S_3$ and $S_4$ have the same form as in eqs.(\ref{S3},\ref{S4}) with
$u$ replaced by
${\mathfrak m}$ and $\langle\!\langle \quad \rangle\!\rangle$ replaced by $
\langle\!\langle \quad \rangle\!\rangle'$. In the same way the kinetic term
remains the same apart from 
\be
\langle\!\langle h_a K^{ab} h_b  \rangle\!\rangle \longrightarrow
\langle\!\langle h_a K^{ab} h_b  \rangle\!\rangle' \label{kineticfinal}
\ee
The symbol $  \langle\!\langle\quad \rangle\!\rangle'$ means that the
integration measure
has changed to
\be
\int d^dx d^d u \equiv {\mathfrak m}^d \int d^dx d^d{\rm u} \longrightarrow
{\mathfrak m}^d\int d^dx d^d{\rm u} \, {\rm u}^{d-2}\label{intmeasure1}
\ee
In other words, apart from this change of measure and the substitution of $u$
replaced by
${\mathfrak m}$, in the expressions $S_3,S_4$ and the kinetic term, nothing has
changed. In particular the dependence on ${\rm u}$ has disappeared from the
integrand.
Since now the integrand is  ${\rm u}$ independent we can factor out the quantity
\be
{\cal V}_d =   {\mathfrak m}^d\int d^d{\rm u} \, {\rm
u}^{d-2}\label{intmeasure2}
\ee
and simplify it with the same factor coming from Remark 1.

So finally we are
simply left with the spacetime action ${\cal S}= {\cal S}_2+  {\cal S}_3+ {\cal
S}_4$:
\be
{\cal S}_2= \int d^dx \sum_{\{\mu\},\{\nu\}} h^a_{ \{\mu\}}(x)
K_{ab}^{\{\mu\}\{\nu\}}(x,{\mathfrak m})  h^b_{ \{\nu\}}(x)\label{kineticnew} 
\ee

\be
{\cal S}_3&=&  {-g}\int d^dx\, \biggl{\{} \partial^a A^b (\partial_\sigma A_a
\chi_b^\sigma - \partial_\sigma A_b \chi_a^\sigma)\label{S3final}\\
&&-\frac 1{{24}} (\partial^a A^b-\partial^b A^a) \bigl(\partial_{\sigma_1}
\partial_{\sigma_2} \partial_{\sigma_3}A_a \, c_b^{\sigma_1\sigma_2\sigma_3}+ 3
\partial_{\sigma_3} b_a^{\sigma_1\sigma_2}\partial_{\sigma_1}
\partial_{\sigma_2}\chi_b^{\sigma_3}\bigr)\0\\
&&+ \frac {{\mathfrak m}^2}{2d} \Bigg(\partial^a A^b \partial_\sigma
b_{a\mu}{}^\mu
\chi_b^\sigma -\partial^a A^b \partial_\sigma b_{b\mu}{}^\mu \chi_a^\sigma\0\\
&&+2\partial^a A^b \partial_\sigma \chi_a^\mu b_{b\mu}{}^{\sigma}- 2\partial^a
A^b \partial_\sigma \chi_b^\mu b_{a\mu}{}^{\sigma}+ \partial^a A^b 
\partial_\sigma A_a c_{b\mu}{}^{\mu\sigma}-\partial^a A^b  \partial_\sigma A_b
c_{a\mu}{}^{\mu\sigma}\0\\
&&+ \partial^a b^{b\mu}{}_\mu \, \partial_\sigma A_a \chi_b^\sigma -  \partial^a
b^{b\mu}{}_\mu \, \partial_\sigma A_b\chi_a^\sigma+2 \partial^a \chi^b_\nu
\bigl(
\partial_\sigma A_a \, b_b{}^{\sigma\nu}- \partial_\sigma A_b \,
b_a{}^{\sigma\nu}\0\\
&& + \partial_\sigma \chi_a^\nu \chi_b^\sigma - \partial_\sigma \chi_b^\nu
\chi_a^\sigma\bigr)\0\\
&&{-\frac 1{24}} \bigg( (\partial^a A^b-\partial_b A^a)\Big( \partial_{\sigma_1}
\partial_{\sigma_2} \partial_{\sigma_3}\chi_a^\mu d_b^{\mu
\sigma_1\sigma_2\sigma_3}{+}\frac 12\partial_{\sigma_1} \partial_{\sigma_2}
\partial_{\sigma_3}A_a f_{b\mu}^{\mu\sigma_1\sigma_2\sigma_3}\0\\
&&{+}\frac 12 \partial_{\sigma_1} \partial_{\sigma_2}
\partial_{\sigma_3}b_{a\mu}^\mu c_b^{\sigma_1\sigma_2\sigma_3}-{\frac 32}
\partial_{\sigma_1} \partial_{\sigma_2} \chi_a^{\sigma_3}\partial_{\sigma_3}
d_{b\mu}^{\mu\sigma_1\sigma_2}-{ 3}\partial_{\sigma_1} \partial_{\sigma_2}
b_{a\mu}^{\sigma_3}\partial_{\sigma_3}c_b^{\mu\sigma_1\sigma_2}\0\\
&&{+\frac 32} \partial_{\sigma_1} \partial_{\sigma_2}
\chi_b^{\sigma_3}\partial_{\sigma_3} d_{a\mu}^{\mu\sigma_1\sigma_2}\Big)+\left(
\partial_a b_{b\mu}^\mu - \partial_b
b_{a\mu}^\mu\right)\bigl(\partial_{\sigma_1} \partial_{\sigma_2}
\partial_{\sigma_3}A_a \, c_b^{\sigma_1\sigma_2\sigma_3}\0\\
&&+ 3 \partial_{\sigma_3} b_a^{\sigma_1\sigma_2}\partial_{\sigma_1}
\partial_{\sigma_2}\chi_b^{\sigma_3}\bigr) \0\\
&&+ \left( \partial^a \chi^{b\mu} - \partial^b
\chi^{a\mu}\right)\Big(\partial_{\sigma_1} \partial_{\sigma_2}
\partial_{\sigma_3}A_a \, d_{b\mu}^{\sigma_1\sigma_2\sigma_3}+
\partial_{\sigma_1} \partial_{\sigma_2} \partial_{\sigma_3}\chi_{a\mu}  \,
c_{b}^{\sigma_1\sigma_2\sigma_3}\0\\
&&-3 \partial_{\sigma_1} \partial_{\sigma_2} \chi_a^{\sigma_3}
\partial_{\sigma_3}c_{b\mu}^{\sigma_1\sigma_2} + 3 
\partial_{\sigma_3}b_{a}^{\sigma_1\sigma_2} \partial_{\sigma_1}
\partial_{\sigma_2} b_{b\mu}^{\sigma_3}\Big)
\bigg)\Bigg)  +  {\cal O}({\mathfrak m}^4,6) \biggr{\}} \0
\ee
where $ {\cal O}({\mathfrak m}^4,6)$ means terms of order at least ${\mathfrak
m}^4$ or containing at least six derivatives, and
\be
{\cal S}_4
&=& -  \frac {g^2}{{2}}\, \int d^dx \biggl{\{}  \bigl(\partial_\sigma A^a
\chi^{b\sigma}- \partial_\sigma A^b \chi^{a\sigma}\bigr) \partial_\tau A_a
\chi_b^\tau\label{S4final}\\
&& + \frac {{\mathfrak m}^2}{d} \Big(\bigl(\partial_\sigma A^a \chi^{b\sigma}-
\partial_\sigma A^b \chi^{a\sigma}\bigr) \bigl(\partial_\tau A_a
c_{b\nu}{}^{\tau\nu}+2\partial_\tau \chi_a^\nu b_{b\nu}{}^\tau +  \partial_\tau
b_{a\nu}{}^\nu\chi_b^\tau\bigr)\0\\
&&+\bigl(  \partial_\sigma A^a b^{b\sigma\nu}-  \partial_\sigma A^b
b^{a\sigma\nu}+
\partial_\sigma\chi^{a\nu} \chi^{b\sigma} - \partial_\sigma\chi^{b\nu}
\chi^{a\sigma} \bigr)
\bigl( \partial_\tau A_a b_{b\nu}{}^\tau+ \partial_\tau\chi_{a\nu}
\chi_{b}^\tau\bigr)\Big)\0\\
&& + {\cal O}({\mathfrak m}^4,4) \biggr{\}}\0
\ee 
where $ {\cal O}({\mathfrak m}^4,4)$ means terms of order at least ${\mathfrak
m}^4$ or containing at least four derivatives.

${\cal S}$ does not have the elegant form of \eqref{YMh}, but it is a good
starting point for quantization.

\subsection{Back to the propagator}
\label{ss:backtoprop}

The kinetic operator in \eqref{kineticnew} is
\be
&&K_{ab}^{\{\mu\}\{\nu\}}(x,{\mathfrak m})=\left( \eta_{ab} \square_x - \frac
{\alpha-1}\alpha
\partial^x_a\partial^x_b\right)N^{\{\mu\}\{\nu\}}({\mathfrak m})\equiv
\EK^x_{ab} N^{\{\mu\}\{\nu\}}({\mathfrak m})\label{Kabx1}
\ee
where  
{\scriptsize
\be
&& N^{\{\mu\}\{\nu\}}({\mathfrak m})=\label{Nm}\\
&&\left(\begin{matrix}
       1 & 0 &\eta^{\nu_1\nu_2} \frac {{\mathfrak m}^2}{2d} &0&
\Pi^{\nu_1\nu_2\nu_3\nu_4} \frac
{{\mathfrak m}^4}{4!d(d+2)}&0\\
0 &\eta^{\mu_1\nu_1} \frac {{\mathfrak m}^2}{d}& 0 &\Pi^{\mu_1\nu_1\nu_2\nu_3}
\frac
{{\mathfrak m}^4}{3!d(d+2)} &0&\ldots\\
\eta^{\mu_1\mu_2} \frac {{\mathfrak m}^2}{2d}& 0& \Pi^{\mu_1\mu_2\nu_1\nu_2}
\frac
{{\mathfrak m}^4}{4d(d+2)}&0&\ldots&\ldots\\
0 &\Pi^{\mu_1\mu_2\mu_3\nu_1} \frac {{\mathfrak m}^4}{3!d(d+2)} &0
&\ldots&\ldots&\ldots\\
 \Pi^{\mu_1\mu_2\mu_3\mu_4} \frac {{\mathfrak m}^4}{4!d(d+2)}&0&\dots
&\ldots&\ldots&\ldots\\
0 &\dots &\ldots&\ldots&\ldots&\ldots\\
 \end{matrix}\right)\0
\ee }
{If the inverse of this matrix exists} the propagator in momentum space is 
 \be
&& \widetilde P^{\{\mu\}\{\nu\}}_{ab}(k,{\mathfrak m}) =-i
 \left(\frac {\eta_{ab}}{k^2} 
+(\alpha-1) \frac {k_ak_b}{k^4} \right)
\,M^{\{\mu\}\{\nu\}}({\mathfrak m})\label{prophahbtilde}
\ee
where $M$ is the inverse of $N$, i.e.
\be
N^{\{\mu\}\{\nu\}}({\mathfrak m}) M_{\{\nu\}\{\lambda\}}({\mathfrak m})=
\delta^{\{\mu\}}_{\{\lambda\}}\0
\ee
$M$ must have the structure
\be
&&M^{\{\mu\}\{\nu\}}({\mathfrak m})=\label{Mm2}\\
&&\left(\begin{matrix}
       1 & 0 &\eta^{\nu_1\nu_2} \frac {a_{0,2}} {{\mathfrak m}^2} &0&
t^{\nu_1\nu_2\nu_3\nu_4} \frac {a_{0,4}}{{\mathfrak m}^4}&0\\
0 &\eta^{\mu_1\nu_1}\frac {a_{1,1}}{{\mathfrak m}^2}& 0
&t^{\mu_1\nu_1\nu_2\nu_3} \frac
{a_{1,3}}{{\mathfrak m}^4} &0&\ldots\\
\eta^{\mu_1\mu_2} \frac {a_{2,0}} {{\mathfrak m}^2}& 0& t^{\mu_1\mu_2\nu_1\nu_2}
\frac{a_{2,2}}
{{\mathfrak m}^4}&0&\ldots&\ldots\\
0 &t^{\mu_1\mu_2\mu_3\nu_1} \frac{a_{3,1}} {{\mathfrak m}^4} &0
&\ldots&\ldots&\ldots\\
 t^{\mu_1\mu_2\mu_3\mu_4}  \frac{a_{4,0} }{{\mathfrak m}^4}&0&\dots
&\ldots&\ldots&\ldots\\
0 &\dots &\ldots&\ldots&\ldots&\ldots\\
 \end{matrix}\right)\0
\ee
where $t^{\{\mu\} \{\nu\}}$ are tensors constructed out of $\eta$, which are
symmetric in $\{\mu\}$ and $\{\nu\}$ separately. $a_{i,j}$ are constants to be
determined, with $a_{i,j}=a_{j,i}$.
\vskip 1cm

{
{\bf Comment}.  As noted in the previous subsection, the propagators in the
frozen momentum background  do not suffer from the pathological feature of
higher and higher inverse powers of $u^2$. On the other hand the move to the
frozen momentum background does not leave the eom's and the HS gauge symmetry unaffected. 
If the momentum $u$ is frozen to
$\mathfrak m$ we cannot anymore partially integrate over it, as it is necessary
to do in order to guarantee the property \eqref{comm0}, which is basic in order
to prove that
\be
\langle\!\langle [G_{ab}\ast G^{ab},\varepsilon]\rangle\!\rangle=0
\label{GabGabepsilon}
\ee
Do we have to conclude that in this phase 
the dynamics is different and the HS gauge symmetry is completely lost? 
We can actually convince ourselves that the symmetry, in {another} form, does survive.
Let  us return to \eqref{deltacalS3}. The rescaling to the frozen momentum
background affects this equation simply by the change of $u^2$ to ${\mathfrak
m}^2$. The latter affects the eom's, because we cannot anymore use the argument that
the master eom has to vanish for any $u$. For instance the equation $\EF_0(x)=0$
is modified to an infinite series
\be
0=\EF_0(x) + \frac{ {\mathfrak m}^2}d \, \EF_{2\mu}{}^\mu +3 \frac{ {\mathfrak
m}^4}{d(d+2)} \EF_{4\mu\nu}{}^{\mu\nu}+\ldots \label{eommodified}
\ee
The second and following terms in the RHS are nothing but traces of the eom's \eqref{eomcomponents}. } Therefore the solutions of \eqref{eomcomponents} are also solutions of the new eom's such as \eqref{eommodified}. Moreover we know that the latter are covariant under HS gauge transformations. Therefore we expect that in some reshuffled form the HS gauge
symmetry persists in the frozen momentum background. In fact this must be so because the
operation of going to the frozen momentum background is invertible. We can start from ${\cal S}$,
invert all the rescalings and reintroduce the integral over $u_\mu$, and finally we recover the
form \eqref{YMh} of the action. Thus, when we go to the frozen momentum background we are simply hiding the HS gauge symmetry, which in ${\cal S}$ takes a highly nonlinear form. 
Needless to say this is reminiscent of what happens in a
spontaneously broken symmetry phase of an ordinary gauge theory.}

\section{Feynman diagrams in the frozen momentum background}

  Let us
return to
\eqref{pertseriestilde} in Appendix  and rewrite it in terms of  components
fields (for two
point functions, for the sake of simplicity)
\be
\langle  \tilde h^{\mu_1\ldots \mu_m} _{a}(q_1)\tilde h_{b}^{\nu_1\ldots \nu_n}
(q_2)\rangle  =\frac {\delta}{\delta \tilde j^{a}_{\mu_1 \ldots \mu_n}(-q_1)}
\frac {\delta}{\delta\tilde j^{b}_{\nu_1\ldots\nu_n}(-q_2)}
\,
e^{i S_{int}\left( \frac {\delta}{\delta\tilde j^a}\right)} 
e^{ \langle\!\langle \tilde j^a \tilde  P_{ab} \tilde j^b 
\rangle\!\rangle}\Big{\vert}_{\tilde j=0}\label{hamhbn}
\ee
The crucial objects are $S_3$, $S_4$ and $\langle\!\langle
\tilde j^a \tilde  P_{ab} \tilde j^b  \rangle\!\rangle$. We have to rewrite them
in the frozen momentum language.
Let us start from the latter. The needed re-definitions are:
\be
u^2 = \mm^2 {\rm u}^2,\quad  x^\mu \to x^{'\mu}=  x^\mu {\rm u}, \quad
k^\mu \to k^{'\mu} = \frac {k^\mu}{\rm u}, \quad h_a^{\mu_1\ldots\mu_s}\to
h_a^{'\mu_1\ldots\mu_s}= {\rm u}^s h_a^{\mu_1\ldots\mu_s} \label{redef1}
\ee
This implies, in particular,
\be
\tilde j^a_{\mu_1\ldots\mu_s}\to\tilde j^{'a}_{\mu_1\ldots\mu_s}= {\rm
u}^{-s}\tilde j^a_{\mu_1\ldots\mu_s}
\label{redef2}
\ee
Taking into account the explicit form of the propagator, see
\eqref{prophahbtilde}, we can decompose the integrand of 
$\langle\!\langle \tilde j^a \tilde  P_{ab} \tilde j^b  \rangle\!\rangle$ into a
$\rm u$-independent factor and a $\mm^d {\rm u}^{d-2}$ factor. Finally, as
before, we can factor out ${\cal V}_d =   {\mathfrak m}^d\int d^d{\rm u} \, {\rm
u}^{d-2}$ and simplify it to get
\be
\langle\!\langle \tilde j^a \widetilde  P_{ab} \tilde j^b  \rangle\!\rangle_0=
\sum_{\footnotesize{\begin{matrix}n,m\\{n+m=even}\end{matrix}}} \frac
1{\mm^{n+m}}\int \frac {d^d k}{(2\pi)^d} \,\tilde j^a_{\mu_1\ldots \mu_m}(k)
\,\widetilde P_{ab}^{\mu_1\ldots\mu_m\nu_1 \ldots \nu_n} (k)\,\tilde
j^b_{\nu_1\ldots\nu_n}(-k)\label{jaPabjbkred}
\ee
where
\be
\widetilde P_{ab}^{\mu_1\ldots\mu_m\nu_1 \ldots \nu_n} (k)=-i\, a_{m,n}\,
t^{\mu_1\ldots\mu_m\nu_1 \ldots \nu_n} \left(\frac {\eta_{ab}}{k^2}+(\alpha-1)
\frac {k_a k_b}{k^4}\right)\label{Pabamntk}
\ee

Next we have to rewrite $S_3$. Here a more subtle representation of $u_\mu$ must
be used: $u_\mu = \mm\,{\rm u}\, n_\mu$, where $n_\mu$ represents the normal
unit vector to the sphere $S^{d-1}$ of radius 1, such that
\be 
\int_{S^{d-1}}dn \,\, n^{\mu_1}\ldots\ldots n^{\mu_{2n}}=\frac{ \Pi^{\mu_1
\ldots\ldots\mu_{2n}}}{d(d+2)\ldots (d+2n-2)} \label{intSd-1}
\ee

 Using \eqref{S3tilde}, rewriting it in components, rescaling as above,   then
factoring out ${\cal V}_d $ and simplifying it as before, we get
{\footnotesize
\be
&&{\cal S}_3=-g \int \frac{d^dk_1}{(2\pi)^d} \frac{d^dk_2}{(2\pi)^d}
\frac{d^dk_3}{(2\pi)^d} \delta(k_1+k_2+k_3)k_{1a}\int_{S^{d-1}}dn \,\,
\sum_{l=0}^\infty \frac {\mm^l}{l!} \frac {\delta}{\delta \tilde
j^b_{\lambda_1\ldots \lambda_l}(k_1)} n_{\lambda_1}\ldots n_{\lambda_l}
 \0\\
&& \times \sum_{n,m=0}^\infty \frac {\mm^{n+m}}{n!m!} \left[\frac
{\delta}{\delta \tilde j^a_{\mu_1\ldots \mu_l}(k_2)} \left(n-\frac
{k_3}2\right)_{\mu_1}\ldots \left(n-\frac {k_3}2\right)_{\mu_m} \frac
{\delta}{\delta \tilde j^b_{\nu_1\ldots \nu_l}(k_3)} \left(n+\frac
{k_2}2\right)_{\nu_1}\ldots \left(n+\frac {k_2}2\right)_{\nu_n}\right.\0\\
&&- \left. \frac {\delta}{\delta \tilde j^a_{\mu_1\ldots \mu_l}(k_2)}
\left(n+\frac {k_3}2\right)_{\mu_1}\ldots \left(n+\frac {k_3}2\right)_{\mu_m}
\frac {\delta}{\delta \tilde j^b_{\nu_1\ldots \nu_l}(k_3)} \left(n-\frac
{k_2}2\right)_{\nu_1}\ldots \left(n-\frac
{k_2}2\right)_{\nu_n}\right]\label{S3tilde1}
\ee}
and a similar expression for ${\cal S}_4$.

As a sample let us compute the two-point function from \eqref{hamhbn}. Let us
recall that formula \eqref{hamhbn} gives the two-point function multiplied on
the left by the propagator $P_{h_a^{(m)}h_a^{(m)}}$ and on the right by
$P_{h_b^{(n)}h_b^{(n)}}$, a piece that contributes to the self-energy. In order
to find the genuine two-point function we have to truncate the two external legs
by multiplying by the respective inverse propagators. The calculation proceeds
in the usual way and the result is ( $n+m=even$, otherwise 0)
\be
&&\langle  \tilde h^{\mu_1\ldots \mu_m} _{a}(q_1)\tilde h_{b}^{\nu_1\ldots
\nu_n} (q_2)\rangle_0= g^2 \,q_1^b\, q_2^a\,\delta(q_1+q_2) \int\frac{
d^dp}{(2\pi)^d} \,\frac 1{p^2 (p-q_1)^2}\frac {\mm^{n+m}}{n!m!}
\label{hamhbn2}\\
&&\times \int_{S^{d-1}}dn_1 \int_{S^{d-1}}dn_2\,\,  n_{1\mu_1}\ldots
n_{1\mu_m}n_{2\nu_1}\ldots n_{2\nu_n}\0\\
&&\quad\quad\quad\quad\sum^\infty_{\footnotesize\begin{matrix}l,r,j,s=0\\{l+r \&
j+s=even}\end{matrix}}\frac 1{l!r! j! s!}\, a_{l,s}\, a_{r,j}\,
t^{\lambda_1\ldots \lambda_l\sigma_1\ldots \sigma_s}\, t^{\rho_1\ldots\rho_r
\tau_1\ldots \tau_s} \0\\
&& \times \left[ \left( n_1 +\frac {p-q_1} {2\mm} \right)_{\lambda_1}\ldots 
\left( n_1 +\frac {p-q_1} {2\mm} \right)_{\lambda_l}\left( n_1 +\frac {p} {2\mm}
\right)_{\rho_1}\ldots  \left( n_1+\frac {p} {2\mm} \right)_{\rho_r}\right.\0\\
&& - \left.  \left( n_1 -\frac {p-q_1} {2\mm} \right)_{\lambda_1}\ldots  \left(
n_1 -\frac {p-q_1} {2\mm} \right)_{\lambda_l}\left( n_1 -\frac {p} {2\mm}
\right)_{\rho_1}\ldots  \left( n_1 -\frac {p} {2\mm} \right)_{\rho_r}\right]\0\\
&& \times \left[ \left( n_2+\frac {p-q_1} {2\mm} \right)_{\sigma_1}\ldots 
\left( n_2 +\frac {p-q_1} {2\mm} \right)_{\sigma_s}\left( n_2 +\frac {p} {2\mm}
\right)_{\tau_1}\ldots  \left( n_2 +\frac {p} {2\mm} \right)_{\tau_j}\right.\0\\
&& - \left.  \left( n_2 -\frac {p-q_1} {2\mm} \right)_{\sigma_1}\ldots  \left(
n_2 -\frac {p-q_1} {2\mm} \right)_{\sigma_s}\left( n_2 -\frac {p} {2\mm}
\right)_{\tau_1}\ldots  \left( n_2 -\frac {p} {2\mm} \right)_{\tau_j}\right]\0
\ee
It is easy to check that the dimension of this 2-pt function is $2+m+n$, as it
should be because it must have the same dimension as the propagator with
opposite sign. The full result is a series in $\frac 1{\mm} $. Thus
$\frac 1{\mm} $ plays a role similar to $\sqrt{\alpha'}$ in the field theory
limit of string theory.

\section{Diagonalizing the propagator}

It is clear that the difficulties met in the previous section could be handled
more effectively were we able to diagonalize the kinetic operator
\eqref{kineticop}. {We would like to show in this section that this can be done
provided we go to a traceless basis for the
component fields}. Since this can in fact be formulated on a general ground we
will use a general notation rather than the specific one of the previous
sections. Let us consider a generic master field
\be
\Phi(x,u) = \sum_{n=0}^\infty \frac 1{n!} \phi^{\mu_1\ldots \mu_n}(x) u_{\mu_1}
\ldots u_{\mu_n} 
\label{Phiux}
\ee
$\Phi$ may have additional indices, like for instance $h_a$, but they will be
understood in the sequel.
Its components can be reshuffled as follows
\be
&&\Phi(x,u)= \sum_{n=0}^\infty \sum_{k=0}^n \frac 1{(2n)!}\left(\begin{matrix}
2n\\ 2k\end{matrix}\right)\frac {(2k-1)!!\,\, u^{2k}}{(d+4(n-1)-2(k-1))\ldots
{(d+4(n-1)-4(k-1))} }\0\\
&&\times\,\,\,\tilde \phi^{[k]\mu_{2k+1}\ldots \mu_{2n}} u_{\mu_{2k+1}}\ldots
u_{\mu_{2n}}\0\\
&+& \sum_{n=0}^\infty \sum_{k=0}^n \frac 1{(2n+1)!}\left(\begin{matrix} 2n+1\\
2k\end{matrix}\right)\frac {(2k-1)!!\,\, u^{2k}}{(d+4(n-1)-2(k-1)+2)\ldots
{(d+4(n-1)-4(k-1)+2)} }\0\\
&&\times\,\,\,\tilde \phi^{[k]\mu_{2k+1}\ldots \mu_{2n+1}} u_{\mu_{2k+1}}\ldots
u_{\mu_{2n+1}}\label{decomposition}
\ee
where $\tilde \phi_0=\phi_0, \tilde \phi_1^\mu = \phi_1^\mu $ and $\tilde
\phi^{[k]\mu_1\ldots {\mu_p} } $ are traceless fields obtained from 
$\phi^{[k]\mu_1\ldots \mu_p} $ by subtracting traces. Here $[k]$ denotes a
$k$-fold trace.

Eq.\eqref{decomposition} is simply a rewriting and we cannot in general assign
any meaning to the coefficients of $u_{\mu_{2k+1}}\ldots u_{\mu_{2n}}$ and
$u_{\mu_{2k+1}}\ldots u_{\mu_{2n+1}}$ because they contain powers of $u^2$.
However in the frozen momentum background things change and, after the
rescalings of section \ref{ss:absorbing}, we can define new component fields
\be
\widehat \phi^{\mu_1 \ldots \mu_{2p} } =\tilde \phi^{\mu_1 \ldots \mu_{2p} } + \sum
_{n=1}^\infty 
\frac 1{(2p)! (2n)!}\frac {(2n-1)!!\,\, {\mathfrak m}^{2n}}{(d+2(n-1)+2p)\ldots
{(d+4p)} }\,\, \tilde\phi^{[n]\mu_{1}\ldots \mu_{2p}}\label{phihateven}
\ee
and
\be
&&\widehat \phi^{\mu_1 \ldots \mu_{2p+1} }\label{phihatodd}\\
& =&\tilde \phi^{\mu_1 \ldots \mu_{2p+1} } +
\sum _{n=1}^\infty 
\frac 1{(2p+1)! (2n)!}\frac {(2n-1)!!\,\, {\mathfrak
m}^{2n}}{(d+2(n-1)+2p+2)\ldots {(d+4p+2)} }\,\, \tilde\phi^{[n]\mu_{1}\ldots
\mu_{2p+1}}\0
\ee
Here are some examples:
{
\be
\widehat \phi_0 &=& \phi_0 + \frac {{\mathfrak m}^2}{2d} \phi_2'+  \frac
{{\mathfrak m}^4}{8d(d+2)} \phi_4''+  \frac {{\mathfrak m}^6}{48d(d+2)(d+4)}
\phi_6'''+\ldots \label{phi0}\\
\widehat \phi_1^\mu &=& \phi_1^\mu + \frac {{\mathfrak m}^2}{2(d+2)}
\phi_3'{}^\mu+  \frac {{\mathfrak m}^4}{8(d+2)(d+4)} \phi_5''{}^\mu+ \ldots
\label{phi1}\\
 \widehat \phi_2^{\mu\nu} &=& \tilde \phi_2^{\mu\nu}+ 
\frac {{\mathfrak m}^2}{2(d+4)}\tilde \phi_4'{}^{\mu\nu}+ 
\frac {{\mathfrak m}^4}{8(d+4)(d+6)}\tilde \phi_6''{}^{\mu\nu}+
\ldots\label{phi2}
\ee}
etc. 

Next let us consider our original kinetic term in the frozen momentum background
\be
h^{Ta}_{\{\mu\}} K_{ab}^{\{\mu\}\{\nu\}}(x,\mathfrak m) h^b_{\{\nu\}}
\label{hakabhb1}
\ee
where
{\footnotesize
\be
&&K_{ab}^{\{\mu\}\{\nu\}}(x,{\mathfrak m})=\left( \eta_{ab} \square_x - \frac
{\alpha-1}\alpha
\partial^x_a\partial^x_b\right)
\label{kineticopm}\\ 
&&\times \left(\begin{matrix}
       1 & 0 &\eta^{\nu_1\nu_2} \frac {{\mathfrak m}^2}{2d} &0&
\Pi^{\nu_1\nu_2\nu_3\nu_4}
\frac
{{\mathfrak m}^4}{4!d(d+2)}&0\\
0 &\eta^{\mu_1\nu_1} \frac {{\mathfrak m}^2}{d}& 0 &\Pi^{\mu_1\nu_1\nu_2\nu_3}
\frac
{{\mathfrak m}^4}{3!d(d+2)} &0&\ldots\\
\eta^{\mu_1\mu_2} \frac {{\mathfrak m}^2}{2d}& 0& \Pi^{\mu_1\mu_2\nu_1\nu_2}
\frac
{{\mathfrak m}^4}{4d(d+2)}&0&\ldots&\ldots\\
0 &\Pi^{\mu_1\mu_2\mu_3\nu_1} \frac {{\mathfrak m}^4}{3!d(d+2)} &0
&\ldots&\ldots&\ldots\\
 \Pi^{\mu_1\mu_2\mu_3\mu_4} \frac {{\mathfrak m}^4}{4!d(d+2)}&0&\dots
&\ldots&\ldots&\ldots\\
0 &\dots &\ldots&\ldots&\ldots&\ldots\\
 \end{matrix}\right)\0
\ee}
and
$h^{Ta}_{\{\mu\}} = (A^a, \chi^a_\mu, b^a_{\mu_1\mu_2},
c^a_{\mu_1\mu_2\mu_3},\ldots)$. 

The claim is that \eqref{hakabhb1} is equal to
\be
\widehat h^{Ta}_{\{\mu\}}\widehat K_{ab}^{\{\mu\}\{\nu\}}(x,\mathfrak m)\widehat
 h^b_{\{\nu\}} \label{hakabhb2}
\ee
where
$\widehat h^{Ta}_{\{\mu\}} = (\widehat A^a,\widehat  \chi^a_\mu,\widehat 
b^a_{\mu_1\mu_2},\widehat  c^a_{\mu_1\mu_2\mu_3},\ldots)$, the component fields
being given by \eqref{phihateven} and \eqref{phihatodd}, and 
\be
&&\widehat K_{ab}^{\{\mu\}\{\nu\}}(x,{\mathfrak m})=\left( \eta_{ab} \square_x -
\frac
{\alpha-1}\alpha
\partial^x_a\partial^x_b\right)
\label{kineticopdiag}\\ 
&&\times \left(\begin{matrix}
       1 & 0 & 0 &0& \ldots\\
0 &\eta^{\mu_1\nu_1} \frac {{\mathfrak m}^2}{d}& 0 &0 &\ldots\\
0& 0&\widetilde \Pi^{\mu_1\mu_2\nu_1\nu_2} \frac
{{\mathfrak m}^4}{4d(d+2)}&0&\ldots\\
0 & 0&0 &\widetilde \Pi^{\mu_1\mu_2\mu_3\nu_1\nu_2\nu_3} \frac
{{\mathfrak m}^4}{36(d+2)(d+4)}&\ldots\\
\ldots &\dots &\ldots&\ldots&\ldots\\
 \end{matrix}\right)\0
\ee
The tensors $\widetilde \Pi^{\{\mu\}\{\nu\}}$ are symmetric and traceless in the
indices $\{\mu\}$ and $\{\nu\}$, separately. For instance
\be
\widetilde  \Pi^{\mu_1\mu_2\nu_1\nu_2}& =& \eta^{\mu_1\nu_1}\eta^{\mu_2\nu_2}+
\eta^{\mu_1\nu_2}\eta^{\mu_2\nu_1} - \frac 2d \eta^{\mu_1\mu_2}\eta^{\nu_1\nu_2}
\label{tildePi4}\\
\widetilde  \Pi^{\mu_1\mu_2\mu_3}{}_{\nu_1\nu_2\nu_3}&=&
\delta^{\mu_1}_{\nu_1}\delta^{\mu_2}_{\nu_2}\delta^{\mu_3}_{\nu_3}+ {\rm
perm(\nu_1,\nu_2,\nu_3})\0\\
&&-\frac 2{d+2} \left( \eta^{\mu_1\mu_2} \delta^{\mu_3}_{\nu_1}
\eta_{\nu_2\nu_3} 
+ {\rm perm}(\mu_1,\mu_2,\mu_3)(\nu_1,\nu_2,\nu_3)\right)\label{tildePi6}
\ee
The terms at the RHS in the second line are 6, while the terms in the third line
are 9.

{When replacing \eqref{phihateven} and \eqref{phihatodd} 
in \eqref{hakabhb2}, one should notice that the fields
$\tilde\phi^{[n]\mu_{1}\ldots \mu_{2p}}$
and $\tilde\phi^{[n]\mu_{1}\ldots\mu_{2p+1}}$ can be replaced by
$\phi^{[n]\mu_{1}\ldots \mu_{2p}}$
and $\phi^{[n]\mu_{1}\ldots\mu_{2p+1}}$, respectively, because the difference
is 
made of trace parts (i.e. they contain at least one $\eta$), saturated with some
traceless
tensor  $\widetilde \Pi^{\{\mu\}\{\nu\}}$. 
The equivalence between \eqref{hakabhb1} and \eqref{hakabhb2} has been
explicitly verified up
for the 4$\times$4 matrix in \eqref{kineticopdiag}. We believe it can be
verified in general, but, as will be seen in a moment, this is not necessary. }

Now comes the surprise. These $\widetilde \Pi$ are projectors
\be
\widetilde  \Pi^{\mu_1\mu_2}{}_{\nu_1\nu_2}\widetilde 
\Pi^{\nu_1\nu_2}{}_{\lambda_1\lambda_2}
=   2\widetilde\Pi^{\mu_1\mu_2}{}_{\lambda_1\lambda_2},\quad\quad \widetilde 
\Pi^{\mu_1\mu_2\mu_3}{}_{\nu_1\nu_2\nu_3}\widetilde 
\Pi^{\nu_1\nu_2}{}_{\lambda_1\lambda_2}
=   6\widetilde\Pi^{\mu_1\mu_2\mu_3}{}_{\lambda_1\lambda_2\lambda_3},
\ldots\label{projectors}
\ee  
Therefore the inverse of the matrix in \eqref{kineticopdiag} does not exist  on the full space of unconstrained component fields. Of course we can invert \eqref{kineticopdiag} on the space of traceless fields. This can be expressed via the restrictions
 \be
b_a'=0,\quad c_a'{}^\mu=0,\quad d_a'^{\mu\nu}=0, \ldots\label{newgaugefixing}
\ee
i.e. all the component fields of $h_a$ being traceless on the $\mu$ indices\footnote{{We remark the fact that the $\widetilde \Pi$'s project out the traceful component fields is a purely algebraic condition, so that the obstruction to the existence of the corresponding propagator is purely algebraic. This is at variance with the  familiar non-invertibility of the kinetic operator for a gauge field, which is dynamical in nature, since the kinetic operator in that case is non-invertible in the direction of the momentum. In fact the propagators of the traceful component fields in the LHS of \eqref{newgaugefixing} do not exist because the corresponding quadratic kinetic terms are absent in the action. Of course such traceful fields may appear in the interaction terms. As will become clear shortly, the exclusion of such fields will be implemented by inserting projectors inside amplitudes. The condition of tracelessness will be one of the conditions to be imposed in order to guarantee the physicality of the latter. This means that the above traceful component fields will never appear in any physical amplitude even without imposing their vanishing.}}.

We can now return
to the kinetic operator \eqref{kineticop} and realize that, in this new gauge
all the non-diagonal terms in the matrix $N^{\{\mu\}\{\nu\}}$ are absent and
only the diagonal ones survive with the replacement
\be
\frac 1{n!}\Pi^{\mu_1\ldots \mu_n}{}_{\nu_1\ldots \nu_n} \rightarrow 
\Delta^{\mu_1\ldots \mu_n}{}_{\nu_1\ldots \nu_n}=\frac 1{n!}\left( 
\delta^{\mu_1}_{\nu_1}\ldots
\delta^{\mu_n}_{\nu_n}+{\rm perm} (\mu_1,\ldots,\mu_n)\right)\label{Delta}
\ee
The summation in the RHS has in total $n!$ different products of Kronecker 
delta's. Let us call
the new matrix
 $\widetilde N^{\{\mu\}\{\nu\}}$. Its inverse  $\widetilde M^{\{\mu\}\{\nu\}}$
is easy to determine 
\be
\widetilde M^{\{\mu\}}_{\{\nu\}}({\mathfrak m})=
\left(\begin{matrix}
       1 & 0 & 0&0&0&\ldots\\
0 &\delta^{\mu_1}_{\nu_1}\frac {1}{{\mathfrak m}^2}& 0 &0 &0&\ldots\\
0& 0& \Delta^{(2)\mu_1\mu_2}_{\nu_1\nu_2}
\frac{1}
{{\mathfrak m}^4}&0&0&\ldots\\
0 &0 &0 &\Delta^{(3)\mu_1\mu_2\mu_3}_{\nu_1\nu_2\nu_3}
\frac{1}
{{\mathfrak m}^6}&0&\ldots\\
0 &0 &0 &0&\Delta^{(4)\mu_1\mu_2\mu_3\mu_4}_{\nu_1\nu_2\nu_3\nu_4}
\frac{1}
{{\mathfrak m}^8}&\ldots\\
\ldots&\dots &\ldots&\ldots&\ldots&\ldots\\
 \end{matrix}\right)\label{tildeMm}
\ee
from which one can extract the values of the coefficients $a_{l,r}$: 
$a_{l,l}= \frac 1{l!}$ and $=0$ otherwise. The tensors $t_{\mu_1\ldots \mu_n
\nu_1\ldots \nu_n}$ are given
by $\Delta^{(n)}_{\mu_1\ldots \mu_n \nu_1\ldots \nu_n}$.
Thus the propagators for the traceless symmetric (in the $\mu$ indices) tensors 
are
\be
  \widetilde P^{\{\mu\}\{\nu\}}_{ab}(k,{\mathfrak m}) =-i\,
 \left(\frac {\eta_{ab}}{k^2} 
+(\alpha-1) \frac {k_ak_b}{k^4} \right) \widetilde 
M^{\{\mu\}\{\nu\}}({\mathfrak 
m})
\label{diagonalpropagators}
\ee
Replacing these in \eqref{hamhbn2} one obtains 
the formula for the 2-pt correlators of any two (traceless in the $\mu$ 
indices) 
components of $h_a^{\mu_1\ldots\mu_n}$. This, in particular, tells us that the 
spectrum of the
HS YM model is made of an infinite set of massless modes with tensor structure
$T^{a\mu_1\ldots\mu_n}$, symmetric and traceless in the $\mu_i$ indices 
  (with positive and negative residues). The  same can be said about the 
components of the 
scalar master field $\Phi$: after quantization and restriction of the domain 
they reduce to traceless symmetric massless modes. 

The perturbative quantum theory is therefore haunted by unphysical states. It 
is 
however possible to show that such states are harmless. In the internal lines 
of 
physical amplitudes only physical states propagate. Showing this is the purpose 
of the next section.

\section{{The no-ghost argument}}
\label{sec:noghost}

It is customary to say that the physical modes of 
$h_a^{\mu_1\ldots\mu_n}$ 
are the transverse ones, i.e. $h_i^{j_1\ldots j_n}$, with 
$i,j_1,\ldots,j_n=2,\ldots,d-1$. Unphysical modes are thus identified with
the temporal and longitudinal modes, { i.e. the ones for which} the indices 
$a, \mu_1,\ldots, \mu_n$ take the value 0 or 1. Although, 
for the sake of simplicity, we will often use similar expressions, this language 
is somewhat inaccurate.
A more precise definition is as follows: by physical modes of 
$h_a^{\mu_1\ldots\mu_n}$
we understand solutions of its free massless equation of motion represented by a 
plane wave multiplied by a polarization, say $\theta_a^{\mu_1\ldots\mu_n}$, with 
non-negative norm: $\theta_a^{\mu_1\ldots\mu_n} \theta^a_{\mu_1\ldots\mu_n}\geq 
0$. To distinguish these {\it physical modes} from the strictly transverse ones,  i.e. $h_i^{j_1\ldots j_n}$,  we shall refer to the latter as {\it strictly physical}.
As we shall see later on, {in order to guarantee Lorentz covariance, 
beside the traceless transverse modes also other zero-norm
traceful modes must contribute.} This fact should be always taken into account.
 In the perturbative expansion the modes propagating in the internal lines are 
usually off-shell. One such mode will be called physical when, cutting the 
internal line where it propagates and putting it on shell, it satisfies the 
previous definition\footnote{{According to the previous definition, among such 
propagating states there are  also modes with zero norm polarization. But, as one can argue from the
path integral approach to perturbative quantization, the latter represent a set of zero 
measure.} }.

  The total number of modes in  $h_a^{\mu_1\ldots\mu_n}$ is 
$d\left(\begin{matrix}d 
+n-1\\n\end{matrix}\right)$.
 As pointed out above
in the HS YM theory there are plenty of unphysical modes. Those
labeled, say, by $a=0, 1$ can be eliminated by means of the ordinary 
FP ghosts\footnote{{It is worth insisting that
this expression is oversimplified: eliminating the modes with $a=0, 1$ 
breaks Lorentz covariance, while the formalism of FP ghosts does the job while preserving it.} }. But 
the temporal and longitudinal modes in the $\mu$ indices require a different 
treatment. The reason is that there is no explicit gauge invariance that 
correspond to these unphysical modes: much like with the local Lorentz 
invariance, the theory, in the present formulation, comes without an explicit 
underlying gauge symmetry. Leaving aside for the moment the problem of how to unfold this 
symmetry in a new formulation of the theory, let us focus on the problem of how the corresponding 
unphysical modes can be excluded from the physical processes. 

The argument is in principle very simple. It consists in decomposing the tensor 
$\Delta^{(n)}_{\mu_1\ldots \mu_n,\nu_1\ldots \nu_n}$, which is the identity in 
the space of symmetric tensors of order $n$, into a sum of orthogonal 
projectors, each corresponding to a representation of  the 
Lorentz group. Only one projector in this sum projects to physical states, 
which 
are traceless and transverse. Therefore replacing in \eqref{tildeMm}  the 
identity  $\Delta^{(n)}_{\mu_1\ldots \mu_n,\nu_1\ldots \nu_n}$ with this unique 
(Lorentz covariant) projector guarantees that, in amplitudes with physical 
states in the external 
legs, only physical states propagate in the internal line, thus ensuring the 
absence of propagating unphysical modes in any physical process.

Let us introduce the elementary projectors
\be
\pi_{\mu\nu}= \eta_{\mu\nu} - \frac {k_\mu k_\nu}{k^2} , \quad 
\quad\omega_{\mu\nu} =\frac {k_\mu k_\nu}{k^2} \label{elemproj}
\ee
with the properties
\be
\pi_{\mu\nu}\, \pi^{\nu}{}_\lambda= \pi_{\mu\lambda},\quad\quad 
\omega_{\mu\nu}\, \omega^{\nu}{}_\lambda= \omega_{\mu\lambda},\quad\quad 
\pi_{\mu\nu}\, \omega^{\nu}{}_\lambda=0  \label{elemproj1}
\ee
$\pi$ is transverse, while $\omega$ is not.

Let us start with the case $n=2$ (the cases $n=0,1$ are treated separately) 
which is well-known, see for instance \cite{VN,SJT}:
\be
\Delta^{(2)}_{\mu_1\mu_2,\nu_1\nu_2} = P^{(2)}_{\mu_1\mu_2,\nu_1\nu_2}+ 
P^{(1)}_{\mu_1\mu_2,\nu_1\nu_2}+ P^{(0)}_{\mu_1\mu_2,\nu_1\nu_2}+ \overline 
P^{(0)}_{\mu_1\mu_2,\nu_1\nu_2}\label{spin2}
\ee
where
\be
P^{(2)}_{\mu_1\mu_2,\nu_1\nu_2}&=& \frac 12 
\left(\pi_{\mu_1\nu_1}\pi_{\mu_2\nu_2} +\pi_{\mu_1\nu_2}\pi_{\mu_2\nu_1} 
\right)-\frac 1{d-1} \pi_{\mu_1\mu_2}\pi_{\nu_1\nu_2}\label{spin2a}\\
P^{(1)}_{\mu_1\mu_2,\nu_1\nu_2}&=& \frac 12 
\left(\pi_{\mu_1\nu_1}\omega_{\mu_2\nu_2} +\pi_{\mu_1\nu_2}\omega_{\mu_2\nu_1}+ 
\omega_{\mu_1\nu_1}\pi_{\mu_2\nu_2} +\omega_{\mu_1\nu_2}\pi_{\mu_2\nu_1}\right) 
\label{spin2b}\\
P^{(0)}_{\mu_1\mu_2,\nu_1\nu_2}&=&\frac 1{d-1} 
\pi_{\mu_1\mu_2}\pi_{\nu_1\nu_2}\label{spin2c}\\
\overline P^{(0)}_{\mu_1\mu_2,\nu_1\nu_2}&=& 
\omega_{\mu_1\mu_2}\omega_{\nu_1\nu_2}\label{spin2d}
\ee
These are projectors
\be
P^{(2)}_{\mu_1\mu_2,\nu_1\nu_2}P^{(2)\nu_1\nu_2,}{}_{\lambda_1\lambda_2}&=&P^{
(2)}_{\mu_1\mu_2,\lambda_1\lambda_2}, \quad\quad {\rm etc.}\label{P2P2}
\ee
orthogonal to one another. $P^{(2)}$ is transverse and traceless.  $P^{(1)}$ is 
traceless but not transverse,  $P^{(0)}$ is transverse but not traceless,  
$\overline P^{(0)}$ is neither transverse nor traceless.

The physical projector for $n=3$ is
\be
P^{(3)}_{\mu_1\mu_2\mu_3,\nu_1\nu_2\nu_3}&=& \frac 16 \Big( 
\pi_{\mu_1\nu_1}\pi_{\mu_2\nu_2} \pi_{\mu_3\nu_3} +  
\pi_{\mu_1\nu_1}\pi_{\mu_2\nu_3} \pi_{\mu_3\nu_2} 
+  \pi_{\mu_1\nu_2}\pi_{\mu_2\nu_3} \pi_{\mu_3\nu_1} \label{spin3a1}\\
&&+ \pi_{\mu_1\nu_2}\pi_{\mu_2\nu_1} \pi_{\mu_3\nu_3} +  
\pi_{\mu_1\nu_3}\pi_{\mu_2\nu_1} \pi_{\mu_3\nu_2}+  
\pi_{\mu_1\nu_3}\pi_{\mu_2\nu_2} \pi_{\mu_3\nu_1}\Big)\0\\
-\frac 1{3(d+1) }&&\!\!\!\!\!\!\!\Big(  \pi_{\mu_1\mu_2}\pi_{\nu_1\nu_2} 
\pi_{\mu_3\nu_3}+
\pi_{\mu_1\mu_3}\pi_{\nu_1\nu_3} 
\pi_{\mu_2\nu_2}+\pi_{\mu_2\mu_3}\pi_{\nu_2\nu_3} \pi_{\mu_1\nu_1}\0\\
&& + \pi_{\mu_1\mu_2}\pi_{\nu_1\nu_3} \pi_{\mu_3\nu_2}+
\pi_{\mu_1\mu_2}\pi_{\nu_2\nu_3} 
\pi_{\mu_3\nu_1}+\pi_{\mu_1\mu_3}\pi_{\nu_1\nu_2} \pi_{\mu_2\nu_3}\0\\
&&+  \pi_{\mu_1\mu_3}\pi_{\nu_2\nu_3} \pi_{\mu_2\nu_1}+
\pi_{\mu_2\mu_3}\pi_{\nu_1\nu_3} 
\pi_{\mu_1\nu_2}+\pi_{\mu_2\mu_3}\pi_{\nu_1\nu_2} \pi_{\mu_1\nu_3}\Big)\0
\ee 
which is tranverse and traceless.  

The other five orthogonal projectors that, together with these, `resolve' 
$\Delta^{(3)}_{\mu_1\mu_2\mu_3,\nu_1\nu_2\nu_3}$, are all non-transverse and/or 
traceful. These two examples are particular cases of a general result, whose 
proof is given in 
Appendix \ref{sec:spinproj}. There we show that
\be
 \Delta^{(n)}_{\mu_1\ldots \mu_n,\nu_1\ldots \nu_n}=\EP^{(n)}_{\mu_1\ldots 
\mu_n,\nu_1\ldots \nu_n}+\sum_{i=1}^{p_n} P^{(i)}_{\mu_1\ldots 
\mu_n,\nu_1\ldots 
\nu_n}
\label{PbarPtrtr}
\ee
where $\EP^{(n)}$ is transverse and traceless, while the remaining  $ P^{(i)}$ 
are 
traceful or non-transverse or both. The projector $ \EP^{(n)}$ is the only one 
that 
projects onto the physical degrees of freedom. All the others project to 
nonphysical modes. 
We stress again that $\EP^{(n)}$ is orthogonal to the subspace of unphysical 
modes. 

All the above projectors are mutually orthogonal. It is worth stressing that 
they are all Lorentz covariant.

Let us go back to the propagator \eqref{diagonalpropagators}
 where $M^{\{\mu\}}_{\{\nu\}}({\mathfrak m})$ is given by \eqref{tildeMm}, 
where 
 $\Delta_{\mu_1\ldots \mu_n,\nu_1\ldots \nu_n}$ is replaced by 
$\EP^{(n)}_{\mu_1\ldots \mu_n,\nu_1\ldots \nu_n}$. Now formula \eqref{hamhbn2} 
can be applied for the two-point function of two tensors representing physical 
modes, with the simplifications
\be
a_{m,n} = \frac 1{{\mathfrak m}^{2n}} \delta_{n,m}, \quad\quad 
t^{\mu_1\ldots\mu_m\nu_1\ldots\nu_n} = 
\EP^{(n)\mu_1\ldots\mu_n\nu_1\ldots\nu_n}\delta_{n,m}\label{anmtnm}
\ee
One finally gets
\be
&&\langle  \tilde h^{\mu_1\ldots \mu_m} _{a}(q_1)\tilde h_{b}^{\nu_1\ldots
\nu_n} (q_2)\rangle_0= g^2 \,q_1^b\, q_2^a\,\delta(q_1+q_2) \int\frac{
d^dp}{(2\pi)^d} \,\frac 1{p^2 (p-q_1)^2}\frac {\mm^{n+m}}{n!m!}
\label{2point}\\
&&\times \int_{S^{d-1}}dn_1 \int_{S^{d-1}}dn_2\,\,  n_{1\mu_1}\ldots
n_{1\mu_m}n_{2\nu_1}\ldots n_{2\nu_n}\0\\
&&\quad\quad\quad\quad\sum^\infty_{l,r=0} \frac 1{(l!)^2(r!)^2}\, \frac 
1{\mm^{2l}}\,\frac 1{\mm^{2r}}\,
\EP^{(l)\lambda_1\ldots \lambda_l\sigma_1\ldots \sigma_l}	(p-q_1)\, 
\EP^{(r)\rho_1\ldots\rho_r
\tau_1\ldots \tau_r}(p) \0\\
&& \times \left[ \left( n_1 +\frac {p-q_1} {2\mm} \right)_{\lambda_1}\ldots 
\left( n_1 +\frac {p-q_1} {2\mm} \right)_{\lambda_l}\left( n_1 +\frac {p} {2\mm}
\right)_{\rho_1}\ldots  \left( n_1+\frac {p} {2\mm} \right)_{\rho_r}\right.\0\\
&& - \left.  \left( n_1 -\frac {p-q_1} {2\mm} \right)_{\lambda_1}\ldots  \left(
n_1 -\frac {p-q_1} {2\mm} \right)_{\lambda_l}\left( n_1 -\frac {p} {2\mm}
\right)_{\rho_1}\ldots  \left( n_1 -\frac {p} {2\mm} \right)_{\rho_r}\right]\0\\
&& \times \left[ \left( n_2+\frac {p-q_1} {2\mm} \right)_{\sigma_1}\ldots 
\left( n_2 +\frac {p-q_1} {2\mm} \right)_{\sigma_s}\left( n_2 +\frac {p} {2\mm}
\right)_{\tau_1}\ldots  \left( n_2 +\frac {p} {2\mm} \right)_{\tau_j}\right.\0\\
&& - \left.  \left( n_2 -\frac {p-q_1} {2\mm} \right)_{\sigma_1}\ldots  \left(
n_2 -\frac {p-q_1} {2\mm} \right)_{\sigma_s}\left( n_2 -\frac {p} {2\mm}
\right)_{\tau_1}\ldots  \left( n_2 -\frac {p} {2\mm} \right)_{\tau_j}\right]\0
\ee
{where it is understood that $\EP^{(0)}=1$ and 
$\EP^{(1)\mu}{}_\nu=\delta^\mu_\nu$. Since the projector $\EP^{(n)}$ are Lorentz 
covariant, the above amplitude is also Lorentz covariant.  }

The highest spin projectors $\EP^{(n)}$, { for $n>1$}, allow 
only the physical modes to propagate (and do not exclude any of them). How this 
happens is explained in subsection \ref{ssec:transverse} below.  Formulas for 
more general amplitudes 
can be easily obtained via the ordinary Feynman rules. The presence of these 
projectors guarantees that only physical degrees of freedom propagate in the 
internal lines. 

{
The discussion concerning the amplitude with two external legs $h^{\mu_1\ldots 
\mu_m} _{a}$ and $h_{b}^{\nu_1\ldots\nu_n}$ so far takes care of the physicality 
relative to the indices labeled by $\mu_i$. The physicality for the modes 
labeled by $a$ and $b$, 
like in ordinary gauge theories, can be 
implemented with the use of FP ghosts. Therefore hereinafter we are going to 
enact the machinery of FP ghosts according to the action 
\eqref{YMhquantumint}}\footnote{{This means 
precisely that equations of motion and BRST symmetry are those relative to 
the eq.\eqref{YMhquantumint}}, {on which the rescaling of $u$ has to 
be carried out according
to the rules at the beginning of sec. 8.3, and that ordinary space-time 
integrated action 
terms must be extracted according to footnote 12.}}.
{  Their propagator is the same as \eqref{kineticop},} 
{ except that $ \eta_{ab} \square_x - \frac
{\alpha-1}\alpha \partial^x_a\partial^x_b$ is replaced by $\square_x$. It has 
therefore the same invertibility problems as \eqref{kineticop}.} 
{Proceeding in the same way we exclude traceful and 
non-transverse components and 
restrict to traceless and transverse  FP ghosts. Eventually the ghost-antighost 
propagator is
\be
  \widetilde P_{\{\mu\}\{\nu\}}(k,{\mathfrak m}) =
 \frac {i}{k^2} 
\widetilde M_{\{\mu\}\{\nu\}}({\mathfrak m})
\label{diagonalpropagatorsghost}
\ee
where, in $M_{\{\mu\}\{\nu\}}$, the identity $ \Delta^{(n)}_{\mu_1\ldots 
\mu_n,\nu_1\ldots \nu_n}$ is replaced by the overall traceless and transverse 
projector ${\EP}^{(n)}_{\mu_1\ldots \mu_n,\nu_1\ldots \nu_n}$. } 
{There are no other traceless and transverse projectors in 
$\Delta^{(n)}$,  and 
${\EP}^{(n)}$ is orthogonal to all the other projectors.}

{Once the machinery of FP ghosts is switched on it takes 
automatically care of the 
unphysical modes of $h_a^{\mu_1\ldots\mu_n}$ in the index $a$, i.e. the temporal 
and longitudinal modes, which can be identified 
with the modes labeled by $a=0,1$.  Formula \eqref{2point},} { as 
well as all amplitude formulas constructed in a similar way, 
are  fit to represent amplitudes among states with transverse and  traceless 
polarizations. 
We stress that the projectors $\EP^{(n)}$ to traceless and transverse states are 
Lorentz covariant.
Therefore all the amplitudes constructed in the same  
way will preserve (global) Lorentz covariance.}

{We stress again that the splitting between traceless 
transverse modes 
and the remaining ones is operated in an algebraic way, by means of mutually 
orthogonal projectors, which preserve Lorentz covariance\footnote{{The algebraic approach to 
physicality of amplitudes, i.e. the use of projectors to project out unphysical states, is allowed by the simple pole structure of the propagators:  in the gauge fixed form they are proportional to $\frac 1{k^2}$ times a projector. This is to be compared with the polar structure   
in various theories of gravity, see for instance \cite{SJT}, where the structure is typically of the form
\be
\frac 1{k^2(k^2-m^2)}P =\frac 1{m^2}\left(-\frac 1{k^2} + \frac 1{k^2-m^2}\right)P\label{pole}
\ee
where $P$ is a projector and $m^2$ is a square mass term which may also be negative. In this situation one can have a tachyon, and, in any case,  a negative residue pole which is algebraically impossible to disentangle from the one with positive residue. 
}}}

\subsection{Physical degrees of freedom and  Lorentz covariance} 
\label{ssec:transverse}

Let us see the counting of degrees of freedom. As already pointed 
out, the HS gauge degrees of freedom contained in $\varepsilon (x,u)$ take care 
of eliminating all the components $h_0^{\mu_1\ldots \mu_n}$ and 
$h_1^{\mu_1\ldots \mu_n}$ via the gauge fixing
$\partial^a h_a(x,u)=0$ and the residual gauge invariance, or, alternatively, 
the formalism of FP ghosts. Once this is done we are left with the modes 
$h_i^{\mu_1\ldots\mu_n}$, $i=2,\ldots, d-1$. Out of these, the unphysical modes 
$h_i^{\mu_1\ldots\mu_n}$ with $n>1$ and $\mu_1,...,\mu_n= 0$ or 1, are 
eliminated by the $\EP^{(n)}$ projectors introduced above. 
This implies the reduction from the n-th order symmetric tensor 
representation of the Lorentz group to  the n-th order symmetric traceless 
tensor representation of its little group, which 
identifies the transverse physical degrees of freedom, plus a few more 
representations of the little group which are needed to guarantee Lorentz 
covariance. 

{ 
Let us start from the simplest transverse projector $\pi_{\mu\nu}$. Let  
$\zeta^\nu$ be a spin 1 generic polarization for a massless on-shell particle. 
Applying $\pi_{\mu\nu}$ to it we obtain a new one which is transverse
\be
\pi_{\mu\nu}\zeta^\nu = \bar \zeta_\mu , \quad\quad \bar \zeta \cdot k =0, \quad 
\quad k^2=0\label{trans1}
\ee
With a well-known argument, by means of a Lorentz transformation, we can reduce $k_\mu$ to the 
form $(k_0,k_1,0,\ldots,0)$, with $k_0=k_1=k$. Then the polarization components, 
due to \eqref{trans1}, must satisfy
$\bar\zeta_0=\bar\zeta_1$, so  that its norm
\be
\bar \zeta_\mu \bar\zeta^\mu = \sum_{i=2}^{d-1} \bar\zeta_i^2\geq 
0\label{trans2}
\ee
A state with such a polarization corresponds to the definition of physical state 
given above. We notice that by a Lorentz transformation it can be reduced to the transverse
form $(0,0,\bar\zeta_2',\ldots ,\bar \zeta_{d-1}')$. Thus the transversality 
condition maps a state in the fundamental representation of the Lorentz group to 
a state in the fundamental representation of its little group SO(d-2). But we see
that a Lorentz transformation does in general switch on non-transverse components
of the polarization.}

Let us see this at work for traceless symmetric tensor representations of order two\footnote{In the rest of this subsection, for simplicity, we disregard the index $a$ (or its trasverse part $i$) since  it is a spectator index.}.
In 
this case we start from a state $\zeta_{\mu\nu} e^{ik \cdot x}$, where $\zeta$ 
is symmetric.
Projecting it with the symmetric transverse traceless projector $\EP^{(2)}\equiv 
P^{(2)}$ (see \eqref{spin2a}) we obtain a symmetric transverse traceless polarization
\be
\bar \zeta_{\mu_1\mu_2} = 
P^{(2)}_{\mu_1\mu_2}{}^{\nu_1\nu_2}\zeta_{\nu_1\nu_2},\quad\quad \bar 
\zeta_{\mu_1\mu_2}k^{\mu_1}= \zeta_{\mu_1\mu_2}k^{\mu_2}=0\label{trans3}
\ee
Setting $k_0=k_1=k$ implies $\bar\zeta_{00}= 
\bar\zeta_{01}=\bar\zeta_{10}=\bar\zeta_{11}$ and 
$\bar\zeta_{0i}=\bar\zeta_{1i}$.
Thus 
\be
\bar\zeta_{\mu_1\mu_2} \bar\zeta^{\mu_1\mu_2} = \sum_{i,j=2}^{d-1} 
\bar\zeta_{ij}^2\geq 0\label{trans4}
\ee
Again the state with such a polarization is physical according to our definition 
because the norm is non-negative. Transversality and tracelessness dramatically reduce the dimension 
of the original symmetric representation of the Lorentz group: the latter
has dimension $\frac {(d+1)d}2$ while the two-index symmetric traceless 
representation of its little  group (where $\zeta_{ij}$ lives) has dimension $\frac {d(d-3)}2$.

In order to preserve a Lorentz covariant formalism, however, we need, beside the 
(strictly physical) symmetric transverse traceless polarization $\zeta_{ij}$, also the zero norm singlet
and the fundamental representations of the little group represented by $\bar\zeta_{00}= 
\bar\zeta_{01} =\bar\zeta_{11}$ and $\bar\zeta_{0i}=\bar\zeta_{1i}$, respectively, which are also zero norm representations.

{It is useful to illustrate the above with afew concrete examples. The symmetric tensor representation both of the Lorentz group and the little group are reducible. For instance in 10 dimensions the symmetric representation of SO(1,9) is the sum of the traceless representation {\bf 54} and the identity {\bf 1} (the trace). The corresponding little group is SO(8), whose symmetric representation splits as a traceless {\bf 35} and {\bf 1}. The strictly physical degrees of freedom are 35. According to the previous analysis to preserve Lorentz covariance we have to add a vector representation {\bf 8} plus an identity one, {\bf 1}. The sum of all these degrees of freedom $35+1+8+1=45$ corresponds to the dimension of the (traceless) adjoint representation of SO(1,9). Another example: in d=9 the little group is SO(7), whose symmetric representation splits into a {\bf 27}+{\bf 1}. If we add to this the vector representation {\bf 7} plus the identity  {\bf 1}, we obtain in  all 36 degrees of freedom, which correspond to the dimension of the adjoint representation of SO(1,8).  In d=4 the little group is SO(2) whose symmetric representation splits into {\bf 2}+{\bf 1}. According to the above scheme we have to add another  {\bf 2}+{\bf 1}. In total 6 dofs, which is the dimension of the adjoint representation of SO(1,3). This is in fact a general scheme. In $d$ dimensions the little group has a symmetric representation of dimension $\frac {d(d-3)}{2} +1$, to which we have to add $(d-2)+1$ dofs to preserve Lorentz. This makes in total $\frac {d(d-1)}2$, which is the dimension of the adjoint representation of SO(1,d-1). Needless to say the adjoint representation is traceless. So to Lorentz covariantize the physical (transverse and traceless)  $\frac {d(d-3)}{2}$ degrees of freedom
we need  $d$ additional degrees of freedom.
}

In the same way with a third order symmetric traceless polarization $\zeta_{\mu\nu\lambda}$, 
transversality implies
\be
\bar\zeta_{0\nu\lambda}=\bar\zeta_{1\nu\lambda}\label{trans5}
\ee
In particular we have $\bar \zeta_{000}=\bar\zeta_{001}=\bar\zeta_{011}=\bar\zeta_{111}$,
$\bar\zeta_{00i}=\bar\zeta_{01i}=\bar\zeta_{11i}$ and $\bar\zeta_{0ij}=\bar\zeta_{1ij}$, 
which induce
\be
\bar \zeta_{\mu\nu\lambda} \bar\zeta^{\mu\nu\lambda} =  
\sum_{i,j,k=2}^{d-1} \bar \zeta_{ijk}^2\geq 0\label{trans6}
\ee
The states obeying the transversality condition are physical. The transverse states defined 
by $\zeta_{ijk}$, which are in the symmetric transverse traceless (strictly physical) 
representation of the little group, need other non-physical representations 
in order to preserve Lorentz covariance (a singlet, vector and symmetric tensor representations)
of the little group. {
For instance the rank three symmetric tensor representation of SO(1,9) splits into a traceless ${\bf 210}$ plus a  vector ${\bf 10}$. The analogous rank three symmetric tensor representation of SO(8) splits into the traceless ${\bf 112}$ plus ${\bf 8}$. According to the previous analysis, in order to preserve Lorentz covariance, we have to add to these the ${\bf 35+1+8+1}$ representations of SO(8). This makes in total 165 dofs. This number splits into the sum ${\bf 45+120}$ of irreducible traceless representations of SO(1,9). The strictly physical dofs are 112. In order to preserve Lorentz covariance we need 53 additional dofs.  In $d$ dimension the rank 3 tensor symmetric representation of SO(1,d-1) splits into the traceless ${\bf \frac 16 d(d-1)(d+4)}$ and the vector ${\bf d}$. The analogous splitting for the little group SO(d-2) is  ${\bf \frac 16 (d-2)(d-3)(d+2)}$ plus ${\bf d-2}$. 
In order to preserve Lorentz covariance we must keep in addition also the representations
${\bf (d-2) +1}$ and ${\bf \frac 13 (d(d-3) +1}$ of  SO(d-2). The total number of dofs that preserve Lorentz covariance are thus
 \be
\left(\frac {d(d-1)(d-2)}6 \right) +\left(\frac {d(d-3)}2 +1\right) +(d-2)+1 = \frac 16 d(d^2-1)\label{dimLCrep}
\ee
The sum splits according to
\be
 \frac 16 d(d^2-1)= \frac 16 d(d-1)(d-2)+\frac 12 d(d-1)\label{dimLCovrep}
\ee
which are the dimensions of two representations of the Lorentz group, the second being the adjoint one. They are of course traceless.}

	The generalization to higher order polarizations is straightforward and we summarize it as follows. {In $d$ dimensions let us start from the symmetric  tensor $n$ representation of the Lorentz group and the corresponding irreducible traceless representation. They have dimension
\be
\left(\begin{matrix} d+n-1\\ n\end{matrix}\right) \quad\quad {\rm and} \quad \quad
\frac{ (d+n-3)(d+n-2)\ldots (d-2)(d-3)}{n!} (d+2n-2),\label{SymrepLor-n}
\ee
respectively.  The symmetric  tensor $n$ representation of the little group and the corresponding irreducible traceless representation have dimension
\be
\left(\begin{matrix} d+n-3\\ n\end{matrix}\right) \quad\quad {\rm and} \quad \quad
\frac{ (d+n-5)(d+n-4)\ldots (d-2)(d-3)}{n!} (d+2n-4),\label{SymirepLit-n}
\ee
respectively. 
The symmetric transverse traceless  representation of the Lorentz 
group, selected by the $\EP^{(n)}$ projector, decomposes into a 
(in general irreducible) symmetric transverse representation of the little group 
characterized by positive norm polarization states, plus a bunch of smaller 
representations with zero norm states. The latter are needed in order to ensure 
Lorentz covariance. As the above argument shows, they are contained in the irrep decomposition
of the little group symmetric $n$-tensor  representation, whose dimension is given by the first number in \eqref{SymirepLit-n}. In addition to this we have to add all the symmetric tensor representation of order less than $n$ and $\geq 2$. It is not difficult to find the sum of all the dofs needed to guarantee the Lorentz covariance by means of the recursion formula 
\be
\left(\begin{matrix} d+p-2\\ p\end{matrix}\right) +\left(\begin{matrix} d+p-2\\ p+1\end{matrix}\right)=\left(\begin{matrix} d+p-1\\ p+1\end{matrix}\right),\label{magicrecur}
\ee
and conclude that such a number, at order $n$, is
\be
\left(\begin{matrix} d+n-2\\ n\end{matrix}\right)\label{magicnumber}
\ee
These states form a (in general reducible) symmetric traceless representation of the Lorentz group. The number of zero norm states is obtained by subtracting from \eqref{magicnumber} the number of positive norm states given by the second entry in  \eqref{SymirepLit-n} }.  {It is of course fitting that such additional zero norm representations 
can be disregarded compared to the negative and positive norm ones. This is the correct way to express physicality for higher massless 
spin modes\footnote{{Unlike 
in the $n=1$ case, 
it is of course impossible in general to reduce, by a Lorentz 
transformation, the state represented by 
$\zeta_{\mu_1\ldots\mu_n}$ to the state represented by $\zeta_{i_1\ldots i_n}$. The analogous 
physicality statement must be expressed in terms of representations of the little group, 
as we have just done.}}} .

{
To conclude this subsection, the number of unphysical states to be eliminated in order to preserve physicality and, simultaneously, Lorentz covariance are
\be
\left(\begin{matrix} d+n-2\\ n-1\end{matrix}\right)\label{magicnumber2}
\ee
Comparing with \eqref{SymrepLor-n}, we see that it is the dimension of the (in general reducible)
symmetric representation of the Lorentz group of order $n-1$, for $n\geq 2$. }

{ Returning to the master field $h_a(x,u)$,
to compensate for the unphysical degrees of freedom \eqref{magicnumber2} we would need a symmetry under local transformations of the master field $h_a$ components, parametrized as follows
\be
\delta h_a^{\mu_1\ldots\mu_n}(x) = \partial^{(\mu_1} \Lambda_a^{\mu_2\ldots\mu_n)}(x)+\ldots, \quad\quad n\geq 2.
\label{newgaugetransf}
\ee
linear in $\Lambda_a^{\mu_2\ldots\mu_n)}$, but with possible additional terms represented by the ellipses. The HS-YM action is clearly not invariant under such transformations. Similarly to local Lorentz invariance, this gauge invariance is completely fixed in the defining action \eqref{YMh}. In the next section we shall see how this additional gauge inveriance can be implemented.
}

\subsection{The case $n=1$}

It remains for us to consider $n=0$ and 1.

The case $n=0$ is in fact a subcase of the one just discussed, corresponding to 
the 
scalar gauge parameter $\epsilon(x)$. The case $n=1$ is not a subcase of the 
previous discussion and requires an {\it ad hoc} analysis. In fact the difference 
between the G and a-G interpretations becomes evident when we go to the 
physical spectrum for the case $n=1$, i.e. the physical modes contained in 
$\chi_a^\mu$.  The initial dofs  are 
$d^2$. In the G case we have
$\frac{d(d-3)}2$ physical modes, in the a-G case the physical modes are 
$(d-2)^2$.
Let us see the two cases in turn.

{ In the a-G case the counting is as follows. From $d^2$ modes 
we subtract $d$ modes corresponding to the gauge parameters $\xi^\mu$, and as 
many modes due to the residual symmetry (this is as usual: the gauge transform 
$ \chi_a^\mu\to \chi_a^\mu + \partial_a \xi^\mu$ eliminates the longitudinal 
modes (in $a$). After the gauge fixing $\partial^a \chi_a ^\mu=0$ we are left  
with the residual symmetry whose parameters satisfy $\square \xi^\mu=0$, they 
eliminate $d$ additional modes, the remaining unphysical modes in the index a.). 
So we are left with $d(d-2)$ modes, say $\chi^\mu_i$ with $i=2,..., d-1$. This 
is not the end because in the propagator \eqref{tildeMm}, \eqref{diagonalpropagators} there is the sum of projectors
$\pi^\mu_\nu+\omega^\mu_\nu$. We have to exclude the projector $\omega$ because 
it allows for negative norm states. We are left with the projector 
$\pi^\mu_\nu$. This is transverse, so it eliminates the non-transverse modes in 
$\mu$, which are 2 for any value of $i$. In conclusion we are left with 
$(d-2)(d-2)$ physical modes (notice that  $\pi^\mu_\nu$ is not traceless, this 
is the reason why the counting is different than with the $\EP$ projectors). }

Let us consider next the G interpretation, in which case we have to take into 
account the 
(recovered) local Lorentz symmetry.  Subtracting the local Lorentz parameters 
we are left with $\left( 
\begin{matrix} d+1 \\ 2\end{matrix}\right)$ modes. Using the gauge parameters 
$\xi^\mu$ in $\varepsilon(x,u)$ and the residual gauge freedom eliminates $2d$ 
parameters, as usual. Finally we are left with $\frac {d(d-3)}2$, which is the 
correct number of physical degrees of freedom for a massles spin 2 particle in 
$d$ dimensions. {In this argument we understand that the local Lorentz symmetry 
has been 
fixed by setting the antisymmetric part of $\chi_a^\mu$ to zero. This point 
perhaps needs a 
(somewhat obvious) formal justification. The introduction of local Lorentz 
invariance changes the 
initial vielbein $E_a^\mu(x)$ to new vielbein $\tilde E_a^\mu(x)$, whose 
transformation 
properties under a LLT are $\delta_\Lambda  \tilde E_a^\mu(x)= \Lambda_a{}^b(x) 
\tilde E_b^\mu(x)$.
The FP method promotes $\Lambda_a{}^b(x)$ to anticommuting ghost field. The 
gauge fixing
$\delta\left(\tilde E_{a\mu}(x)-\tilde E_{\mu a}(x)\right)$ induces 
in the action a term 
$\int d^dx\, \beta_\mu^a(x)\, \Lambda_a{}^b(x) \left(\tilde E_{b\mu}(x)-\tilde 
E_{\mu b}(x)\right)$, 
where $\beta_\mu^a(x)$ is the
antighost (anti-commuting) field. Endowing $\Lambda_a{}^b(x)$ with the 
transformation property
$\delta_\Lambda \Lambda_a{}^b= \Lambda_a{}^c \Lambda_c{}^b$ the transformation 
$\delta_\Lambda$ 
becomes a nilpotent BRST transformation and the above action term invariant, 
provided
$\delta_\Lambda \beta_\mu^a =0$.}

{
In conclusion, the stage is now ready for an analysis of renormalization of HS-YM theories, 
which will be carried out elsewhere.} Here we limit ourselves to
remark that, generally, many contributions like \eqref{hamhbn2} are IR and UV
divergent, the degree of divergence increasing with the power of $\frac 1{\mm}$.
Although these integrals can be dealt with for instance via dimensional
regularization (they are similar to the ones explicitly calculated in \cite{BCDGLS,BCDGS}), {the IR ones may represent a severe challenge.}

In the $n=1$ argument the local Lorentz gauge freedom is essential. We recall 
that in the original model the Lorentz gauge is fixed and we had to enlarge the 
theory by introducing inertial frames and connections in order to recover the 
LL symmetry. Perhaps a similar scheme could explain the presence of the 
unphysical modes   $h_i^{\mu_1\ldots\mu_n}$ with $n>1$ and $\mu_1,...,\mu_n= 0$ 
or 1. By suitably enlarging the theory it may be possible to introduce an 
additional gauge symmetry that accounts for them. {As shown in the next section
one such possibility exists, at the cost of  introducing non-localities in the theory.}

\section{Unfolding the hidden gauge symmetry}
\label{s:hiddengauge}

{
In the previous section we have shown that by inserting the projectors $\EP^{(n)}$ in any propagator of the field $h_a^{\mu_1\ldots \mu_n}$  for $n\geq 2$, we eliminate the unphysical degrees of freedom that, otherwise, would propagate along the same Feynman diagram line. These are the unphysical dofs that it is necessary to exclude in order to preserve at the same time physicality and Lorentz covariance. The price we pay is that the Feynman diagrams we obtain are complicated by the presence of the projectors $\EP^{(n)}$. It is  natural to inquire whether it is possible to avoid such a complication, at the expense of introducing additional degrees of freedom. We have seen above that the unphysical dofs would be naturally eliminated should the theory be endowed with a gauge symmetry like \eqref{newgaugetransf}. We have already noticed above that our action \eqref{YMh} is not invariant under such transformations as \eqref{newgaugetransf}. Much like in the case of local Lorentz symmetry, this would-be gauge symmetry is completely broken in the original formulation of the theory. But we have seen in section \ref{s:LLI} that the local Lorentz symmetry can be completely restored by adding appropriate additional degrees of freedom. In this section we would like to show that the same is actually possible also for this gauge symmetry, which, for convenience, we will refer to henceforth as {\it the hidden gauge symmetry}. To show this we first incorporate the projectors in the original theory. Then we show that we obtain in this way a nonlocal action symmetric under the simple transformations
\be
\delta h_a^{\mu_1\ldots\mu_n}(x) = n\,\partial^{(\mu_1} \Lambda_a^{\mu_2\ldots\mu_n)}(x), \quad\quad n\geq 2.
\label{hiddengaugetransf}
\ee
Here non-local means that the action contains terms with inverse powers of $\square$. }

\subsection{Non-local hidden-gauge-symmetric theory}
{
The first step is rather simple. It consists in replacing everywhere in the action any component field $h_a^{\mu_1\ldots\mu_n}$ with $n\geq 2$ with its projection by the appropriate $\EP^{(n)}$ in configuration space. This means that $\pi_{\mu\nu}$ is replaced by 
\be
\check \pi_{\mu\nu}(\partial) = \eta_{\mu\nu} - \frac{\partial_\mu\partial_\nu}{\square} \label{tildepi}
\ee
For instance
\be
b_a^{\mu_1\mu_2} \longrightarrow \check b_a^{\mu_1\mu_2}=\left( \frac 12 \left( \check \pi^{\mu_1}{}_{\nu_1} \check \pi^{\mu_2}{}_{\nu_2} 
+\check \pi^{\mu_1}{}_{\nu_2} \check \pi^{\mu_2}{}_{\nu_1}\right)- \frac 1{d-1}\check\pi^{\mu_1\mu_2} \check \pi_{\nu_1\nu_2} \right) b_a^{\nu_1\nu_2} \label{bamumu}
\ee
It is convenient at this stage to adopt the compact notation of \cite{FS} for symmetric tensors. So, in particular the transformation \eqref{hiddengaugetransf} becomes
\be
\delta_{\Lambda} h_a = \partial \Lambda_a 
\ee
for any component field. For instance,  the new $b_a$ field takes the form
\be
\check b_a= b_a-\frac {\partial}{\square}\, \partial\!\cdot \!b_a + \frac {\partial^2}{\square^2}\partial\!\cdot\!\partial\!\cdot \! b_a -\frac 1{d-1} \left( \eta - \frac {\partial^2} {\square} \right) \left(b_a'-\frac 1{\square} \partial\! \cdot\!\partial\!\cdot\! b_a\right) \label{tildeba}
\ee
Similarly the component field $c_a^{\mu_1\mu_2\mu_3}$ is replaced by
\be
\check c_a& =& c_a -\frac {\partial}{\square}\, \partial\!\cdot \! c_a+ \frac {\partial^2}{\square^2}\partial\!\cdot\!\partial\!\cdot \! c_a - \frac {\partial^3}{\square^3}\partial\! \cdot\!\partial\!\cdot\!  
\partial\! \cdot\! c_a\label{ca}\\
&&- \frac 1{d+1} \left( \eta - \frac {\partial^2} {\square} \right)\left(c_a'-\frac {\partial}{\square}\, \partial\!\cdot \!c'_a - \frac 1{\square}\partial\!\cdot\!\partial \!\cdot\!c_a + \frac {\partial}{\square^2}\partial\!\cdot\!\partial\!\cdot \! \partial \!\cdot\!c_a \right)\0
\ee
and so on. We recall that in these formulas a dot, $\cdot$, denotes index contraction, a prime $'$ denotes a trace, free $\partial$ a gradient,  and symmetrization of the free indices is understood.}
{ 
After the above replacements \eqref{YMh}} { becomes a nonlocal action, which is, however, automatically invariant under \eqref{hiddengaugetransf}. This follows from the fact that all $\EP^{(n)}$ are transverse to the momentum and from the form of  \eqref{hiddengaugetransf} where a  derivative always factors out. }

In \cite{FS} { in order to realize an unrestricted gauge symmetry, similar to \eqref{hiddengaugetransf}}, {the authors decided to give up locality, and succeeded in writing fully covariant Fronsdal equations of any spin, in which inverse powers of $\square$ feature. What we have done above is somewhat similar, although what we obtain here is a non-local action invariant under the hidden gauge symmetry \eqref{hiddengaugetransf}.}  {In our case, however, we have to take care of an additional important symmetry: the invariance under the HS gauge transformations \eqref{deltahxp}. }{Replacing all the component fields $h_a^{\mu_1\ldots\mu_n}$ according to the above recipe with new fields $\check h_a^{\mu_1\ldots\mu_n}$, non-locally related to the previous ones for  $n\geq 2$, we effectively replace $h_a(x,u)$ with a new superfield $\check h_a(x,u)$ which insures the invariance of the action under the hidden symmetry. We obtain in this way a new action ${\cal Y}{\cal M}({\bf {\check h}})$, which has the same form as \eqref{YMh}. An important remark is that, since the action is form invariant, it remains invariant also under the new HS gauge transformations 
\be
\delta_{\hat\varepsilon} \check h_a(x,u)= \partial_a {\check \varepsilon}(x,u)     -i [\check h_a(x,u) \stackrel{\ast}{,} \check \varepsilon(x,u)] \label{newHSgaugetransf}
\ee   
So the new action is invariant both under these HS gauge transformations and under \eqref{hiddengaugetransf}.
The relation between $\check \varepsilon$ and the $\varepsilon$ of the original HS gauge transform \eqref{deltahxp} is complicated, and, in particular, field dependent and nonlocal. However what matters here is that such a symmetry exists in the new action and allows us to write down the corresponding Ward identities and study renormalization.}
{
In conclusion  the hidden gauge symmetry can be made manifest in the action at the price of introducing non-local terms in it.} One can turn around this conclusion by saying that the non-locality is a gauge artifact, because it disappears if we return to the initial formulation of the theory.

{At this point we should start the analysis of ${\cal Y}{\cal M}({\bf {\check h}})$ from scratch, introduce the FP ghosts for all the symmetries and perturbatively quantize it in the frozen momentum frame. However in \cite{FS}} {it
was shown that, by introducing suitable auxiliary fields, one can recover locality and that the Fronsdal-type equations are Lagrangian. It may be possible to do the same in our case, i.e. introduce appropriate auxiliary fields and render the new action local. Redoing the quantization in this new formulation of the theory would be much more interesting. This is what we would like to implement in  a forthcoming research.}

\section{No-go theorems}

Let us pause to consider the results obtained so far for HS YM theories.
The eom's  are well defined in a Minkowski background in
any dimensions! They are {\it perturbatively local }, i.e. the number of terms 
with a fixed number of derivatives is finite.
They encompass a Maxwell or YM eom, gravity, etc. They are
interacting eom's, which include up to third order (fourth order in the action) 
interactions (infinite order in the metric fluctuation). They are characterized
by a unique coupling $g$.
We have also introduced a perturbative quantization, {in which 
only physical degrees of freedom propagate in physical processes.}

As anticipated in the introduction, this at first cannot be but surprising,
for there exist in the literature no-go theorems forbidding massless HS
particles in flat spacetime. Let us briefly review this issue, relying on the
nice review \cite{Bekaert}

The argument goes as follows (in a 4d Minkowski spacetime).
Particles with spin $s\leq 2$ are known to couple minimally to gravity. Weinberg's
equivalence principle (based on an S-matrix argument), \cite{Weinberg}, states
that all particles of
whatever spin must
as well couple minimally to gravity at low energy  (if we want a non-vanishing
emission of such particles, Weinberg's soft emission theorem). But
the Weinberg-Witten theorem, \cite{WW}, and its generalizations say that HS
particles cannot
couple minimally to gravity. As a consequence HS particles decouple from
low spin ones at low energies, which means that an action containing LS and HS
particles
split into two non-interacting pieces (at low energy). 

These theorems on a general ground are based on the existence of the S-matrix,
which requires in particular the existence of asymptotic states in the full
range of energy; more specifically they are based on a lemma, which in
Lagrangian language, can be formulated as follows: any local polynomial which is
at least quadratic in a spin $s$
massless field, non-trivial on-shell and gauge invariant, contains at least 2$s$
derivatives.

The consequence of this lemma is that any perturbatively local
theory with a Lorentz covariant and gauge invariant energy-momentum tensor
cannot have spin higher than 2. The reason is that the energy-momentum tensor is
assumed to contain two derivatives. Now since the em tensor is quadratic in an
HS field $h^{(s)}$ (the
coupling of $h^{(s)}$ to gravity is in accord with the scheme
$g\! -\! h^{(s)}\! - \! h^{(s)}$), according to the lemma,  it must contain at
least $2s$
derivatives, which is impossible.

In \cite{Bekaert}, from the above no-go theorems  the following
conclusions are drawn
for local cubic vertices in flat space including at least one massless particle:
\begin{enumerate}
\item the number of derivatives in any consistent local cubic vertex is at least
equal to the highest spin  in the vertex;
\item a local cubic vertex containing at least one massless field with spin
higher than 2 contains at least three derivatives;
\item massless higher spin particle couple non-minimally to low-spin particles.
\end{enumerate}

None of these is true for HS YM models. Looking at \eqref{S3final} we see that
in
the
third line the coupling $A\chi b$ contains two derivatives (not three, like the
spin of $b$). This disagrees with 1 and 2.
As for 3, comparing the first and third line, one sees that the coupling
$AA\chi$, which is minimal, has the same structure as the coupling $A\chi b$,
where $b$ has spin 3.

So the question is: where is the bug? Let us observe that there are several ways
to evade the hypotheses on which no-go theorems rely:
\begin{itemize}
\item The em tensor (like in the HS YM case) may not be a polynomial but an
infinite series, like in \eqref{Jmu1mun}.
\item The coupling to gravity via the em tensor is non-minimal, that is it
contains more than two derivatives (for instance, the coupling $\chi bb$ in
\eqref{S3} has four derivatives).
\item No-go theorems always understand Einstein-Hilbert gravity: 
the coupling to gravity is implemented by replacing simple derivatives with
covariant ones. In HS YM models, for instance, this is not the case,
{while covariance is nevertheless implemented.}
\item No-go theorems are based on 
the existence of asymptotic free particle state. The question is: do these
states always exist? or, at least, do they exist in the full range of energy?
For instance, the escape for
Vasilev's models is that they hold in AdS spaces where asymptotic states and
S-matrix do not exist (globally).
\item {Some no-go statements originate from the attempt to implement the 
Fronsdal linearized equations at the interaction level\footnote{{This is the 
case for }\cite{CJM},{ although in the light-cone formalism.}}. In HS YM models 
the linearized 
equations of motion are not the Fronsdal ones.}
\end{itemize}

A particular mention should be perhaps made for \cite{Bekaert2}, which is still 
another type of no-go theorem: it is a clear statement about the impossibility 
of extending the BBvD idea, \cite{BBvD},  to the quartic vertex. However, the 
analysis is limited to the BBvD scheme, where one starts from Fronsdal's free 
theory. As a consequence, its validity seems to be limited to that 
construction. 
In the HS YM models considered above, for instance,  we do not start from 
Fronsdal's 
free theory and do not reproduce it in the free case. The covariance of the 
eom's is attained thanks to the $L_\infty$ structure of the model, see eq. 
(4.28).  In summary, the no-go theorem in \cite{Bekaert2} is certainly 
valuable, 
but it is not clear how far one can extend the range of its applicability.

Summarizing: we do not have a general statement of the no-go theorems, 
their applicability has still to be decided case by case. What we can say about 
the HS YM models is that there are important differences
with one or more hypotheses of the different no-go theorems:
\begin{itemize}
\item an infinite number of fields is involved;
\item the coupling of HS fields to gravity is non-minimal;
\item the em tensor and the other conserved currents are non-polynomial;
\item the covariantization procedure and the gravity that emerges are not the 
conventional ones;
\item { the linearized equations of motion are not the Fronsdal's ones.}
\end{itemize}
{  These specifications certainly cut out many no-go theorems. We are not sure they cut out all of them. {On the other hand there seems to be in the literature a general consensus that the set of existing no-go theorems excludes the possibility of a perturbatively local massless HS theory in a flat background.} However there is another feature of HS-YM models which seems to be decisive in this sense. In perturbation theory we  have developed precise formulas for any type of amplitudes at any loop order in which only physical modes propagate in the internal lines. To do so we have had to introduce appropriate physical projectors for the internal lines. Such projectors are non-local and introduce an effective non-locality in the theory, in the sense that if, in order to unfold the hidden symmetry of the HS YM theory, we incorporate these projectors in the initial HS YM theory  we end up with a more symmetric theory characterized by non-localities (meaning inverse powers of $\square$). This non-locality definitely seems to put offside all the no-go theorems demostrated so far. One may be led to conclude that it sets offside also the HS-YM models. For, as for instance the author of \cite{Taronna:2017wbx} makes it clear, once strict locality is relaxed, the Noether procedure by which gauge invariance is imposed allows in principle for an infinite number of solutions. However the main point our paper is making is that in the HS YM the upshot is not chaos: it is still possible to lay out a well-defined procedure that allows to select some of these theories and to compute sensible physical amplitudes out of them. The effective non-locality is a pure gauge artifact that can be circumvented by the use of physical projectors.  Said another way, the original contribution of our paper is that it has a HS gauge invariance on the frame index which is used to construct the vertices, but  the absence of ghosts relies not only on it  but also on projectors to physical states. The latter are a remnant of a hidden gauge symmetry over the non-frame indices.}

\section{Conclusions}

In this paper we have {introduced classical} massless HS theories exist in a flat spacetime.
In fact, inspired by the effective action method, by which integrating out
fermion matter fields one can derive HS models, we have constructed HS YM-type
theories in any dimension and CS-like theories in any odd dimension. We have
defined their actions and found their equations of motion, as well as their
conserved currents. These theories are perturbatively local. They are
characterized by a HS gauge symmetry which includes in particular ordinary gauge
transformations and diffeomorphisms. {We have presented two different interpretations,
a gravitational (G) and an a-gravitational (a-G).}
On the same footing we have also introduced
HS scalar type theories. Focusing in particular on HS YM theories we have shown 
that, with the addition of ghosts and auxiliary fields, they can be easily BRST 
quantized. It is
possible to reproduce the  Higgs phenomenon, by which the HS potentials acquire
a mass. Finally we have shown how to recover local Lorentz
covariance in all of these HS models.

Then we have taken on the problem of perturbative quantization of HS YM-like
models (without matter). {Expanding around $u=0$ we have seen 
that a 
perturbative expansion and the relevant Feynman diagrams can be defined.} We 
have seen
that this leads to the appearance of a mass parameter, although the HS particles
remain massless. A perturbative series can indeed be defined and used for 
calculations.
In this context a crucial issue is represented by the unphysical modes. In a 
gauge theory
they are unavoidable, and may lead to unitarity violations; but good theories 
contain the 
necessary antidotes. This seems to be the case also for the 
HS YM-like theories. We have shown that the quantum perturbative series for 
physical amplitudes can be formulated in such a way as to exclude unphysical 
modes and allow only the propagation of the physical ones\footnote{{This seems to be in keeping with the results of \cite{Steinacker1}}.}. This remarkable
result has been obtained by using both the traditional FP ghosts and a system 
of projectors to the transverse and traceless modes for the non-frame indices. {Whether this is fully satisfactory for unitarity is an open problem. It requires showing in particular that the renormalization procedure works, at least in 4d, in the presence of IR singularities increasing with spin.}

{It is worth adding that
this seems to fit in well with a classical result derived in a forthcoming paper. There it will be shown that, for instance, in 
4d (but a similar argument can be repeated in any dimension) the overall density functional of the YM-like theory in the quadratic approximation has the form
\be
\sim \int d^3x \int d^4u \left(\left({\stackrel 
{\rightarrow}{\EE}(x,u)}\right)^2 +\left({\stackrel 
{\rightarrow}{\EF}}(x,u)\right)^2 \right)+ \dots \label{E2F2}
\ee
where ${\stackrel {\rightarrow}{\EE}}(x,u)$ and ${\stackrel 
{\rightarrow}{\EF}}(x,u)$ are vectors with components $G_{0i}(x,u)$ and 
$G_{ij}(x,u)$, $i,j=1,2,3$, respectively, like in ordinary Maxwell theory, and 
dots denote cubic and quartic terms in the fields. The first term in the RHS of} 
\eqref{E2F2} { is positive and vanishes only when $h_a(x,u)=0$. 
In the limit of slowly varying fields it 
represents the `sum' of all the kinetic terms. This means the system is classically stable.}

One last issue we have {discussed} is the gauge origin of the unphysical modes of the HS YM theory. In its initial formulation there is no trace of such a symmetry, in the sense that if it exists it is completely fixed (we called this putative gauge symmetry, the hidden gauge symmetry). On the other hand also the local Lorentz symmetry turns out to be fixed 
in the simplest formulation of HS YM-like models. Yet we have shown that this symmetry 
can be recovered by enlarging the model. Thus it was not unnatural to expect that something similar
would happen also with hidden symmetry. We have shown that unfolding the latter in the action is possible but leads to incorporating in it nonlocal terms. However such non-localities are a gauge artifact. They are produced if we force the theory to be symmetric under the hidden gauge transformations \eqref{hiddengaugetransf}.} {To summarize, the main point of our paper is that even if the model is effectively non-local, due to such artifact, it is still possible to single out a well-defined procedure that allows to compute sensible physical amplitudes. The effective non-locality can be circumvented by the use of physical projectors.  Said another way, the original contribution of our paper is that it has a HS gauge invariance on the frame index which is used to construct the vertices, but  the absence of ghosts relies not only on it  but also on projectors to physical states. The latter are a remnant of an implicit gauge symmetry over the non-frame indices. Whether, like in  \cite{FS}, the HS YM theories may be made local by a suitable choice of auxiliary fields is an interesting problem we defer to future research.

\vskip 1cm
{\bf Acknowledgements.}

S.G. would like to thank Massimo Taronna and Dario Francia for illuminating discussion of topics 
relevant to the paper. {L.B. and S.G. would like to thank  the 
Galileo Galilei Institute for Theoretical Physics and INFN for hospitality and 
partial support during the workshop "String Theory from a worldsheet 
perspective" where part of this work has been done.} S.G. would also like to thank the Asia Pacific Center for Theretical Physics for hosting the workshop "Higher Spin Gravity: Chaotic, Conformal and Algebraic Aspects", which provided a favourable environment for the discussion of some issues related to the present work.
This research has been supported by the Croatian Science Foundation under the 
project No.~8946 and
by the University of
Rijeka under the research support No.~13.12.1.4.05 and project 
uniri-prirod-18-256. The research of S.G. has been supported by the Israel 
Science Foundation (ISF), grant No. 244/17.

\appendix

\section{Master field expansions}

In this Appendix we collect some $u$ expansions of various master fields
referred to in the text.

\subsection{Curvature components}
\label{ss:curvatureexp}

Here are the expansions referred to in section \ref{s:eoms}:
\be
F_{ab} &=& \partial_a A_b - {\partial_b A_a + \partial_\sigma A_a
\chi_b^\sigma
-\partial_\sigma A_b \chi_a^\sigma}- {\frac 1{24}} \Big(
\partial_{\sigma_1}\partial_{\sigma_2}\partial_{\sigma_3} A_a  \,
c_b^{\sigma_1\sigma_2\sigma_3}\label{Fab}\\
&& -\partial_{\sigma_1}\partial_{\sigma_2}\partial_{\sigma_3} A_b \,
c_a^{\sigma_1\sigma_2\sigma_3}+3 \partial_{\sigma_3} b_a^{\sigma_1\sigma_2}
\partial_{\sigma_1}\partial_{\sigma_2}\chi_b^{\sigma_3}-  3 \partial_{\sigma_3}
b_b^{\sigma_1\sigma_2}
\partial_{\sigma_1}\partial_{\sigma_2}\chi_a^{\sigma_3}\Big)\0\\
X_{ab}^\mu &=& \partial_a \chi_b^\mu - \partial_b \chi_a^\mu + {\partial_\sigma
A_a\, b_b{}^{\sigma\mu} -\partial_\sigma A_b\, b_a{}^{\sigma\mu}-\chi_a^\sigma\,
\partial_\sigma \chi_b^\mu + \chi_b^\sigma \,\partial_\sigma
\chi_a^\mu}\label{Xab}\\
&&  -  {\frac 1{24}} \Big(
\partial_{\sigma_1}\partial_{\sigma_2}\partial_{\sigma_3}\chi_a^\mu \,
c_b^{\sigma_1\sigma_2\sigma_3}-\partial_{\sigma_1}\partial_{\sigma_2}\partial_{
\sigma_3}\chi_b^\mu \,
c_a^{\sigma_1\sigma_2\sigma_3}+ \partial_{\sigma_1}\partial_{\sigma_2}\partial_{
\sigma_3} A_a  \, d_b^{\mu\sigma_1\sigma_2\sigma_3}\0\\
&&\quad-\partial_{\sigma_1}\partial_{\sigma_2}\partial_{\sigma_3} A_b  \,
d_a^{\mu\sigma_1\sigma_2\sigma_3}-{3}
\partial_{\sigma_1}\partial_{\sigma_2}\chi_a^{\sigma_3}\partial_{\sigma_3}
c_b^{\sigma_1\sigma_2\mu} +{3}
\partial_{\sigma_1}\partial_{\sigma_2}\chi_b^{\sigma_3}\partial_{\sigma_3}
c_a^{\sigma_1\sigma_2\mu}\0\\
&&\quad+ {3} \partial_{\sigma_3} b_a^{\sigma_1\sigma_2} 
\partial_{\sigma_1}\partial_{\sigma_2}b_b ^{\sigma_3\mu} - {3}
\partial_{\sigma_3} b_b^{\sigma_1\sigma_2} 
\partial_{\sigma_1}\partial_{\sigma_2}b_a ^{\sigma_3\mu} \Big) +\ldots\0\\
B_{ab}{}^{\mu\nu} &=& \partial_a b_b{}^{\mu\nu} -  \partial_b
b_a{}^{\mu\nu}+{2\partial_\sigma \chi_a^{(\mu}
b_b{}^{\nu)\sigma}-2\partial_\sigma \chi_b^{(\mu} b_a{}^{\nu)\sigma}}\0\\
&&+
{ \partial_\sigma b_a{}^{\mu\nu} \chi_b^\sigma - \partial_\sigma b_b{}^{\mu\nu}
\chi_a^\sigma +\partial_\sigma A_a c_b^{\sigma \mu\nu}- \partial_\sigma A_b
c_a^{\sigma \mu\nu}}
+\ldots\label{Bab}\\
&&  -{ \frac 1{24}} \Big(2
\partial_{\sigma_1}\partial_{\sigma_2}\partial_{\sigma_3}\chi_a^{(\mu }\,
{d}_b^{\nu)\sigma_1\sigma_2\sigma_3}-2\partial_{\sigma_1}\partial_{\sigma_2}
\partial_{\sigma_3}\chi_b^{(\mu }\,
{d}_a^{\nu)\sigma_1\sigma_2\sigma_3}+\partial_{\sigma_1}\partial_{\sigma_2}
\partial_{\sigma_3} A_a  \, f_b^{\mu\nu\sigma_1\sigma_2\sigma_3}\0\\
&&\quad-\partial_{\sigma_1}\partial_{\sigma_2}\partial_{\sigma_3} A_b  \,
f_a^{\mu\nu\sigma_1\sigma_2\sigma_3}
+\partial_{\sigma_1}\partial_{\sigma_2}\partial_{\sigma_3} b_a^{\mu\nu}
c_b^{\sigma_1\sigma_2\sigma_3}-
\partial_{\sigma_1}\partial_{\sigma_2}\partial_{\sigma_3} b_b^{\mu\nu}
c_a^{\sigma_1\sigma_2\sigma_3}\0\\
&&\quad-
{3}\partial_{\sigma_1}\partial_{\sigma_2}\chi_a^{\sigma_3}\partial_{\sigma_3}
d_b^{\sigma_1\sigma_2\mu\nu} + {3}
\partial_{\sigma_1}\partial_{\sigma_2}\chi_b^{\sigma_3}\partial_{\sigma_3}
d_a^{\sigma_1\sigma_2\mu\nu}
-{6} \partial_{\sigma_3} c_b^{\sigma_1\sigma_2(\nu} 
\partial_{\sigma_1}\partial_{\sigma_2}b_a ^{{\mu)\sigma_3}}\0\\
&&\quad  {+6} \partial_{\sigma_3} c_a^{\sigma_1\sigma_2(\nu} 
\partial_{\sigma_1}\partial_{\sigma_2}b_b ^{{\mu)\sigma_3}}+{3}{
\partial_{\sigma_3}
b_a^{\sigma_1\sigma_2}
\partial_{\sigma_1}\partial_{\sigma_2}c_b ^{\sigma_3\mu\nu}-
{3}\partial_{\sigma_3}
b_b^{\sigma_1\sigma_2}
\partial_{\sigma_1}\partial_{\sigma_2}c_a ^{\sigma_3\mu\nu}}\Big)
+\ldots\0
\ee
\be
C_{ab}^{\mu\nu\lambda} &=&  \partial_a c_b{}^{\mu\nu\lambda} -  \partial_b
c_a{}^{\mu\nu\lambda}+{ \partial_\sigma A_a d_b^{\sigma
\mu\nu\lambda}
- \partial_\sigma A_b d_a^{\sigma \mu\nu\lambda}+ 3\partial_\sigma
\chi_a^{(\mu}c_b^{\nu\lambda)\sigma}
-3 \partial_\sigma \chi_b^{(\mu}c_a^{\nu\lambda)}}\0\\
&&{+3\partial_\sigma b_a^{(\mu\nu} b_b^{\lambda)\sigma}
-   3\partial_\sigma b_b^{(\mu\nu} b_a^{\lambda)\sigma}
+\partial_\sigma c_a^{\mu\nu\lambda} \chi_b^\sigma-
\partial_\sigma c_b^{\mu\nu\lambda} \chi_a^\sigma }\label{Cab}\\
&&-{\frac 1{24} }\Big( 
\partial_{\sigma_1}\partial_{\sigma_2}\partial_{\sigma_3}
A_a  \, g_b^{\mu\nu\lambda\sigma_1\sigma_2\sigma_3}-
\partial_{\sigma_1}\partial_{\sigma_2}\partial_{\sigma_3} A_b  \,
g_a^{\mu\nu\lambda\sigma_1\sigma_2\sigma_3}+ {3} 
{\partial_{\sigma_1}\partial_{\sigma_2}\partial_{\sigma_3}\chi_a^{(\mu }\,
f_b^{\nu\lambda)\sigma_1\sigma_2\sigma_3}}\0\\
&&- {3}   \partial_{\sigma_1}\partial_{\sigma_2}\partial_{\sigma_3}\chi_b^{(\mu
}\,
f_a^{\nu\lambda)\sigma_1\sigma_2\sigma_3}}
+ {3} \partial_{\sigma_1}\partial_{\sigma_2}
\partial_{\sigma_3} b_a^{(\mu\nu}
d_b^{\nu)\sigma_1\sigma_2\sigma_3}
- {3} \partial_{\sigma_1}\partial_{\sigma_2}
\partial_{\sigma_3} b_b^{(\mu\nu} d_a^{\nu)\sigma_1\sigma_2\sigma_3\0\\
&&+\partial_{\sigma_1}\partial_{\sigma_2}\partial_{\sigma_3} c_a^{\mu\nu\lambda}
c_b^{\sigma_1\sigma_2\sigma_3} -
\partial_{\sigma_1}\partial_{\sigma_2}\partial_{\sigma_3} c_b^{\mu\nu\lambda}
c_a^{\sigma_1\sigma_2\sigma_3} -3
\partial_{\sigma_1}\partial_{\sigma_2}\chi_a^{\sigma_3}\partial_{\sigma_3}
f_b^{\sigma_1\sigma_2\mu\nu\lambda}\0\\
&&+3
\partial_{\sigma_1}\partial_{\sigma_2}\chi_b^{\sigma_3}\partial_{\sigma_3}
f_a^{\sigma_1\sigma_2\mu\nu\lambda}
- {9}
\partial_{\sigma_1}\partial_{\sigma_2}c_a^{\sigma_3(\mu\nu}\partial_{\sigma_3}
c_b^{\lambda)\sigma_1\sigma_2} 
 + {9}
\partial_{\sigma_1}\partial_{\sigma_2}c_b^{\sigma_3(\mu\nu}\partial_{\sigma_3}
c_a^{\lambda)\sigma_1\sigma_2}\0\\
&& - {9}\partial_{\sigma_1}\partial_{\sigma_2}b_a
^{\sigma_3(\mu}\partial_{\sigma_3} d_b^{\nu\lambda)\sigma_1\sigma_2}
 + {9}
\partial_{\sigma_1}\partial_{\sigma_2}b_b ^{\sigma_3(\mu}\partial_{\sigma_3}
d_a^{\nu\lambda)\sigma_1\sigma_2}\0\\
&&-3\partial_{\sigma_1}\partial_{\sigma_2}d_a
^{\sigma_3\mu\nu\lambda}\partial_{\sigma_3} b_b^{\sigma_1\sigma_2}
 + 3
\partial_{\sigma_1}\partial_{\sigma_2}d_b
^{\sigma_3\mu\nu\lambda}\partial_{\sigma_3}
b_a^{\sigma_1\sigma_2}\Big)\Big)+\ldots\0
\ee
where the ellipses in the RHS refer to terms containing at least five
derivatives.

\subsection{$\delta \Phi$}
\label{ss:deltaPhi}

Here we consider the transformation of the complex scalar field $\Phi$
introduced in sec.\ref{ss:higgs}. Under \eqref{deltaPhi} its components
transform as 
\be
\delta_\varepsilon\varphi_0 &=& i\epsilon \,\varphi_0-\frac 12
\xi\!\cdot\!\partial \varphi_0 +\frac 12 \varphi_1 ^\mu
\partial_\mu\varepsilon-{\frac i8} \left(\partial^\mu\partial^\nu\varepsilon\,
\varphi_{2\mu\nu}+\partial^\mu\partial^\nu\varphi_0\, \Lambda_{\mu\nu} -2
\partial^\mu\xi^\nu\partial_\mu \varphi_{1\nu}\right)+\ldots\0\\
\delta_\varepsilon \varphi_1^\lambda &=&  i\varepsilon \,\varphi_1^\lambda+i
\varphi_0 \xi^\lambda    - \frac 12\left( \varphi_2 ^{\mu\lambda}
\partial_\mu\varepsilon  + \varphi_1 ^\mu \partial_\mu\xi^\lambda -\xi\!\cdot\!
\varphi_1^\lambda  -\Lambda ^{\mu\lambda} \partial_\mu\varphi_0\right)\0\\
&&+{\frac i4} \left( \partial^\mu\xi^\nu\partial_\mu \varphi_{2\nu}{}^\lambda
+\partial^\mu \Lambda^{\nu\lambda} \partial_\mu \varphi_{1\nu}\right)\0\\
&&- {\frac i8}\left(\partial^\mu\partial^\nu\varepsilon
\varphi_{3\mu\nu}{}^\lambda+ \partial^\mu\partial^\nu\xi^\lambda 
\varphi_{2\mu\nu}+\partial^\mu\partial^\nu\varphi_0\, \Sigma_{\mu\nu}^\lambda +
\partial^\mu\partial^\nu\varphi_1^\lambda \Lambda_{\mu\nu}\right)+
\ldots\0\\
\delta_\varepsilon \varphi_2^{\lambda\rho} &=&  i\varepsilon
\,\varphi_2^{\lambda\rho}+ i \varphi_1^{(\lambda} \xi^{\rho)} +i\varphi_0\,
\Lambda^{\lambda\rho} {-\frac 12}\left( \partial^\mu\varepsilon\,
\varphi_{3\mu}{}^{\lambda\rho}+2 \partial_\mu \xi^{(\lambda}
\varphi_2^{\rho)\mu}+ \varphi_1^\mu \partial_\mu \Lambda^{\lambda\rho}
\right)\0\\
&&+{\frac 12}\left( \xi\!\cdot\!\partial\varphi_2^{\lambda\rho}
+2\Lambda^{\mu(\lambda}\partial_\mu \varphi_1^{\rho)}+ \partial_\mu
\varphi_0\,\Sigma^{\mu\lambda\rho} \right)\0\\
&&+\frac i4 \left( \partial_\mu \xi^\nu \partial_\mu
\varphi_{3\nu}{}^{\lambda\rho} + 2 \partial^\mu \Lambda^{\nu(\lambda}
\partial_\mu \varphi_{2\nu}{}^{\rho)}+ \partial^\mu\Sigma^{\nu\lambda\rho}
\partial_\mu \varphi_{1\nu}\right)\0\\
&&-\frac i8 \left( \partial^\mu\partial^\nu\varepsilon\,
\varphi_{4\mu\nu}{}^{\lambda\rho}+ 2\partial^\mu\partial^\nu\xi^{(\lambda} 
\varphi_{3\mu\nu}{}^{\rho)}
+\partial^\mu\partial^\nu\varphi_0\, P_{\mu\nu}^{\lambda\rho}  +
2\partial^\mu\partial^\nu\varphi_1^{(\lambda}
\,\Sigma_{\mu\nu}^{\rho)}\right.\0\\
&&\left. + \partial^\mu\partial^\nu \Lambda^{\lambda\rho}\varphi_{2\mu\nu} +
\partial^\mu\partial^\nu \varphi_2^{\lambda\rho}\Lambda_{\mu\nu}\right)+
\ldots\label{deltavarphi}
\ee
\be
\delta_\varepsilon \varphi_3^{\mu\nu\lambda} &=& i\varepsilon
\,\varphi_3^{\mu\nu\lambda}+\frac 32 i \xi^{(\mu} \varphi_2^{\nu\lambda)}+ \frac
32 i \Lambda^{(\mu\nu} \varphi_1^{\lambda)}+i \Sigma^{\mu\nu\lambda}
\varphi_0\label{deltavaphi3}\\
&&{-\frac 12}\left(
\partial_\sigma \Sigma^{\mu\nu\lambda} \varphi_1^\sigma+3  \partial_\sigma
\Lambda^{(\mu\nu} \varphi_2^{\lambda)\sigma}+3   \partial_\sigma\xi^{(\mu}
\varphi_3^{\nu\lambda)\sigma}+   \partial_\sigma \epsilon\,
\varphi_4^{\mu\nu\lambda\sigma}\right)\0\\
&&+\frac 12 \left( \xi^\sigma \partial_\sigma\varphi_3 ^{\mu\nu\lambda}+3
\Lambda^{\sigma(\mu} \partial_\sigma \varphi_2^{\nu\lambda)} +3
\Sigma^{\sigma(\mu\nu} \partial_\sigma \varphi_1^{\lambda)}
+P^{\sigma\mu\nu\lambda} \partial_\sigma \varphi_0\right)\0\\
&& - {\frac i8} \left( \partial_\sigma\partial_\tau \Sigma^{\mu\nu\lambda}
\varphi_2^{\sigma\tau}+ 3 \partial_\sigma\partial_\tau
\Lambda^{(\mu\nu}\varphi_3^{\lambda)\sigma\tau} +3  \partial_\sigma\partial_\tau
\xi^{(\mu} \varphi_4 ^{\nu\lambda)\sigma\tau} + 
\partial_\sigma\partial_\tau\epsilon\,
\varphi_5^{\mu\nu\lambda\sigma\tau}\right)  \0 \\
&&-{ \frac i8} \left( \partial_\sigma\partial_\tau \varphi_0\,
\Omega^{\mu\nu\lambda\sigma\tau}+3  \partial_\sigma\partial_\tau
\varphi_1^{(\mu}P^{\nu\lambda)\sigma\tau} +3  \partial_\sigma\partial_\tau
\varphi_2^{(\mu\nu}\Sigma^{\lambda)\sigma\tau} ++3  \partial_\sigma\partial_\tau
\varphi_3^{\mu\nu\lambda}\Lambda^{\sigma\tau} \right)\0\\
&&+ \frac i4 \left(\partial_\sigma \xi^{\tau}\partial_\tau
\varphi_4^{\sigma\mu\nu\lambda} +3 \partial_\sigma
\Lambda^{\tau(\mu}\partial_\tau \varphi_3^{\nu\lambda)\sigma} +3 \partial_\sigma
\Sigma^{\tau(\mu\nu}\partial_\tau \varphi_2^{\lambda)\sigma}+ \partial_\sigma
P^{\tau\mu\nu\lambda}\partial_\tau \varphi_1^{\sigma}\right)+\ldots\0
\ee

\subsection{$S_3$ and $S_4$}
\label{ss:S3S4}

Here are the explicit expressions for the lowest order terms of $S_3$ and
$S_4$. They are computed by stripping $\langle\!\langle G_{ab}
G^{ab}\rangle\!\rangle$ of the integration over $u$. 
\be
S_3&=& -i g\,  \langle\!\langle\partial_a h_b \ast [h^a\stackrel{\ast}{,} h^b] 
\rangle\!\rangle\label{S3}\\
&=& { {-}g}\langle\!\langle \partial^a A^b (\partial_\sigma A_a
\chi_b^\sigma - \partial_\sigma A_b \chi_a^\sigma)\0\\
&&-\frac 1{{24}} (\partial^a A^b-\partial^b A^a) \bigl(\partial_{\sigma_1}
\partial_{\sigma_2} \partial_{\sigma_3}A_a \, c_b^{\sigma_1\sigma_2\sigma_3}+ 3
\partial_{\sigma_3} b_a^{\sigma_1\sigma_2}\partial_{\sigma_1}
\partial_{\sigma_2}\chi_b^{\sigma_3}\bigr)\0\\
&&+ \frac 1{2d} \Bigg( \partial^a A^b \partial_\sigma b_{a\mu}{}^\mu
\chi_b^\sigma - \partial^a A^b \partial_\sigma b_{b\mu}{}^\mu \chi_a^\sigma\0\\
&&+ {2}\partial^a A^b \partial_\sigma \chi_a^\mu
b_{b\mu}{}^{\sigma}-{2}\partial^a
A^b \partial_\sigma \chi_b^\mu b_{a\mu}{}^{\sigma}+   \partial^a A^b 
\partial_\sigma A_a c_{b\mu}{}^{\mu\sigma}- \partial^a A^b  \partial_\sigma A_b
c_{a\mu}{}^{\mu\sigma}\0\\
&&+ \partial^a b^{b\mu}{}_\mu \, \partial_\sigma A_a \chi_b^\sigma -  \partial^a
b^{b\mu}{}_\mu \, \partial_\sigma A_b\chi_a^\sigma+  {2}\partial^a \chi^b_\nu
\bigl(
\partial_\sigma A_a \, b_b{}^{\sigma\nu}- \partial_\sigma A_b \,
b_a{}^{\sigma\nu}\0\\
&& + \partial_\sigma \chi_a^\nu \chi_b^\sigma - \partial_\sigma \chi_b^\nu
\chi_a^\sigma\bigr)\0\\
&&{-\frac 1{24}} \bigg( (\partial^a A^b-\partial_b A^a)\Big( \partial_{\sigma_1}
\partial_{\sigma_2} \partial_{\sigma_3}\chi_a^\mu d_b^{\mu
\sigma_1\sigma_2\sigma_3} {+}\frac 12\partial_{\sigma_1} \partial_{\sigma_2}
\partial_{\sigma_3}A_a f_{b\mu}^{\mu\sigma_1\sigma_2\sigma_3}\0\\
&&{+}\frac 12 \partial_{\sigma_1} \partial_{\sigma_2}
\partial_{\sigma_3}b_{a\mu}^\mu c_b^{\sigma_1\sigma_2\sigma_3}-{\frac 32}
\partial_{\sigma_1} \partial_{\sigma_2} \chi_a^{\sigma_3}\partial_{\sigma_3}
d_{b\mu}^{\mu\sigma_1\sigma_2}-{3}\partial_{\sigma_1} \partial_{\sigma_2}
b_{a\mu}^{\sigma_3}\partial_{\sigma_3}c_b^{\mu\sigma_1\sigma_2}\0\\
&&{+\frac 32} \partial_{\sigma_1} \partial_{\sigma_2}
\chi_b^{\sigma_3}\partial_{\sigma_3} d_{a\mu}^{\mu\sigma_1\sigma_2}\Big)+\left(
\partial_a b_{b\mu}^\mu - \partial_b
b_{a\mu}^\mu\right)\bigl(\partial_{\sigma_1} \partial_{\sigma_2}
\partial_{\sigma_3}A_a \, c_b^{\sigma_1\sigma_2\sigma_3}\0\\
&&+ 3 \partial_{\sigma_3} b_a^{\sigma_1\sigma_2}\partial_{\sigma_1}
\partial_{\sigma_2}\chi_b^{\sigma_3}\bigr) \0\\
&&+ \left( \partial^a \chi^{b\mu} - \partial^b
\chi^{a\mu}\right)\Big(\partial_{\sigma_1} \partial_{\sigma_2}
\partial_{\sigma_3}A_a \, d_{b\mu}^{\sigma_1\sigma_2\sigma_3}+
\partial_{\sigma_1} \partial_{\sigma_2} \partial_{\sigma_3}\chi_{a\mu}  \,
c_{b}^{\sigma_1\sigma_2\sigma_3}\0\\
&&-3 \partial_{\sigma_1} \partial_{\sigma_2} \chi_a^{\sigma_3}
\partial_{\sigma_3}c_{b\mu}^{\sigma_1\sigma_2} + 3 
\partial_{\sigma_3}b_{a}^{\sigma_1\sigma_2} \partial_{\sigma_1}
\partial_{\sigma_2} b_{b\mu}^{\sigma_3}\Big)
\bigg)\Bigg) u^2 + {\cal O}(u^4,6) \rangle\!\rangle \0
\ee
where $ {\cal O}(u^4,6)$ means terms of order at least $u^4$ or containing at
least six derivatives, and
\be
S_4 &=& \frac {g^2}{{4}}\, \langle\!\langle [h^a\stackrel{\ast}{,} h^b] \ast 
  [h_a\stackrel{\ast}{,} h_b]  \rangle\!\rangle\label{S4}\\
&=& - \frac {g^2}2\, \langle\!\langle  \bigl(\partial_\sigma A^a
\chi^{b\sigma}- \partial_\sigma A^b \chi^{a\sigma}\bigr) \partial_\tau A_a
\chi_b^\tau\0\\
&& + \frac 1{d} \Big(\bigl(\partial_\sigma A^a \chi^{b\sigma}- \partial_\sigma
A^b \chi^{a\sigma}\bigr) \bigl(\partial_\tau A_a
c_{b\nu}{}^{\tau\nu}+2\partial_\tau \chi_a^\nu b_{b\nu}{}^\tau +  \partial_\tau
b_{a\nu}{}^\nu\chi_b^\tau\bigr)\0\\
&&+  \bigl(  \partial_\sigma A^a b^{b\sigma\nu}-  \partial_\sigma A^b
b^{a\sigma\nu}+
\partial_\sigma\chi^{a\nu} \chi^{b\sigma} - \partial_\sigma\chi^{b\nu}
\chi^{a\sigma} \bigr)
\bigl( \partial_\tau A_a b_{b\nu}{}^\tau+ \partial_\tau\chi_{a\nu}
\chi_{b}^\tau\bigr)\Big)u^2\0\\
&& + {\cal O}(u^4,4) \rangle\!\rangle\0
\ee

\section{Compatibility of LL and HS gauge transformations}
\label{s:compatible}

In this Appendix we answer the question: after the introduction of the inertial
frame $e_a^\mu$ and connection $\EA_\mu$, does the HS gauge symmetry still hold?
For instance, is \eqref{SpsihEA} still invariant under \eqref{deltaLPsi}?
Let us consider first 
\be
\delta_\varepsilon S_2&=& \delta_\varepsilon \langle\!\langle \overline \Psi
\gamma^a h_a \ast\Psi \rangle\!\rangle=\delta_\varepsilon
\langle\!\langle \overline \Psi\ast \gamma^a h_a \ast\Psi \rangle\!\rangle\0\\
&=&i  \langle\!\langle \overline \Psi \ast\gamma^a h_a \ast\varepsilon \ast\Psi
\rangle\!\rangle
-i  \langle\!\langle \overline \Psi\ast\varepsilon\ast \gamma^a h_a \ast\Psi
\rangle\!\rangle+   \langle\!\langle \overline \Psi \ast \gamma^a
\left(\partial_a \varepsilon -i [h_a\stackrel{\ast}{,}
\varepsilon]\right)\ast\Psi \rangle\!\rangle\0\\
&=& \langle\!\langle \overline \Psi \ast \gamma^a \partial_a \varepsilon
\ast\Psi \rangle\!\rangle=  \langle\!\langle \overline \Psi \ast \gamma^a D_a
\varepsilon \ast\Psi \rangle\!\rangle\label{deltaepsS2}
\ee
where it is understood that $\partial_a = e_a^\mu \partial_\mu=  e_a^\mu  D_\mu$
and in the $\ast$ products the ordinary spacetime derivatives are replaced by
inertial covariant ones. This is possible because, not only $[D_\mu, D_\nu]=0$,
but also $[D_a, D_b]=0$ due of \eqref{metriclike}.

Next 
\be
\delta_\varepsilon S_1&=&\delta_\varepsilon\langle\!\langle \overline \Psi
\gamma^a e_a^\mu \left(i\partial_\mu+\frac 12 \EA_\mu\right) \Psi
\rangle\!\rangle = \delta_\varepsilon  \langle\!\langle \overline \Psi \ast
\gamma^aD_a \Psi \rangle\!\rangle\0\\
&=&  \langle\!\langle \overline \Psi\ast \varepsilon \ast \gamma^aD_a \Psi
\rangle\!\rangle-
   \langle\!\langle \overline \Psi \ast \gamma^aD_a\left(\varepsilon\ast
\Psi\right) \rangle\!\rangle\0\\
&=& -   \langle\!\langle \overline \Psi \ast \gamma^aD_a \varepsilon\ast \Psi
\rangle\!\rangle\label{deltaepsS1}
\ee
where we have used $D_a\left(\varepsilon\ast \Psi\right) = D_a \varepsilon \ast
\Psi + \varepsilon \ast D_a \Psi$, which is possible because we have inserted
the inertial covariant derivative in the $\ast$ product, and because
$[D_a,D_b]=0$, as already pointed out.

In conclusion the HS gauge invariance of $S(\Psi,h,\EA)$ still holds. Let us
remark that, in order to achieve invariance,  the inertial frame and connection
must not transform under HS gauge transformations. 

As another example let us consider the transformation of 
\be
G_{ab}= D_a h_b- D_b h_a -i[h_a\stackrel{\ast}{,} h_b]\label{Gabcov}
\ee
It follows immediately
\be
\delta G_{ab} =-i [G_{ab}\stackrel {\ast}{,} \varepsilon] \0
\ee
provided one remarks that, once we replace ordinary spacetime derivatives with
inertial covariant ones in the $\ast$ product, the inertial covariant derivative
commutes with the $\ast$ product. For instance:
\be
D_a (h_b \ast \varepsilon )= D_a h_b \ast \varepsilon + h_b\ast D_a
\varepsilon\0
\ee
and $D_a\varepsilon = \partial_a\varepsilon= e_a^\mu \partial_\mu\varepsilon\0$.

\section{Functional calculus}
\label{app:funct}

In the functional integral manipulations of the perturbative approach the 
conjugate variables are
\be
h_a(x,u)=\sum_{n=0}^\infty \frac 1{n!}\, 
h_a^{\mu_1\ldots\mu_n} (x)\, u_{\mu_1}\ldots u_{\mu_n},\label{haxu}
\ee 
and
\be
j^a(x,u)
&=& \sum_{n=0}^\infty\, (-1)^n j^a_{\mu_1 \ldots \mu_{n}}(x)\,\, \frac
{\partial^{n}}{\partial u_{\mu_1} \ldots \partial
u_{\mu_{n}}}\delta(u)\label{jaxu}
\ee
so that
\be
\langle\!\langle j_a \, h^a \rangle\!\rangle = \int d^dx \, j^a_{\mu_1 \ldots
\mu_{n}}(x)\, h_a ^{\mu_1 \ldots \mu_{n}}(x)\label{jaha2}
\ee
In the presence of the factor $e^{i \langle\!\langle j_a \, h^a
\rangle\!\rangle}$ we can represent
\be
h_a(x,u)=\frac {\delta }{\delta j^a(x,u)}\langle\!\langle j_a \, h^a
\rangle\!\rangle= \sum_{n=0}^\infty \frac 1{n!}  u_{\mu_1}\ldots u_{\mu_n}\frac
{\delta}{\delta j^a_{ {\mu_1}\ldots {\mu_n}}(x)}\langle\!\langle j_a \, h^a
\rangle\!\rangle\label{ha=deltaja}
\ee
It follows that
\be
\frac {\delta }{\delta j^a(x,u)} j^b(y,v)& =& \sum_{n=0}^\infty \frac 1{n!} 
u_{\mu_1}\ldots u_{\mu_n}\frac {\delta}{\delta j^a_{ {\mu_1}\ldots
{\mu_n}}(x)}\sum_{l=0}^\infty \frac {(- 1)^l}{l!} j^b_{\nu_1 \ldots
\nu_{l}}(y)\,\, \frac
{\partial^{l}}{\partial v_{\nu_1} \ldots \partial v_{\nu_{l}}}\delta(v)\0\\
&=&\delta_a^b\, \delta(x-y)  \sum_{n=0}^\infty \frac {(-1)^n}{n!} 
u_{\mu_1}\ldots u_{\mu_n} \frac
{\partial^{n}}{\partial v_{\mu_1} \ldots \partial v_{\mu_{n}}}\delta(v)\0\\
&=&\delta_a^b\, \delta(x-y)  \delta(u-v)\label{haja1}
\ee
since
\be
\sum_{n=0}^\infty \frac {(-1)^n}{n!}  u_{\mu_1}\ldots u_{\mu_n} \frac
{\partial^{n}}{\partial v_{\mu_1} \ldots \partial v_{\mu_{n}}}\delta(v)=
\delta(u-v)\label{haja2}\ee
for one can show, integrating by parts, that 
\be
\int \frac{ d^d v }{(2\pi)^d}\, \sum_{n=0}^\infty \frac {(- 1)^n}{n!} 
u_{\mu_1}\ldots u_{\mu_n} \frac
{\partial^{n}}{\partial v_{\mu_1} \ldots \partial v_{\mu_{n}}}\delta(v) f(v)
=f(u)\label{haja3}
\ee

Similarly one can prove
\be
\frac {\delta }{\delta h_a(x,u)} h_b(y,v)= \delta^a_b  \delta(x-y) 
\delta(u-v)\label{haja4}
\ee

In momentum representation  
\be
h_a(x,u) = \int  \frac{ d^d k }{(2\pi)^d}
e^{ik\cdot x}\,\tilde h_a(k,u) ,\quad \quad j^a(x,u) = \int  \frac{ d^d k
}{(2\pi)^d}
e^{ik\cdot x}\,\tilde j^a(k,u)\0
\ee
\be
h_{a\mu_1\ldots \mu_n}(x) = \int  \frac{ d^d k }{(2\pi)^d}
e^{ik\cdot x}\,\tilde h_{a\mu_1\ldots \mu_n}(k) ,\quad \quad j^a_{\mu_1\ldots
\mu_n}(x) = \int  \frac{ d^d k }{(2\pi)^d}
e^{ik\cdot x}\,\tilde j^a_{\mu_1\ldots \mu_n}(k)\0
\ee
and
\be
\langle\!\langle j_a (x,u)\, h^a(x,u) \rangle\!\rangle = \int d^dx \,
j^a_{\mu_1\ldots \mu_n}(x)h_a^{\mu_1\ldots \mu_n}(x)= \int  \frac{ d^d k
}{(2\pi)^d}\, \tilde j^{a\mu_1\ldots \mu_n}(k)\,\tilde h_{a\mu_1\ldots
\mu_n}(-k) \0
\ee

\subsection{Perturbative series for master fields}

Let us try  to develop a perturbative treatment for master fields 
(instead of simple space-time fields). First let us introduce external currents
$j_a(x,u)$ and define the coupling
\be
 \langle\!\langle h^a(x,u) \ast j_a (x,u) \rangle\!\rangle=
\int d^dx \int \frac {d^du}{(2\pi )^d} \, h^a(x,u) \ast j_a (x,u).\label{jaha}
\ee
Then we define the generating functional
\be
Z[h_a;j_a] &=& \int {\cal D}h_a\, e^{i\Big( S_0+ \langle\!\langle h^a(x,u) \ast
j_a (x,u) \rangle\!\rangle\Big)}
e^{i S_{int}}\label{Zjaha}
\ee
where
\be
S_0=-\frac 12  \langle\!\langle h^a(x,u) \ast K_{ab}(x) h^b (x,u)
\rangle\!\rangle\label{S0ab}
\ee
and
\be
S_{int} =S_3+S_4\0
\ee
see Appendix \ref{ss:S3S4}.

The (unnormalized) n-point function of the master field $h_a$ is defined by
\be
&&\langle  h_{a_1}(x_1,u_1)\ldots\ldots h_{a_n} (x_n,u_n)\rangle \0\\
&=& \int {\cal D}h_a\,   h_{a_1}(x_1,u_1)\ldots\ldots h_{a_n} (x_n,u_n) 
e^{i\Big( S_0+ \langle\!\langle h^a(x,u) \ast
j_a (x,u) \rangle\!\rangle\Big)}
e^{i S_{int}}\label{npthahb}
\ee

Going through the usual process (completing the square and integrating over
$h_a$) one gets
\be
\langle   h_{a_1}(x_1,u_1)\ldots h_{a_n} (x_n,u_n)\rangle  =\frac
{\delta}{\delta j^{a_1}(x_1,u_1)} \dots\ldots\frac {\delta}{\delta
j^{a_n}(x_n,u_n)}
\,
e^{i S_{int}\left( \frac {\delta}{\delta j^a}\right)} 
e^{ \langle\!\langle   j^a    P_{ab} j^b 
\rangle\!\rangle}\Big{\vert}_{j=0}\label{pertseries}
\ee
where $P_{ab}$ is the propagator  \eqref{jaPabjb}. It is convenient to express
everything in terms of Fourier transforms:
\be
\tilde h_a(k,u)= \int d^dx \, e^{ik\cdot x} h_a(x,u), \quad\quad  
\tilde j_a(k,u)= \int d^dx \, e^{ik\cdot x} j_a(x,u)\0
\ee

For instance $S_3$ can be rewritten 
\be
S_3&=&-g \int \frac{d^dk_1}{(2\pi)^d} \frac{d^dk_2}{(2\pi)^d}
\frac{d^dk_3}{(2\pi)^d} \delta(k_1+k_2+k_3)
\int \frac{d^du}{(2\pi)^d}\label{S3tilde}\\
&&\times k_{1a}\, \tilde h_b(k_1,u) \left[ \tilde h^a\left( k_2,u-\frac
{k_3}2\right) 
 \tilde h^b\left( k_3,u+\frac {k_2}2\right)-  \tilde h^a\left( k_2,u+\frac
{k_3}2\right) 
\tilde h^b\left( k_3,u-\frac {k_2}2\right)  \right]\0
\ee

Next 
\be
\langle  h_{a_1}(x_1,u_1)\ldots\,h_{a_n} (x_n,u_n)\rangle = \int
\frac{d^dq_1}{(2\pi)^d} \ldots\frac{d^dq_n}{(2\pi)^d}\,
e^{i(q_1\cdot x_1+\ldots  +q_n\cdot x_n)}\langle \tilde h_{a_1}(q_1
,u_1)\ldots\,\tilde h_{a_n}(q_n,u_n)\rangle\0
\ee

Since 
\be 
\tilde h_a(k,u) =\frac {\delta}{\delta \tilde j^a(-k,u)}= \int d^d x\,  e^{-i
k\cdot x} \frac {\delta}{\delta  j^a(x,u)}\0
\ee
we can write
\be
\langle  \tilde h_{a_1}(q_1,u_1)\ldots \tilde h_{a_n} (q_n,u_n)\rangle  =\frac
{\delta}{\delta \tilde j^{a_1}(-q_1,u_1)} \dots\frac {\delta}{\delta\tilde
j^{a_n}(-q_n,u_n)}
\,
e^{i S_{int}\left( \frac {\delta}{\delta\tilde j^a}\right)} 
e^{-\langle\!\langle \tilde j^a \tilde  P_{ab} \tilde j^b 
\rangle\!\rangle}\Big{\vert}_{\tilde j=0}\label{pertseriestilde}
\ee
{
Now one could apply the machinery of Feynman diagrams. But there are apparently insurmountable obstacles. First, we do not have an accessible interpretation for the $u$ dependence of the propagator in $\langle\!\langle \tilde j^a \tilde P_{ab} \tilde j^b\rangle\!\rangle$, see
\eqref{prophahb} and \eqref{Mm}.}  { Therefore we are obliged to go to the frozen momentum frame. So this term  can only be expressed in terms of component fields. Therefore, although so far in this Appendix we have formulated  everything in terms of master fields, at least as far as the perturbative approach is concerned, we can rely only in a component field formulation.}

\section{Spin projectors}
\label{sec:spinproj}

The purpose of this Appendix is to `resolve' the operators 
\be
 \Delta^{\mu_1\ldots \mu_n}{}_{\nu_1\ldots \nu_n}=\frac 1{n!}\left( 
\delta^{\mu_1}_{\nu_1}\ldots
\delta^{\mu_n}_{\nu_n}+{\rm perm} (\mu_1,\ldots,\mu_n)\right)\label{Deltaapp}
\ee
They are identity operators in the space of symmetric tensors of order $n$. We 
wish to express them as a sum of projectors, each of which projects onto a 
representation of the little group of the Lorentz group.
We will basically use the elementary projectors \eqref{elemproj} and their 
properties \eqref{elemproj1}, together with 
\be
\pi_\mu{}^\mu =  d-1,\quad\quad \omega_\mu{}^\mu=1,\quad\quad 
\pi_{\mu\nu}\pi^{\mu\nu}=d-1\label{traceelemproj}
\ee
Let us start we a few simple examples.

\subsection{spin 1, $n=1$}

The spin 1 case is given by $\Delta^\mu{}_\nu = \delta_\mu^\nu $. The 
decomposition is 
given by
\be
\Delta_{\mu\nu} = P^{(1)}_{\mu,\nu}+ P^{(0)}_{\mu,\nu} \label{spin1}
\ee
where
\be
 P^{(1)}_{\mu,\nu}=\pi_{\mu\nu}, \quad\quad 
P^{(0)}_{\mu,\nu}=\omega_{\mu\nu}\label{spin1a}
\ee
$P^{(1)}$ is traceless

\subsection{spin 2, $n=2$}

The $n=2$ case has already been reported in section \ref{sec:noghost}.

\subsection{spin 3, $n=3$}

In the $n=3$ case one has
\be
\Delta_{\mu_1\mu_2\mu_3,\nu_1\nu_2\nu_3}& =& 
P^{(3)}_{\mu_1\mu_2\mu_3,\nu_1\nu_2\nu_3}+ 
P^{(2)}_{\mu_1\mu_2\mu_3,\nu_1\nu_2\nu_3} \0\\
& &+ P^{(0)}_{\mu_1\mu_2\mu_3,\nu_1\nu_2\nu_3}+ 
P^{(0a)}_{\mu_1\mu_2\mu_3,\nu_1\nu_2\nu_3}+ 
P^{(0b)}_{\mu_1\mu_2\mu_3,\nu_1\nu_2\nu_3}+\overline 
P^{(0)}_{\mu_1\mu_2\mu_3,\nu_1\nu_2\nu_3} \label{spin3}
\ee
where $P^{(3)}$  is recorded in \eqref{spin3a1} 
The next relevant one is
\be
P^{(2)}_{\mu_1\mu_2\mu_3,\nu_1\nu_2\nu_3}&=& \frac 13 \Big( 
P^{(2)}_{(\mu_1\mu_2,\nu_1\nu_2}\,\omega_{\mu_3)\nu_3} + 
P^{(2)}_{\mu_1\mu_2,\nu_1\nu_3}\,\omega_{\mu_3\nu_2} +  
P^{(2)}_{\mu_1\mu_2,\nu_2\nu_3}\,\omega_{\mu_3\nu_1}\0\\
&&+  P^{(2)}_{\mu_1\mu_3,\nu_1\nu_2}\,\omega_{\mu_2\nu_3} + 
P^{(2)}_{\mu_1\mu_3,\nu_1\nu_3}\,\omega_{\mu_2\nu_2} +  
P^{(2)}_{\mu_1\mu_3,\nu_2\nu_3}\,\omega_{\mu_2\nu_1}\0\\
&&+  P^{(2)}_{\mu_2\mu_3,\nu_1\nu_2}\,\omega_{\mu_1\nu_3} + 
P^{(2)}_{\mu_2\mu_3,\nu_1\nu_3}\,\omega_{\mu_1\nu_2} +  
P^{(2)}_{\mu_2\mu_3,\nu_2\nu_3}\,\omega_{\mu_1\nu_1}\Big)\label{spin3a2}
\ee
which is traceless but not transverse. The remaining projectors are
\be
P^{(0)}_{\mu_1\mu_2\mu_3,\nu_1\nu_2\nu_3}&=&
\frac 1{3(d+1) }\Big(  \pi_{\mu_1\mu_2}\pi_{\nu_1\nu_2} \pi_{\mu_3\nu_3}+
\pi_{\mu_1\mu_3}\pi_{\nu_1\nu_3} 
\pi_{\mu_2\nu_2}+\pi_{\mu_2\mu_3}\pi_{\nu_2\nu_3} \pi_{\mu_1\nu_1}\0\\
&& + \pi_{\mu_1\mu_2}\pi_{\nu_1\nu_3} \pi_{\mu_3\nu_2}+
\pi_{\mu_1\mu_2}\pi_{\nu_2\nu_3} 
\pi_{\mu_3\nu_1}+\pi_{\mu_1\mu_3}\pi_{\nu_1\nu_2} \pi_{\mu_2\nu_3}\0\\
&&+  \pi_{\mu_1\mu_3}\pi_{\nu_2\nu_3} \pi_{\mu_2\nu_1}+
\pi_{\mu_2\mu_3}\pi_{\nu_1\nu_3} 
\pi_{\mu_1\nu_2}+\pi_{\mu_2\mu_3}\pi_{\nu_1\nu_2} 
\pi_{\mu_1\nu_3}\Big)\label{spin3a4}
\ee 
\be
P^{(0a)}_{\mu_1\mu_2\mu_3,\nu_1\nu_2\nu_3}&=&\frac 13 \Big( 
\pi_{\mu_1\nu_1}\omega_{\mu_2\mu_3} \omega_{\nu_2\nu_3} +
\pi_{\mu_1\nu_2}\omega_{\mu_2\mu_3} 
\omega_{\nu_1\nu_3}+\pi_{\mu_1\nu_3}\omega_{\mu_2\mu_3} \omega_{\nu_1\nu_2}\0\\
&&+ \pi_{\mu_2\nu_1}\omega_{\mu_1\mu_3} \omega_{\nu_2\nu_3} +
\pi_{\mu_2\nu_2}\omega_{\mu_1\mu_3} 
\omega_{\nu_1\nu_3}+\pi_{\mu_2\nu_3}\omega_{\mu_1\mu_3} \omega_{\nu_1\nu_2}\0\\
&&+ \pi_{\mu_3\nu_1}\omega_{\mu_1\mu_2} \omega_{\nu_2\nu_3} +
\pi_{\mu_3\nu_2}\omega_{\mu_1\mu_2} 
\omega_{\nu_1\nu_3}+\pi_{\mu_3\nu_3}\omega_{\mu_1\mu_2} 
\omega_{\nu_1\nu_2}\Big) 
\label{spin3a5}
\ee 
\be
P^{(0b)}_{\mu_1\mu_2\mu_3,\nu_1\nu_2\nu_3}&=&\frac 1{3(d-1)} 
\Big(\pi_{\mu_1\mu_2}\pi_{\nu_1\nu_2}\,\omega_{\mu_3\nu_3} 
+\pi_{\mu_1\mu_2}\pi_{\nu_1\nu_3}\,\omega_{\mu_3\nu_2} + 
\pi_{\mu_1\mu_2}\pi_{\nu_2\nu_3}\,\omega_{\mu_3\nu_1}\0\\
&&+  \pi_{\mu_1\mu_3}\pi_{\nu_1\nu_2}\,\omega_{\mu_2\nu_3} + 
\pi_{\mu_1\mu_3}\pi_{\nu_1\nu_3}\,\omega_{\mu_2\nu_2} +  \pi_{\mu_1\mu_3} 
\pi_{\nu_2\nu_3}\,\omega_{\mu_2\nu_1}\0\\
&&+  \pi_{\mu_2\mu_3}\pi_{\nu_1\nu_2}\,\omega_{\mu_1\nu_3} + 
\pi_{\mu_2\mu_3}\pi_{\nu_1\nu_3}\,\omega_{\mu_1\nu_2} +  \pi_{\mu_2\mu_3} 
\pi_{\nu_2\nu_3}\,\omega_{\mu_1\nu_1}  \Big) \label{spin3a6}
\ee 

\be
\overline P^{(0)}_{\mu_1\mu_2\mu_3,\nu_1\nu_2\nu_3}&=&\frac 19 \Big( 
\omega_{\mu_1\nu_1}\omega_{\mu_2\mu_3} \omega_{\nu_2\nu_3} +
\omega_{\mu_1\nu_2}\omega_{\mu_2\mu_3} 
\omega_{\nu_1\nu_3}+\omega_{\mu_1\nu_3}\omega_{\mu_2\mu_3} 
\omega_{\nu_1\nu_2}\0\\
&&+ \omega_{\mu_2\nu_1}\omega_{\mu_1\mu_3} \omega_{\nu_2\nu_3} +
\omega_{\mu_2\nu_2}\omega_{\mu_1\mu_3} 
\omega_{\nu_1\nu_3}+\omega_{\mu_2\nu_3}\omega_{\mu_1\mu_3} 
\omega_{\nu_1\nu_2}\0\\
&&+\omega_{\mu_3\nu_1}\omega_{\mu_1\mu_2} \omega_{\nu_2\nu_3} +
\omega_{\mu_3\nu_2}\omega_{\mu_1\mu_2} 
\omega_{\nu_1\nu_3}+\omega_{\mu_3\nu_3}\omega_{\mu_1\mu_2} 
\omega_{\nu_1\nu_2}\Big) \label{spin3a7}
\ee
$P^{(0)}, P^{(0a)}, P^{(0b)},\overline P^{(0)}$ are traceful. 

The six projectors are orthogonal to one another. Only $P^{(3)}$ is transverse 
and traceless .

\subsection{Arbitrary $n$}

From the previous examples it is clear that, in order to consider the general 
$n$ case, it is necessary to use a more compact notation.  Let us start from 
\be
\Delta^{\mu_1\ldots \mu_n}{}_{\nu_1\ldots \nu_n}=\; 
\delta^{(\mu_1}_{\nu_1}\ldots
\delta^{\mu_n)}_{\nu_n}
\ee
where the round brackets denote symmetrization over the $\mu$ indices 
normalized 
to 1.
Now we change the notation such that on the right hand side we strip $i$ from 
$\mu_i$ and symmetrize over all $\mu_i$ indices. We do the same with $\nu$ 
indices. (Repeated indices at the same level are symmetrized so that the weight 
of the symmetrized expression is 1.)
\be
 \Delta^{\mu_1\ldots \mu_n}{}_{\nu_1\ldots \nu_n}
=\delta^{\mu}_{\nu}\ldots\delta^{\mu}_{\nu} 
= (\delta^{\mu}_{\nu})^n 
=(\pi^{\mu}_{\nu}+\omega^{\mu}_{\nu})^n
= \sum_{k=0}^n \frac{n!}{k!(n-k)!} (\pi^{\mu}_{\nu})^k 
(\omega^{\mu}_{\nu})^{n-k} \label{Delta2}
\ee
For symmetric tensors $A$ and $B$, identities like
\be
A_{(\rho\mu_2\mu_3}B_{\mu_4\mu_5)} = 
\frac{3}{2+3}A_{\rho(\mu_2\mu_3}B_{\mu_4\mu_5)} + 
\frac{2}{2+3}A_{(\mu_2\mu_3\mu_4}B_{\mu_5)\rho}\0
\ee
hold. We write them in general as follows
\be
A_{(\rho\mu_2\ldots\mu_{A}}B_{\mu_{A+1}\ldots\mu_{A+B})} = 
\frac{A}{A+B}A_{\rho\mu\ldots\mu}B_{\mu\ldots\mu} + 
\frac{B}{A+B}A_{\mu\ldots\mu}B_{\rho\mu\ldots\mu}\0
\ee
Applying the same identity twice we get
\be
A_{(\rho\sigma\mu_3\ldots\mu_{A}}B_{\mu_{A+1}\ldots\mu_{A+B})} 
= \frac{A}{A+B} \frac{A-1}{A+B-1} A_{\rho\sigma\mu\ldots\mu} B_{\mu\ldots\mu} 
+ \frac{A}{A+B} \frac{B}{A+B-1} A_{\rho\mu\ldots\mu} B_{\sigma\mu\ldots\mu} \0\\
+ \frac{B}{A+B} \frac{A}{A+B-1} A_{\sigma\mu\ldots\mu}B_{\rho\mu\ldots\mu}
+ \frac{B}{A+B} \frac{B-1}{A+B-1} A_{\mu\ldots\mu}B_{\rho\sigma\mu\ldots\mu}\0
\ee
In this notation we get for the trace
\be
\tr(A_{\mu\ldots\mu}B_{\mu\ldots\mu}) &=& \eta^{\rho\sigma} 
A_{(\rho\sigma\mu_3\ldots\mu_{A}}B_{\mu_{A+1}\ldots\mu_{A+B})} 
\0\\
&=& \frac{1}{A+B} \frac{1}{A+B-1} 
\left( A(A-1) \eta^{\rho\sigma} A_{\rho\sigma\mu\ldots\mu} B_{\mu\ldots\mu} 
+ 2 A B  \eta^{\rho\sigma} A_{\rho\mu\ldots\mu} B_{\sigma\mu\ldots\mu} 
\right.\0\\
&& \qquad \qquad \qquad \qquad \left.
+ B (B-1) \eta^{\rho\sigma} A_{\mu\ldots\mu}B_{\rho\sigma\mu\ldots\mu} \right) 
\label{traceAB}
\ee
We are interested in the trace that acts only on the $\mu$ indices of the 
following expression
\be
\pi_{\nu\nu}^l \pi_{\mu\mu}^m \pi_{\nu\mu}^n\0
\ee
We have
\be
\tr(\pi_{\nu\nu}^l \pi_{\mu\mu}^m \pi_{\nu\mu}^n) &=& \pi_{\nu\nu}^l 
\tr(\pi_{\mu\mu}^m \pi_{\nu\mu}^n)\0
\ee
\be
\tr(\pi_{\mu\mu}^m \pi_{\nu\mu}^n) &=& \frac{1}{(2m+n)(2m+n-1)}
\left( 2 m \; \eta^{\rho\sigma} \pi_{\rho\sigma} \pi_{\mu\mu}^{m-1} 
\pi_{\nu\mu}^n \right.\0\\
&& \qquad \qquad \qquad \qquad \left. + 4 m (m-1) \; \eta^{\rho\sigma} 
\pi_{\mu\rho}\pi_{\mu\sigma} \pi_{\mu\mu}^{m-2} \pi_{\nu\mu}^{n} \right.\0\\
&& \qquad \qquad \qquad \qquad \left. + 4 m n \; \eta^{\rho\sigma} 
\pi_{\mu\rho} 
\pi_{\mu\mu}^{m-1} \pi_{\nu\sigma} \pi_{\nu\mu}^{n-1} \right.\0\\
&& \qquad \qquad \qquad \qquad \left. + n (n-1) \eta^{\rho\sigma} 
\pi_{\mu\mu}^{m} \pi_{\nu\rho}\pi_{\nu\sigma} \pi_{\nu\mu}^{n-2} \right)
\0\\
&=& \frac{1}{(2m+n)(2m+n-1)}
\left( 2 m \; (d-1) \pi_{\mu\mu}^{m-1} \pi_{\nu\mu}^n \right.\0\\
&& \qquad \qquad \qquad \qquad \left. + 4 m (m-1) \;   \pi_{\mu\mu}^{m-1} 
\pi_{\nu\mu}^{n} \right.\0\\
&& \qquad \qquad \qquad \qquad \left. + 4 m n \; \pi_{\mu\mu}^{m-1}  
\pi_{\nu\mu}^{n} \right.\0\\
&& \qquad \qquad \qquad \qquad \left. + n (n-1) \pi_{\nu\nu}  \pi_{\mu\mu}^{m} 
\pi_{\nu\mu}^{n-2} \right)\0
\ee
Therefore,
\be
\tr(\pi_{\nu\nu}^l \pi_{\mu\mu}^m \pi_{\nu\mu}^n) 
=
A_{m,n}^{(d-1)}
\pi_{\nu\nu}^l \pi_{\mu\mu}^{m-1}  \pi_{\nu\mu}^{n} + 
B_{m,n}
\pi_{\nu\nu}^{l+1}  \pi_{\mu\mu}^{m} \pi_{\nu\mu}^{n-2} \0
\ee
where
\be
A_{m,n}^{(d-1)}=\frac{2m\left(  d-3 + 2 (m+n)  \right)}{(2m+n)(2m+n-1)}\0\\
B_{m,n}=\frac{n (n-1) }{(2m+n)(2m+n-1)}\0
\ee
The trace of $\pi_{\nu\nu}^{l} \pi_{\mu\mu}^{m} \pi_{\nu\mu}^{n}$ contains
$\pi_{\nu\nu}^l \pi_{\mu\mu}^{m-1}  \pi_{\nu\mu}^{n}$ and
$\pi_{\nu\nu}^{l+1}  \pi_{\mu\mu}^{m} \pi_{\nu\mu}^{n-2}$.\\
The trace of $\pi_{\nu\nu}^{l+1} \pi_{\mu\mu}^{m+1} \pi_{\nu\mu}^{n-2}$ contains
again $\pi_{\nu\nu}^{l+1} \pi_{\mu\mu}^{m}  \pi_{\nu\mu}^{n-2}$ and also
$\pi_{\nu\nu}^{l+2}  \pi_{\mu\mu}^{m+1} \pi_{\nu\mu}^{n-4}$.
So, forming the sum
\be
S_{lmn}^{(d-1)}(\pi) = \sum_{p=0}^{\bar{p}} c_{lmnp}^{(d-1)} \pi_{\nu\nu}^{l+p} 
\pi_{\mu\mu}^{m+p} \pi_{\nu\mu}^{n-2p}\label{Slmn}
\ee
and putting the right coefficients we can cancel all terms except the first and 
the last in the trace of the sum. The result is then
\be
\tr\, S_{lmn}^{(d-1)}(\pi) = c_{lmn0}^{(d-1)} A_{mn}^{(d-1)} \pi_{\nu\nu}^{l} 
\pi_{\mu\mu}^{m-1} \pi_{\nu\mu}^{n} + 
c_{lmnp}^{(d-1)} B_{m+\bar{p},n-2\bar{p}} \pi_{\nu\nu}^{l+\bar{p}+1} 
\pi_{\mu\mu}^{m+\bar{p}} \pi_{\nu\mu}^{n-2\bar{p}-2} \label{trsum1}
\ee
The condition for the cancellation that leads to \eqref{trsum1} is 
\be
c_{lmnp}^{(d-1)} A_{m+p,n-2p}^{(d-1)} + c_{lmn(p-1)}^{(d-1)} B_{m+p-1,n-2(p-1)} 
= 0\0
\ee
that is
\be
c_{lmnp}^{(d-1)} = -c_{lmn(p-1)}^{(d-1)} \frac{(n-2p+2) (n-2p+1) }{2(m+p)\left( 
d-3 + 2 (m+n-p)  \right)}\0
\ee
\be
c_{lmnp}^{(d-1)} = 
\frac{(n-2p+2) (n-2p+1) }{2(m+p)\left(  d-3 + 2 (m+n-p)  \right)} 
\frac{(n-2p+4) (n-2p+3) }{2(m+p-1)\left(  d-3 + 2 (m+n-p+1)  \right)}
\ldots\0\\\quad\ldots
\frac{(n-2) (n-3) }{2(m+2)\left(  d-3 + 2 (m+n-2)  \right)}
\frac{n (n-1) }{2(m+1)\left(  d-3 + 2 (m+n-1)  \right)}(-)^p c_0^{(d-1)}\0
\ee
or
\be
c_{lmnp}^{(d-1)} = (-)^p \frac{\frac{n!}{(n-2p)!}}{2^p\frac{(m+p)!}{m!}  
\frac{\left(  d-3 + 2 (m+n-1)  \right)!!}{\left(  d-3 + 2 (m+n-p-1)  
\right)!!}} 
\0
\ee
where we used $c_{lmn0}^{(d-1)} = 1$.
When $m=0$ (and $n\geq 2$) we have $A_{0,n} = 0$, and the first term in 
\eqref{trsum1} drops out
\be
\tr\, S_{00n}^{(d-1)}(\pi) = c_{00n\bar{p}}^{(d-1)} B_{\bar{p},n-2\bar{p}} 
\pi_{\nu\nu}^{\bar{p}+1} \pi_{\mu\mu}^{\bar{p}} \pi_{\nu\mu}^{n-2\bar{p}-2} 
\label{trsum2}
\ee
If the upper bound $\bar{p}$ of the sum is big enough, the numerator of 
$B_{\bar{p},n-2\bar{p}}$ becomes zero
\be
(n-2\bar{p}) (n-2\bar{p}-1) = 0\label{np0}
\ee
This happens when $\bar{p} = n/2$ for $n$ even, or $\bar{p} = (n-1)/2$ for $n$ 
odd i.e.:
\be
\bar{p} = \lfloor n/2 \rfloor \label{barp}
\ee
Thus we conclude that $S_{00n}^{(d-1)}(\pi)$ with $\bar{p} = \lfloor n/2 
\rfloor$ is traceless
\be
\tr S_{00n}^{(d-1)}(\pi) = 0 \label{trsum3}
\ee
Let us split it into the first term plus the rest which we denote 
$R_{00n}^{(d-1)}$
\be
S_{00n}^{(d-1)}(\pi) &=& \pi_{\mu\nu}^n 
- \frac{n(n-1)}{2(d+2n-5)}\pi_{\mu\mu}\pi_{\mu\nu}^{n-2}\pi_{\nu\nu}
+ 
\frac{n(n-1)(n-2)(n-3)}{8(d+2n-5)(d+2n-7)}\pi_{\mu\mu}^2\pi_{\mu\nu}^{n-4}\pi_{
\nu\nu}^2
+\ldots\0\\
&\equiv & \pi_{\mu\nu}^n + R_{00n}^{(d-1)}(\pi)\label{S00n}
\ee
We repeat the same procedure to form traceless sums containing 
$\omega_{\mu\nu}^n$. Here, $\pi \rightarrow \omega$, and $d-1 \rightarrow 1$. 
\be
S_{00n}^{(1)}(\omega) &=& \omega_{\mu\nu}^n 
- \frac{n(n-1)}{2(2n-3)}\omega_{\mu\mu}\omega_{\mu\nu}^{n-2}\omega_{\nu\nu}
+ 
\frac{n(n-1)(n-2)(n-3)}{8(2n-3)(2n-5)}\omega_{\mu\mu}^2\omega_{\mu\nu}^{n-4}
\omega_{\nu\nu}^2
+\ldots\0\\
&\equiv & \omega_{\mu\nu}^n + R_{00n}^{(1)}(\omega)\0
\ee
We note that 
\be
\omega_{\mu\mu}\omega_{\nu\nu} = \frac{k_{\mu}k_{\mu}k_{\nu}k_{\nu}}{k^4} = 
\omega_{\mu\nu}^2 \label{omega2}
\ee
and therefore
$\omega_{\mu\mu}^m\omega_{\mu\nu}^{n-2m}\omega_{\nu\nu}^m = 
\omega_{\mu\nu}^{n}$ 
so in this case we find
\be
S_{00n}^{(1)}(\omega) &=& \omega_{\mu\nu}^n \left( 1\!
-\! \frac{n(n-1)}{2(2n-3)}\!
+ \!\frac{n(n-1)(n-2)(n-3)}{8(2n-3)(2n-5)}\! +\!\ldots \right)
= \begin{cases} 
1 & n=0 \\
\omega_{\mu\nu} & n=1 \\
0 & n \geq 2 
\end{cases} \label{explS}
\ee
\be
R_{00n}^{(1)}(\omega) &=& \begin{cases} 
0 & n=0 \\
0 & n=1 \\
-\omega_{\mu\nu}^n & n \geq 2 
\end{cases} \label{explR}
\ee

We now use these traceless expressions in \eqref{Delta2} i.e.\ we add and 
subtract $R$:
{\scriptsize
\be
\Delta_{\mu_1\ldots \mu_n}{}_{\nu_1\ldots \nu_n}
&=& \sum_{k=0}^n \frac{n!}{k!(n-k)!} 
(\pi_{\mu\nu}^k + R_{00k}^{(d-1)}(\pi) - R_{00k}^{(d-1)}(\pi)) 
(\omega_{\mu\nu}^{n-k} + R_{0,0,n-k}^{(1)}(\omega) -  R_{0,0,n-k}^{(1)}(\omega))
\0\\
&=& \sum_{k=0}^n \frac{n!}{k!(n-k)!} 
(S_{00k}^{(d-1)}(\pi) S_{0,0,n-k}^{(1)}(\omega)
- \pi_{\mu\nu}^k R_{0,0,n-k}^{(1)}(\omega)
- R_{00k}^{(d-1)}(\pi) S_{0,0,n-k}^{(1)}(\omega))\0\\
&\equiv & \sum_{k=0}^n  (P^{(k,n)}_1 + P^{(k,n)}_2 + P^{(k,n)}_3)\label{P1P2P3}
\ee}
where in the last line we defined three quantities $P^{(k,n)}_i$
\be
P^{(k,n)}_1 &=& \frac{n!}{k!(n-k)!} S_{00k}^{(d-1)}(\pi) 
S_{0,0,n-k}^{(1)}(\omega) \label{P1kn}\\
P^{(k,n)}_2 &=& - \frac{n!}{k!(n-k)!} \pi_{\mu\nu}^k 
R_{0,0,n-k}^{(1)}(\omega)\label{P2kn}\\
P^{(k,n)}_3 &=& - \frac{n!}{k!(n-k)!} R_{00k}^{(d-1)}(\pi) 
S_{0,0,n-k}^{(1)}(\omega)\label{P3kn}
\ee
We note that $P^{(k,n)}_1$ is traceless
\be
\tr P^{(k,n)}_1 
=
c_1 \left(\tr S_{00k}^{(d-1)}(\pi)\right) S_{0,0,n-k}^{(1)}(\omega) +
c_2 S_{00k}^{(d-1)}(\pi) \tr\left(S_{0,0,n-k}^{(1)}(\omega)\right) = 
0\label{trP1kn}
\ee
since the mixed term in \eqref{traceAB} vanishes because of the orthogonality 
$\eta^{\rho\sigma} \pi_{\rho\mu} \omega_{\sigma\nu} = 0$.

Using \eqref{explS} and \eqref{explR} for $n\geq 2$ we see that the expression 
for $\Delta$ simplifies
\be
\Delta_{\mu_1\ldots \mu_n}{}_{\nu_1\ldots \nu_n}
&=& S_{00n}^{(d-1)}(\pi) + n \, S_{00(n-1)}^{(d-1)}(\pi) \, \omega_{\mu\nu}
\0\\ && 
+ \sum_{k=0}^{n-2} \frac{n!}{k!(n-k)!} \, \pi_{\mu\nu}^k \, 
\omega_{\mu\nu}^{n-k}
\0\\ && 
- R_{00n}^{(d-1)}(\pi) - n \, R_{00(n-1)}^{(d-1)}(\pi) \, 
\omega_{\mu\nu}\label{newDelta}
\ee
That is
\be
P^{(n,n)}_1 &=& S_{00n}^{(d-1)}(\pi) , \0\\
P^{(n-1,n)}_1 &=&  n \, S_{00(n-1)}^{(d-1)}(\pi) \, \omega_{\mu\nu}, \0\\
P^{(k,n)}_1 &=& 0, \qquad k \leq n-2 , \0\\
P^{(n,n)}_2 &=&0, \0\\ 
P^{(n-1,n)}_2 &=& 0, \0\\
P^{(k,n)}_2 &=&  \frac{n!}{k!(n-k)!} \, \pi_{\mu\nu}^k \, 
\omega_{\mu\nu}^{n-k}, 
\qquad  k \leq n-2 , \0\\
P^{(n,n)}_3 &=& - R_{00n}^{(d-1)}(\pi), \0\\
P^{(n-1,n)}_3 &=&- n \, R_{00(n-1)}^{(d-1)}(\pi) \, \omega_{\mu\nu}, \0\\ 
P^{(k,n)}_3 &=& 0, \qquad k \leq n-2 . \label{Ps}
\ee
Putting numbers
\be
S_{000}^{(d-1)}(\pi) &=& 1\label{S000}
\ee
\be
S_{001}^{(d-1)}(\pi) &=& \pi_{\mu\nu}\label{S001}
\ee
\be
S_{002}^{(d-1)}(\pi) &=& \pi_{\mu\nu}^2 - 
\frac{1}{d-1}\pi_{\mu\mu}\pi_{\nu\nu}\label{S002}
\ee
\be
S_{003}^{(d-1)}(\pi) &=& \pi_{\mu\nu}^3 - 
\frac{3}{d+1}\pi_{\mu\mu}\pi_{\mu\nu}\pi_{\nu\nu}\label{S003}
\ee
\be
S_{004}^{(d-1)}(\pi) &=& \pi_{\mu\nu}^4 
- \frac{6}{d+3}\pi_{\mu\mu}\pi_{\mu\nu}^{2}\pi_{\nu\nu}
+ \frac{3}{(d+3)(d+1)}\pi_{\mu\mu}^2\pi_{\nu\nu}^2\label{S004}
\ee
we can connect these P's to the P's from the beginning of the Appendix. For 
$n=1$
\be
P^{(1,1)}_1 &=& S_{001}^{(d-1)}(\pi) = \pi_{\mu\nu} =  P^{(1)}_{\mu_1,\nu_1}, 
\0\\
P^{(0,1)}_1 &=&  S_{000}^{(d-1)}(\pi) \, \omega_{\mu\nu} = \omega_{\mu\nu} =  
P^{(0)}_{\mu_1,\nu_1}, \0\\
P^{(1,1)}_2 &=&0, \0\\ 
P^{(0,1)}_2 &=& 0, \0\\
P^{(1,1)}_3 &=& - R_{001}^{(d-1)}(\pi) = 0, \0\\
P^{(0,1)}_3 &=&- R_{000}^{(d-1)}(\pi) \, \omega_{\mu\nu} = 
0.\label{n=1}\label{Pn=1}
\ee
For $n=2$
\be
P^{(2,2)}_1 &=& S_{002}^{(d-1)}(\pi) = \pi_{\mu\nu}^2 - 
\frac{1}{d-1}\pi_{\mu\mu}\pi_{\nu\nu} = P^{(2)}_{\mu_1\mu_2,\nu_1\nu_2} , \0\\
P^{(1,2)}_1 &=&  2 \, S_{001}^{(d-1)}(\pi) \, \omega_{\mu\nu} = 
2\pi_{\mu\nu}\omega_{\mu\nu} = P^{(1)}_{\mu_1\mu_2,\nu_1\nu_2} , \0\\
P^{(0,2)}_1 &=& 0, \0\\
P^{(2,2)}_2 &=& 0, \0\\ 
P^{(1,2)}_2 &=& 0, \0\\
P^{(0,2)}_2 &=& \omega_{\mu\nu}^2 = \bar{P}^{(0)}_{\mu_1\mu_2,\nu_1\nu_2} 
\text{ 
because of \eqref{omega2}}, \0\\
P^{(2,2)}_3 &=& - R_{002}^{(d-1)}(\pi) = \frac{1}{d-1}\pi_{\mu\mu}\pi_{\nu\nu} 
= 
P^{(0)}_{\mu_1\mu_2,\nu_1\nu_2}, \0\\
P^{(1,2)}_3 &=& - 2 \, R_{001}^{(d-1)}(\pi) \, \omega_{\mu\nu}  = 0 \text{ 
because } R_{001}^{(d-1)}(\pi)=0, \0\\ 
P^{(0,2)}_3 &=& 0. \label{Pn=2}
\ee
For $n=3$
\be
P^{(3,3)}_1 &=& S_{003}^{(d-1)}(\pi) 
= \pi_{\mu\nu}^3 - \frac{3}{d+1}\pi_{\mu\mu}\pi_{\mu\nu}\pi_{\nu\nu}
= P^{(3)}_{\mu_1\mu_2\mu_3,\nu_1\nu_2\nu_3},  \0\\
P^{(2,3)}_1 &=&  n \, S_{002}^{(d-1)}(\pi) \, \omega_{\mu\nu} 
= 3 \left( \pi_{\mu\nu}^2 - \frac{1}{d-1}\pi_{\mu\mu}\pi_{\nu\nu} \right)  \, 
\omega_{\mu\nu} 
= P^{(2)}_{\mu_1\mu_2\mu_3,\nu_1\nu_2\nu_3},\0\\
P^{(k,3)}_1 &=& 0, \qquad k \leq 1 , \0\\
P^{(3,3)}_2 &=&0, \0\\ 
P^{(2,3)}_2 &=& 0, \0\\
P^{(1,3)}_2 &=&  3 \pi_{\mu\nu}^1 \, \omega_{\mu\nu}^{2}
=P^{(0a)}_{\mu_1\mu_2\mu_3,\nu_1\nu_2\nu_3} , \0\\
P^{(0,3)}_2 &=&    \omega_{\mu\nu}^{3} = 
\bar{P}^{(0)}_{\mu_1\mu_2\mu_3,\nu_1\nu_2\nu_3}, \0\\
P^{(3,3)}_3 &=& - R_{003}^{(d-1)}(\pi) = 
\frac{3}{d+1}\pi_{\mu\mu}\pi_{\mu\nu}\pi_{\nu\nu}
 = P^{(0)}_{\mu_1\mu_2\mu_3,\nu_1\nu_2\nu_3}, \0\\
P^{(2,3)}_3 &=& -3 \, R_{00(n-1)}^{(d-1)}(\pi) \, \omega_{\mu\nu} = 
\frac{3}{d-1}\pi_{\mu\mu}\pi_{\nu\nu}\omega_{\mu\nu}
= P^{(0b)}_{\mu_1\mu_2\mu_3,\nu_1\nu_2\nu_3}, \0\\ 
P^{(k,n)}_3 &=& 0, \qquad k \leq 1 , \label{Pn=3}
\ee

For $n=4$
\be
P^{(4,4)}_1 &=& S_{004}^{(d-1)}(\pi) = \pi_{\mu\nu}^4 
- \frac{6}{d+3}\pi_{\mu\mu}\pi_{\mu\nu}^{2}\pi_{\nu\nu}
+ \frac{3}{(d+3)(d+1)}\pi_{\mu\mu}^2\pi_{\nu\nu}^2, \0\\
P^{(3,4)}_1 &=&  4 \, S_{003}^{(d-1)}(\pi) \, \omega_{\mu\nu}
 = 4\left(\pi_{\mu\nu}^3 - 
\frac{3}{d+1}\pi_{\mu\mu}\pi_{\mu\nu}\pi_{\nu\nu}\right) \, \omega_{\mu\nu}, 
\0\\
P^{(k,4)}_1 &=& 0, \qquad k \leq 2 , \0\\
P^{(4,4)}_2 &=& 0, \0\\ 
P^{(3,4)}_2 &=& 0, \0\\
P^{(2,4)}_2 &=&  6 \, \pi_{\mu\nu}^2 \, \omega_{\mu\nu}^2, \0\\
P^{(1,4)}_2 &=&  4 \, \pi_{\mu\nu}^1 \, \omega_{\mu\nu}^3, \0\\
P^{(0,4)}_2 &=&  \omega_{\mu\nu}^4, \0\\
P^{(4,4)}_3 &=& - R_{004}^{(d-1)}(\pi)
= \frac{6}{d+3}\pi_{\mu\mu}\pi_{\mu\nu}^{2}\pi_{\nu\nu}
-\frac{3}{(d+3)(d+1)}\pi_{\mu\mu}^2\pi_{\nu\nu}^2, \0\\
P^{(3,4)}_3 &=&- 4 \, R_{003}^{(d-1)}(\pi) \, \omega_{\mu\nu} = 
\frac{12}{d+1}\pi_{\mu\mu}\pi_{\mu\nu}\pi_{\nu\nu} \, \omega_{\mu\nu}, \0\\
P^{(k,4)}_3 &=& 0, \qquad k \leq 2 , \label{Pn=4}
\ee

We now check orthonormality of the $P$'s with the same $n$'s. First we check 
orthogonality of the product
\be
P^{(k,n)}_i \circ P^{(k',n)}_{i'}\label{PknPk'n}
\ee
where $\circ$ denotes the contraction of the second group of indices of the 
first factor with the first group of the second factor.
Since $k$ and $k'$ determine the number of $\omega_{\mu\nu}$ factors, and since 
both $P$'s have the same  indices, if $k \neq k'$, it must be that in 
contractions $\omega$ hits $\pi$ and the result is zero. So, it remains to check
\be
P^{(k,n)}_i \circ P^{(k,n)}_{i'}\label{PkniPkni'}
\ee
If $i\neq i'$ and if one of the $i$'s is $2$ the result is zero as can be seen 
from \eqref{Ps}. If none of the $i$'s is $2$ we have
\be
P^{(k,n)}_1 \circ P^{(k,n)}_{3}\label{Pkn1Pkn3}
\ee
which is zero because $P_3$ contains at least one $\pi_{\nu\nu}$, and 
$S_{\mu\ldots\mu,\nu\ldots\nu} \pi^{\nu\nu} = \tr^{(\nu)} S = 0$ (where 
$\tr^{(\nu)}$ denotes a trace with respect to the indices $\nu$). We conclude 
that the $P$'s with the same $n$ are orthogonal.

Now we consider the square
\be
P^{(k,n)}_i \circ P^{(k,n)}_{i}\label{PkniPkni}
\ee
Let us  denote 
\be
\left( \pi_0^m\pi_1^n \pi_2^l \right)_{\mu\ldots\mu,\nu\ldots\nu} \equiv 
\left( \pi_{\mu\mu}\right)^m
\left( \pi_{\mu\nu}\right)^n
\left( \pi_{\nu\nu}\right)^l\0
\ee
We see that $\pi_1^{n}$ acts as an identity in the sense that
\be
\pi_0^m \pi_1^n \pi_2^l \circ \pi_1^{n+2l} = \pi_0^m \pi_1^n \pi_2^l\0
\ee
So we have
\be
S_{00n}^{(d-1)} \circ \pi_1^{n} = S_{00n}^{(d-1)}\label{S00nS00n}
\ee
As mentioned
\be
\left(S_{00n}^{(d-1)}\right)_{\mu\ldots\mu,\nu\ldots\nu} \pi^{\nu\nu} = 
\tr^{(\nu)} \left(S_{00n}^{(d-1)}\right)_{\mu\ldots\mu,\nu\ldots\nu} = 0, 
\label{S00nd-1}
\ee
from which it follows that
\be
S_{00n}^{(d-1)} \circ R_{00n}^{(d-1)} = 0\label{S00nRoon}
\ee
since all terms in $R$ contain at least one $\pi_{\nu\nu}$. We conclude
\be
S_{00n}^{(d-1)} \circ S_{00n}^{(d-1)} = S_{00n}^{(d-1)} \circ \left(\pi_1^{n} + 
R_{00n}^{(d-1)}\right) = S_{00n}^{(d-1)}\label{S00nS00n1}
\ee
i.e.
\be
P_1^{(n,n)} \circ P_1^{(n,n)} = P_1^{(n,n)}\label{P1nnP1nn}
\ee
and
\be
\left(-R_{00n}^{(d-1)}\right) \circ \left(-R_{00n}^{(d-1)}\right) 
= \left(S_{00n}^{(d-1)} - \pi_1^{n}\right) \circ R_{00n}^{(d-1)}
= - \pi_1^{n} \circ R_{00n}^{(d-1)}
= - R_{00n}^{(d-1)}\label{R00nR00nd-1}
\ee
i.e.
\be
P_3^{(n,n)} \circ P_3^{(n,n)} = P_3^{(n,n)}\label{P3nnP3nn}
\ee
Similarly one finds the remaining equations $P_2^{(k,n)} \circ P_2^{(k,n)} = 
P_2^{(k,n)}$, 
$P_1^{(n-1,n)} \circ P_1^{(n-1,n)} = P_1^{(n-1,n)}$ and
$P_3^{(n-1,n)} \circ P_3^{(n-1,n)} = P_3^{(n-1,n)}$.

In summary $\Delta$ is the sum of all the P's i.e.
\be
\Delta_{\mu_1\ldots \mu_n}{}_{\nu_1\ldots \nu_n}
&=& \sum_{i=1}^3 \sum_{k=0}^n P^{(k,n)}_i
\ee
where only the $P^{(k,n)}_1$ are traceless. Two of $P^{(k,n)}_1$'s are 
non-zero: 
$P^{(n-1,n)}_1$ and $P^{(n,n)}_1$. Of $P^{(k,n)}_2$'s, $n-2$ are non-zero, and 
$2$ of $P^{(k,n)}_3$'s are non-zero. The $P$'s are orthonormal
\be
P_i^{(k,n)} \circ P_{i'}^{(k',n)} = \delta_{i,i'} \delta^{k,k'} P_i^{(k,n)}
\ee
Of all these projectors {\it only $P^{(n,n)}_1$ is transverse and traceless}. In 
the text it will be denoted $\EP^{(n)}$.



\begin{thebibliography}{99}

\bibitem{HSgeneral} {\it Higher-Spin Gauge Theories}, Proceedings of the First
Solvay Workshop, held in Brussels on May 12-14, 2004, eds. R. Argurio, G.
Barnich, G. Bonelli and M. Grigoriev (Int. Solvay Institutes, 2006).


\bibitem{Maldacena}
X.~O.~Camanho, J.~D.~Edelstein, J.~Maldacena and A.~Zhiboedov,
{\it Causality Constraints on Corrections to the Graviton Three-Point
Coupling,}
  JHEP \textbf {1602} (2016) 020
   [arXiv:1407.5597 [hep-th]].

\bibitem{Kundu}
  N.~Afkhami-Jeddi, S.~Kundu and A.~Tajdini,
{\it A Bound on Massive Higher Spin Particles,}
  arXiv:1811.01952 [hep-th].
\bibitem{Steinacker} 
  H.~C.~Steinacker,
  {\it On the quantum structure of space-time, gravity, and higher spin,'}
  arXiv:1911.03162 [hep-th].
 

\bibitem{I}
  L.~Bonora, M.~Cvitan, P.~Dominis Prester, S.~Giaccari and T.~Stemberga,
  {\it HS in flat spacetime. The effective action method.}
[arxiv:1811.04847[hep-th]]

\bibitem{Taronna}
C.~Sleight and M.~Taronna,
{\it Higher-Spin Gauge Theories and Bulk Locality,}
  Phys.\ Rev.\ Lett.\  {\bf 121}, no. 17, 171604 (2018)
  doi:10.1103/PhysRevLett.121.171604
  [arXiv:1704.07859 [hep-th]].

\bibitem{3d} H.~A.~Gonzalez, J.~Matulich, M.~Pino and R.~Troncoso,
{\it Asymptotically flat spacetimes in three-dimensional higher spin gravity,}
  JHEP {\bf 1309} (2013) 016
  doi:10.1007/JHEP09(2013)016
  [arXiv:1307.5651 [hep-th]].

H.~Afshar, A.~Bagchi, R.~Fareghbal, D.~Grumiller and J.~Rosseel,
{\it Spin-3 Gravity in Three-Dimensional Flat Space,}
  Phys.\ Rev.\ Lett.\  {\bf 111}, no. 12, 121603 (2013)
  doi:10.1103/PhysRevLett.111.121603
  [arXiv:1307.4768 [hep-th]].



\bibitem{BBvD} F.A.Berends, G.J.H. Burgers and H. Van Dam 
{\it On the theoretical problems in constructing interactions involving
higher-spin masslass particles}
Nucl. Phys. {\bf B260} (1985) 295-322.



\bibitem{FV} E. S. Fradkin and M. A. Vasiliev, {\it On the Gravitational
Interaction of Massless Higher Spin
Fields}, Phys. Lett. {\bf B189} (1987) 89.

E. S. Fradkin and M. A. Vasiliev, {\it Cubic Interaction in Extended Theories of
Massless Higher Spin Fields}, Nucl. Phys. {\bf B291} (1987) 141.

\bibitem{Vasiliev}  M.~A.~Vasiliev,
{\it Consistent equation for interacting gauge fields of all spins in
(3+1)-dimensions,}
  Phys.\ Lett.\ B \textbf {243} (1990) 378;
   {\it Properties of equations of motion of interacting gauge fields of all
spins in (3+1)-dimensions,}
  Class.\ Quant.\ Grav.\  \textbf {8} (1991) 1387;
  {\it Algebraic aspects of the higher spin problem,}
  Phys.\ Lett.\ B \textbf {257} (1991) 111;
{\it More on equations of motion for interacting massless fields of all
spins in (3+1)-dimensions,}
  Phys.\ Lett.\ B \textbf {285} (1992) 225.

\bibitem{lightcone1}
 
  R.~R.~Metsaev,
  {\it Poincare invariant dynamics of massless higher spins: Fourth order 
analysis on mass shell,}
  Mod.\ Phys.\ Lett.\ A {\bf 6}, 359 (1991).
  doi:10.1142/S0217732391000348
  

D.~Ponomarev and E.~D.~Skvortsov,
  {\it Light-Front Higher-Spin Theories in Flat Space,}
  J.\ Phys.\ A {\bf 50}, no. 9, 095401 (2017)
  doi:10.1088/1751-8121/aa56e7
  [arXiv:1609.04655 [hep-th]].


E.~D.~Skvortsov, T.~Tran and M.~Tsulaia,
{\it Quantum Chiral Higher Spin Gravity,}
  Phys.\ Rev.\ Lett.\  {\bf 121} (2018) no.3,  031601
  doi:10.1103/PhysRevLett.121.031601
  [arXiv:1805.00048 [hep-th]].

 
\bibitem{Sleight:2016xqq}
  C.~Sleight and M.~Taronna,
  {\it Higher-Spin Algebras, Holography and Flat Space,}
  JHEP {\bf 1702} (2017) 095
  doi:10.1007/JHEP02(2017)095
  [arXiv:1609.00991 [hep-th]].
  


\bibitem{3vertex} 
 A. K. H. Bengtsson, I. Bengtsson, and L. Brink, 
{\it Cubic interaction terms for arbitrary spin,}
Nucl. Phys. {\bf B227} (1983) 31.

A. K. H. Bengtsson, I. Bengtsson, and L. Brink, 
{\it Cubic interaction terms for arbitrarily terms for arbitrary extended
supermultiplets,}
Nucl. Phys. {\bf B227} (1983) 41.

A. K. H. Bengtsson, I. Bengtsson, and N. Linden,
{\it Interacting higher spin gauge fields on the light front}
Class. Quant. Grav. {\bf 4} (1987) 1333.

F.~A.~Berends, G.~J.~H.~ Burgers and H~. Van Dam 
{\it On spin 3 selfinteractions}
Z.\ Phys.\ {\bf C24} (1984) 247.

F.~A.~Berends, G.~J.~H.~ Burgers and H~. Van Dam
{\it On the theoretical problems in constructing interactions involving higher
spin massless particles ,} Nucl. Phys. {\bf B260} (1985) 295.


E.~ S.~ Fradkin and R. ~R.~ Metsaev, 
{\it A Cubic interaction of totally symmetric massless representations of the
Lorentz group in arbitrary dimensions,} 
Class. Quant. Grav. {\bf 8} (1991) L89-L94.

R.~R.~Metsaev,
 {\it Cubic interaction vertices for fermionic and bosonic arbitrary spin
fields,}  Nucl.\ Phys.\ B {\bf 859} (2012) 13
   [arXiv:0712.3526 [hep-th]].

R.~ R.~ Metsaev, 
{\it Generating function for cubic interaction vertices of higher spin fields in
any dimension,} 
Mod. Phys. Lett. {\bf A8} (1993) 2413-2426.

R.~Manvelyan, K.~Mkrtchyan and W.~Ruhl,
 {\it Off-shell construction of some trilinear higher spin gauge field
interactions,}
  Nucl.\ Phys.\ B {\bf 826} (2010) 1
  doi:10.1016/j.nuclphysb.2009.07.007
  [arXiv:0903.0243 [hep-th]].

R.~Manvelyan, K.~Mkrtchyan and W.~Ruehl,
 {\it Direct Construction of A Cubic Selfinteraction for Higher Spin gauge
Fields,}
  Nucl.\ Phys.\ B {\bf 844} (2011) 348
  doi:10.1016/j.nuclphysb.2010.11.009
  [arXiv:1002.1358 [hep-th]].

 R.~Manvelyan, K.~Mkrtchyan and W.~Ruhl,
 {\it General trilinear interaction for arbitrary even higher spin gauge
fields,}
  Nucl.\ Phys.\ B {\bf 836} (2010) 204
  doi:10.1016/j.nuclphysb.2010.04.019
  [arXiv:1003.2877 [hep-th]].

 
  A.~Sagnotti and M.~Taronna,
  {\it String Lessons for Higher-Spin Interactions,}
  Nucl.\ Phys.\ B {\bf 842}, 299 (2011)
  doi:10.1016/j.nuclphysb.2010.08.019
  [arXiv:1006.5242 [hep-th]].
   
  A.~Fotopoulos and M.~Tsulaia,
  {\it On the Tensionless Limit of String theory, Off - Shell Higher Spin 
Interaction Vertices and BCFW Recursion Relations,}
  JHEP {\bf 1011}, 086 (2010)
  doi:10.1007/JHEP11(2010)086
  [arXiv:1009.0727 [hep-th]].
  
 
  M.~Taronna,
  {\it Higher-Spin Interactions: four-point functions and beyond,}
  JHEP {\bf 1204}, 029 (2012)
  doi:10.1007/JHEP04(2012)029
  [arXiv:1107.5843 [hep-th]].
   
 
  M.~Taronna,
  {\it Higher-Spin Interactions: three-point functions and beyond,}
  PhD thesis
  arXiv:1209.5755 [hep-th].
   

 
Y.~M.~Zinoviev,
 {\it On spin 3 interacting with gravity,}
  Class.\ Quant.\ Grav.\  {\bf 26} (2009) 035022
  doi:10.1088/0264-9381/26/3/035022
  [arXiv:0805.2226 [hep-th]].

Y.~M.~Zinoviev,
 {\it Spin 3 cubic vertices in a frame-like formalism,}
  JHEP {\bf 1008} (2010) 084
  doi:10.1007/JHEP08(2010)084
  [arXiv:1007.0158 [hep-th]].

 
\bibitem{Taronna:2017wbx} 
  M.~Taronna,
  {\it On the Non-Local Obstruction to Interacting Higher Spins in Flat Space,}
  JHEP {\bf 1705}, 026 (2017)

  
\bibitem{Roiban}
  R.~Roiban and A.~A.~Tseytlin,
 {\it On four-point interactions in massless higher spin theory in flat space,}
  JHEP {\bf 1704}, 139 (2017)
  doi:10.1007/JHEP04(2017)139
  [arXiv:1701.05773 [hep-th]].



\bibitem{BCDGPS}
  L.~Bonora, M.~Cvitan, P.~Dominis Prester, S.~Giaccari, M.~Paulisic and
T.~Stemberga,
  {\it Worldline quantization of field theory, effective actions and $L_\infty$
structure,}
arXiv:1802.02968 [hep-th].

\bibitem{strassler}
  M.~J.~Strassler,
 {\it Field theory without Feynman diagrams: One loop effective actions,}
  Nucl.\ Phys.\ B {\bf 385} (1992) 145
  doi:10.1016/0550-3213(92)90098-V
  [hep-ph/9205205].

\bibitem{segal}
  A.~Y.~Segal,
{\it Conformal higher spin theory,}
  Nucl.\ Phys.\ B {\bf 664} (2003) 59
  doi:10.1016/S0550-3213(03)00368-7
  [hep-th/0207212].

A.~Y.~Segal, 
{\it Point particle in general background fields versus gauge theories of
traceless symmetric tensors,}
  Int.\ J.\ Mod.\ Phys.\ A {\bf 18} (2003) 4999
  doi:10.1142/S0217751X03015830
  [hep-th/0110056].


\bibitem{Tseytlin} A. A. Tseytlin, {\it On limits of superstring in AdS(5) x
S**5,} Theor.\ Math. \ Phys. {\bf 133}  (2002) 1376 [Teor. Mat. Fiz.{\bf 133}
(2002) 69  ] [hep-th/0201112].

\bibitem{bekaert}  X.~Bekaert, E.~Joung and J.~Mourad,
{\it Effective action in a higher-spin background,}
JHEP {\bf 1102}, 048 (2011), [arXiv:1012.2103 [hep-th]].

\bibitem{schmidt}
  M.~G.~Schmidt and C.~Schubert,
 {\it The Worldline path integral approach to Feynman graphs,}
  hep-ph/9412358.

 M.~G.~Schmidt and C.~Schubert,
{\it Worldline Green functions for multiloop diagrams,}
  Phys.\ Lett.\ B {\bf 331} (1994) 69
  doi:10.1016/0370-2693(94)90944-X
  [hep-th/9403158].

 
\bibitem{dai}
  P.~Dai and W.~Siegel,
 {\it Worldline Green Functions for Arbitrary Feynman Diagrams,}
  Nucl.\ Phys.\ B {\bf 770} (2007) 107
  doi:10.1016/j.nuclphysb.2007.02.004
  [hep-th/0608062].

\bibitem{Bekaert1}
  X.~Bekaert, E.~Joung and J.~Mourad,
 {\it On higher spin interactions with matter,}
  JHEP {\bf 0905} (2009) 126
  [arXiv:0903.3338 [hep-th]].
    
\bibitem{Bekaert2}
  X.~Bekaert, E.~Joung and J.~Mourad,
{\it Effective action in a higher-spin background,}
  JHEP {\bf 1102} (2011) 048
  [arXiv:1012.2103 [hep-th]].

\bibitem{bonezzi}
  R.~Bonezzi,
{\it Induced Action for Conformal Higher Spins from Worldline Path Integrals,}
  Universe {\bf 3} (2017) no.3,  64
  doi:10.3390/universe3030064
  [arXiv:1709.00850 [hep-th]].


\bibitem{BCLPS} L.~Bonora, M.~Cvitan, P.~Dominis Prester, B.~Lima de Souza and
I.~Smoli\'{c},
  {\it Massive fermion model in 3d and higher spin currents,}
  JHEP {\bf 1605} (2016) 072
    [arXiv:1602.07178 [hep-th]].

 \bibitem{BCDGLS} L.~Bonora, M.~Cvitan, P.~Dominis Prester, S.~Giaccari, B.~Lima
de Souza and T.~\v{S}temberga,
{\it One-loop effective actions and higher spins,}
  JHEP {\bf 1612} (2016) 084,
  [arXiv:1609.02088 [hep-th]].

 \bibitem{BCDGS} L.~Bonora, M.~Cvitan, P.~Dominis Prester, S.~Giaccari,  and
T.~\v{S}temberga,
{\it One-loop effective actions and higher spins. II}
  JHEP {\bf 1801} (2018) 080,
  [arXiv:1709.01738 [hep-th]].

 
\bibitem{Savvidy}
  G.~Savvidy,
{\it Non-Abelian Tensor Gauge Fields,}
  Proc.\ Steklov Inst.\ Math.\  {\bf 272}, no. 1, 201 (2011)
  doi:10.1134/S0081543811010196
  [arXiv:1004.4456 [hep-th]].
  doi:10.1007/JHEP05(2017)026
  [arXiv:1701.05772 [hep-th]].
 

 \bibitem{Fronsdal} C.~Fronsdal,
{\it Massless Fields with Integer Spin,} 
  Phys.\ Rev.\ D {\bf 18} (1978) 3624.

J.~Fang and C.~Fronsdal,
{\it Massless Fields with Half Integral Spin,}
  Phys.\ Rev.\ D {\bf 18} (1978) 3630.


\bibitem{FS} 
 D.~Francia and A.~Sagnotti,
  {\it On the geometry of higher spin gauge fields,}
  Class.\ Quant.\ Grav.\  {\bf 20} (2003) S473,
   [Comment.\ Phys.\ Math.\ Soc.\ Sci.\ Fenn.\  {\bf 166} (2004) 165],
   [PoS JHW {\bf 2003} (2003) 005], 
  [hep-th/0212185].
  
D.~Francia and A.~Sagnotti,
{\it Free geometric equations for higher spins,}
  Phys.\ Lett.\ B {\bf 543} (2002) 303,  
  [hep-th/0207002].
 
\bibitem{HZ}
  O.~Hohm and B.~Zwiebach,
 {\it $L_{\infty}$ Algebras and Field Theory,}
  Fortsch.\ Phys.\  {\bf 65} (2017) no.3-4,  1700014
  doi:10.1002/prop.201700014
  [arXiv:1701.08824 [hep-th]].

\bibitem{Linfmath} T.~Lada,
{\it $L_\infty$ algebra representations},
Applied Categorical Structures {\bf 12} (2004) 29-34.

 
 H.~Kajiura and J.~Stasheff,
{\it Homotopy algebras inspired by classical open-closed string field theory,}
  Commun.\ Math.\ Phys.\  {\bf 263} (2006) 553
  doi:10.1007/s00220-006-1539-2
  [math/0410291 [math-qa]].

 
  T.~Lada and J.~Stasheff,
{\it Introduction to SH Lie algebras for physicists,}
  Int.\ J.\ Theor.\ Phys.\  {\bf 32} (1993) 1087
  doi:10.1007/BF00671791
  [hep-th/9209099].

 
T.~Lada and M.~Markl,
{\it Strongly homotopy Lie algebras,}
  hep-th/9406095.

\bibitem{Linfappl}  M.~R.~Gaberdiel and B.~Zwiebach,
{\it Tensor constructions of open string theories. 1: Foundations,}
  Nucl.\ Phys.\ B {\bf 505} (1997) 569
  doi:10.1016/S0550-3213(97)00580-4
  [hep-th/9705038].

 B.~Zwiebach,
{\it Oriented open - closed string theory revisited,}
  Annals Phys.\  {\bf 267} (1998) 193
  doi:10.1006/aphy.1998.5803
  [hep-th/9705241].
 
  G.~Barnich, R.~Fulp, T.~Lada and J.~Stasheff,
{\it The sh Lie structure of Poisson brackets in field theory,}
  Commun.\ Math.\ Phys.\  {\bf 191} (1998) 585
  doi:10.1007/s002200050278
  [hep-th/9702176].

  R.~Blumenhagen, M.~Fuchs and M.~Traube,
{\it W algebras are L$_{\infty}$ algebras,}
  JHEP {\bf 1707} (2017) 060
  doi:10.1007/JHEP07(2017)060
  [arXiv:1705.00736 [hep-th]].

R.~Blumenhagen, M.~Fuchs and M.~Traube,
 {\it On the Structure of Quantum L$_\infty$ algebras,}
  JHEP {\bf 1710} (2017) 163
  doi:10.1007/JHEP10(2017)163
  [arXiv:1706.09034 [hep-th]].

\bibitem{III} to appear.


\bibitem{teleparallel} R.~Aldrovandi and J.~G.~Pereira {\it Teleparallel
gravity. An introduction}, 
Springer Dordrecht Heidelberg New York London 2013, and references therein.


\bibitem{Weinberg} S.~Weinberg, {\it Photons and gravitons in a matrix theory:
derivation of charge conservation and equality of gravitational and inertial
mass}, Phys.Rev. {\bf 135} (1964) B1049.

\bibitem{WW} S.~ Weinberg and E.~Witten, {\it Limits on massless particles},
Phys. Lett. {\bf B96} (1980) 59.

\bibitem{Porrati} M.~Porrati, {\it Universal Limits on massless High-Spin
Particles}, Phys. Rev. {\bf D78} (2008) 065016. [ArXiv: 1108.3078 [hep-th]].

\bibitem{Bekaert3}

 X.\ Bekaert, N.\ Boulanger and S.\ Leclercq, 
{\it Strong obstruction of the Berends-Burgers-van Dam spin-3 vertex,}
J. Phys. {\bf A 43} (2010) 185401 [arXiv:1002.0289 [hep-th]].



\bibitem{Bekaert}
  X.~Bekaert, N.~Boulanger and P.~Sundell,
{\it How higher-spin gravity surpasses the spin two barrier: no-go theorems 
versus yes-go examples,}
  Rev.\ Mod.\ Phys.\  {\bf 84} (2012) 987
  doi:10.1103/RevModPhys.84.987
  [arXiv:1007.0435 [hep-th]].


\bibitem{ColMand} S.~ R.~ Coleman and J.~ Mandula, {\it All possible symmetries
of the s matrix,}
Phys. Rev. {\bf 159} (1967) 1251.

   
\bibitem{benin} P.~ Benincasa and F.~ Cachazo, {\it Consistency Conditions on
the S-Matrix of Massless Particles,}
arXiv:0705.4305 [hep-th].

\bibitem{Ponomarev}
D.~Ponomarev,
{\it Chiral Higher Spin Theories and Self-Duality,}
  JHEP {\bf 1712} (2017) 141
  doi:10.1007/JHEP12(2017)141
  [arXiv:1710.00270 [hep-th]].

\bibitem{CJM}
  E.~Conde, E.~Joung and K.~Mkrtchyan,
 {\it Spinor-Helicity Three-Point Amplitudes from Local Cubic Interactions,'}
  JHEP {\bf 1608} (2016) 040
  doi:10.1007/JHEP08(2016)040
  [arXiv:1605.07402 [hep-th]].

\bibitem{gates} I.~L.~Buchbinder, S.~J.~Gates and K.~Koutrolikos,
 {\it Integer superspin supercurrents of matter supermultiplets,}
  arXiv:1811.12858 [hep-th].


 
\bibitem{VN} P.~van Nieuwenhuizen, {\it On ghost-free tensor Lagrangians and 
linearized gravitation}, Nucl.Phys. {bf B 60} (1973) 478.

\bibitem{SJT}
K.~S.~Stelle, 
{\it Renormalization of higher-derivative quantum gravity}, Phys. Rev. {\bf D 
16} (1977) 953.

J.~Julve and M.~Tonin, {\it Quantum gravity with higher derivative terms }, 
Nuovo Cimento {\bf 46} (1978) 137. 


\bibitem{Steinacker1}
  H.~C.~Steinacker,
 {\it Higher-spin kinematics and no ghosts on quantum space-time in Yang-Mills matrix models,}
  arXiv:1910.00839 [hep-th].



\end{thebibliography}
\end{document}